\documentclass[aps,floatfix,pra,longbibliography,superscriptaddress,reprint,showpacs,10pt,preprintnumbers]{revtex4-1}
\usepackage[utf8]{inputenc}
\usepackage[dvips]{graphicx}
\usepackage{subfigure}
\usepackage{float}
\usepackage{amssymb}
\usepackage{amsmath}  
\usepackage{dsfont}
\usepackage{color}
\usepackage{array}
\usepackage[locale=US]{siunitx}
\usepackage{bm}
\usepackage{mathrsfs}
\usepackage{pifont}
\usepackage{multirow}
\usepackage{tensor}
\usepackage[pdftex,
            pdftitle={SU(2) no gauge links},
            pdfauthor={Stefan Kuehn},
            bookmarks,
            colorlinks,
            linkcolor=black,
            citecolor=magenta,
            menucolor=black,
            urlcolor=blue,
            plainpages=false,
            pdfpagelabels,
            hypertexnames=false]{hyperref}

\newcommand{\dlink}{\ensuremath{d_\mathrm{link}}}
\newcommand{\jmax}{\ensuremath{j_\mathrm{max}}}

\newcommand{\cg}[6]{\ensuremath{{\tensor*{C}{*^{#1\,\,}_{#2\,\,}^{#3\,\,}_{#4\,\,}^{#5}_{#6}}}}}

\newcommand{\tr}{\ensuremath{\text{tr}}}
\newcommand{\ket}[1]{\mbox{$ | #1 \rangle $}}
\newcommand{\bra}[1]{\mbox{$ \langle #1 | $}}
\newcommand{\Id}{\ensuremath{\mathds{1}}}
\newcommand{\Mloc}{\ensuremath{\mathcal{M}_\mathrm{loc}}}

\newcommand{\Sdist}{\ensuremath{S_\mathrm{dist}}}
\newcommand{\Srep}{\ensuremath{S_\mathrm{rep}}}
\newcommand{\Sclass}{\ensuremath{S_\mathrm{class}}}

\begin{document}
\title{Efficient Basis Formulation for (1+1)-Dimensional SU(2) Lattice Gauge Theory: Spectral calculations with matrix product states}
\author{Mari Carmen  Ba\~nuls}
\affiliation{Max-Planck-Institut f\"ur Quantenoptik, Hans-Kopfermann-Straße 1, 85748 Garching, Germany}
\author{Krzysztof Cichy}
\affiliation{Goethe-Universit\"at Frankfurt am Main, Institut für Theoretische Physik, Max-von-Laue-Stra{\ss}e 1, 60438 Frankfurt am Main, Germany}
\affiliation{Faculty of Physics, Adam Mickiewicz University, Umultowska 85, 61-614 Pozna\'{n}, Poland}
\author{J. Ignacio Cirac}
\affiliation{Max-Planck-Institut f\"ur Quantenoptik, Hans-Kopfermann-Straße 1, 85748 Garching, Germany}
\author{Karl Jansen}
\affiliation{NIC, DESY Zeuthen, Platanenallee 6, 15738 Zeuthen, Germany}
\author{Stefan K\"uhn}
\affiliation{Max-Planck-Institut f\"ur Quantenoptik, Hans-Kopfermann-Straße 1, 85748 Garching, Germany}

\date{\today}
\begin{abstract}
We propose an explicit formulation of the physical subspace for a (1+1)-dimensional SU(2) lattice gauge theory, where the gauge degrees of freedom are integrated out. Our formulation is completely general, and might be potentially suited for the design of future quantum simulators. Additionally, it allows for addressing the theory numerically with matrix product states. We apply this technique to explore the spectral properties of the model and the effect of truncating the gauge degrees of freedom to a small finite dimension. In particular, we determine the scaling exponents for the vector mass. Furthermore, we also compute the entanglement entropy in the ground state and study its scaling towards the continuum limit.
\end{abstract}

\preprint{DESY 17-108}

\maketitle
\section{Introduction}
Gauge theories play a central role in our understanding of modern particle physics, with the Standard Model being one of the most prominent examples. In the usual Lagrangian or Hamiltonian formulation, the local gauge symmetry is ensured by introducing additional degrees of freedom (d.o.f.) in the form of a gauge field. However, this also leads to redundant d.o.f.\ in the theory. As the physical observables are strictly gauge invariant, the only relevant subspace is the one spanned by the gauge-invariant states, which is in general much smaller than the full Hilbert space of the theory. Because of the absence of transversal directions for the special case of (1+1) dimensions, the gauge fields are not genuinely independent d.o.f. Therefore, it is possible to remove them by integrating the Gauss law. This long-known fact, to the best of our knowledge, has been explicitly exploited in practice only for the Abelian case of the Schwinger model~\cite{Coleman1976,Melnikov2000,Hamer1997}.

Although this renders (1+1)-dimensional gauge theories seemingly simple, nevertheless, they often cannot be solved analytically, in particular, in the nonperturbative regime. A fundamental tool for the numerical study of gauge models is lattice gauge theory (LGT)~\cite{Wilson1974}. Recently, the tensor network (TN) approach to LGT has proven itself as a promising tool for this task, in particular, in the Hamiltonian formulation. Originally developed in the context of quantum information theory, TNs are efficient ans\"atze for the many-body ground-state wave function as well as low-lying excitations. Besides theoretical progress in developing gauge invariant TNs suitable for LGT~\cite{Tagliacozzo2011,Rico2013,Buyens2013,Silvi2014,Zohar2015,Zohar2015b,Zohar2015c,Tagliacozzo2013,Tagliacozzo2014,Zohar2016}, their power for computing mass spectra~\cite{Banuls2013,Buyens2013,Buyens2014,Buyens2015a,Banuls2016c} and thermal states~\cite{Banuls2015,Banuls2016,Saito2014,Saito2015,Buyens2016} has already been demonstrated. Contrary to the conventional Monte Carlo approach to LGT, methods based on TNs are free from the sign problem~\cite{Troyer2005}, and they enable the study of real-time dynamics~\cite{Buyens2013,Kuehn2015,Pichler2015,Buyens2016b} as well as phase diagrams at nonzero chemical potential~\cite{Silvi2016,Banuls2016a,Banuls2016b,Banuls2016c} for certain gauge models. Moreover, variational TN methods explicitly yield the wave function at the end of the computation and, hence, allow for access to all kinds of (local) observables~\cite{Kuehn2015,Banuls2016a,Banuls2016b,Zapp2017}. Another advantage of TNs is that one can easily study the entanglement structure of the state~\cite{Pichler2015,Buyens2015,Silvi2016}, thus, opening up new possibilities for characterizing LGT problems. A different approach to the Hamiltonian lattice formulation explored during recent years is quantum simulation of gauge theories~\cite{Zohar2011,Banerjee2012,Banerjee2013,Zohar2012,Zohar2013c,Zohar2013a,Zohar2013,Rico2013,Marcos2014,Wiese2014,Zohar2015,Zohar2015a,Mezzacapo2015,Zohar2016b,Zohar2016c,Brennen2016,Muschik2016,Gonzalez-Cuadra2017}. Already experimentally realized for a small system~\cite{Martinez2016}, this route is promising for the future, as it is free from purely numerical limitations.

Despite these encouraging prospects, there are also some limitations. In particular, the Hilbert spaces for the gauge d.o.f.\ are typically infinite dimensional. Hence, in cases for which the gauge fields cannot be integrated out, they typically have to be truncated to a finite dimension to allow for a TN approach or a potential implementation in a quantum simulator. Previous works therefore resorted to the truncation methods from Refs.~\cite{Zohar2015,Milsted2016} to achieve a finite dimension while simultaneously preserving gauge invariance. A different type of finite-dimensional gauge models explored in that context are quantum link models~\cite{Horn1981,Orland1990,Chandrasekharan1997}, where the gauge d.o.f.\ are replaced by discrete spins. However, these truncated models do not necessarily correspond to the continuum theory in the limit of vanishing lattice spacing, or might not have a continuum limit at all~\cite{Brower1999}. 

Here, we address these questions for a (1+1)-dimensional SU(2) lattice gauge theory. In a first step, we show how, starting from a color-neutral basis developed in Ref.~\cite{Hamer1982a}, the gauge d.o.f.\ can be integrated out on a lattice with open boundary conditions (OBC). The resulting formulation allows for truncating the color-electric flux at an arbitrary value. These truncated models can be efficiently addressed with TN. In principle, since the maximum flux on a finite lattice with OBC is upper bounded, this enables an exact treatment of the model. In practice, to solve the model with matrix product states (MPS), we limit the number of color-flux sectors and compute the low-lying spectrum for this family of truncated SU(2) gauge models and investigate truncation effects in a systematic manner. In particular, here we explore such effects in the closing of the mass gaps as we approach criticality, and in the entanglement entropy of the ground state. Recent developments in the context of the holographic principle have suggested a deep connection between entanglement and emergent geometry~\cite{Ryu2006,VanRaamsdonk2010}, and have rekindled interest in understanding the peculiarities of entanglement in gauge theories~\cite{Casini2014,Ghosh2015,Soni2016,VanAcoleyen2016,Aoki2017}. Being especially well suited to compute entanglement entropies, TNs allow us to study how the truncation of the flux alters the entanglement of the vacuum in the approach to the continuum.

In our study, we are interested in various aspects. On the one hand, the basis we develop efficiently describes the physical subspace and can in principle be used with other analytical or numerical methods. Using TNs, we demonstrate the suitability of our formulation for addressing LGT problems as they appear in high-energy and also condensed-matter physics. Because of the vastly reduced number of basis states compared to the full basis, we are able to explore much larger values for the maximum color-electric flux than has been achieved in previous studies of the model~\cite{Kuehn2015,Silvi2016}. On the other hand, the questions we explore are also relevant for quantum simulation of gauge theories. The corresponding, Abelianized, Hamiltonian in our basis is nonlocal, similar to the one recently realized in trapped ions for the Schwinger model~\cite{Martinez2016}. Hence, our formulation might have potential applications for the design of future quantum simulators. Some proposals for quantum simulation of gauge models~\cite{Zohar2011,Banerjee2012,Banerjee2013,Zohar2012,Zohar2013c,Zohar2013a,Zohar2013,Marcos2014,Wiese2014,Zohar2015,Zohar2015a,Mezzacapo2015,Zohar2016b,Zohar2016c,Brennen2016,Gonzalez-Cuadra2017} rely on the representation of the gauge variables by finite-dimensional d.o.f. For those cases, a truncation of the color flux will be required, and our formulation provides a tool for the systematic study of the effect on various observables.

The rest of the paper is organized as follows. In Sec. \ref{sec:model}, we introduce the model we are studying. After a brief review of the color-neutral basis developed in Ref.~\cite{Hamer1982a}, we present our new formulation for systems with OBC where the gauge field is integrated out in Sec. \ref{sec:basis_no_links}. Furthermore, we explain how this formulation readily allows for a truncation of the link Hilbert spaces in a gauge-invariant manner to a finite dimension. In Sec. \ref{sec:results}, we briefly review the MPS methods we are applying and present our results for the low-lying spectrum and the entanglement properties of the ground state while approaching the continuum limit. Finally, we conclude in Sec. \ref{sec:conclusion}.

\section{Model\label{sec:model}}
The model we are studying is a (1+1)-dimensional SU(2) lattice gauge theory. We use a Hamiltonian lattice formulation with Kogut-Susskind staggered fermions~\cite{Kogut1975} in the temporal gauge, given by
\begin{align}
\begin{aligned}
H&=\frac{1}{2a}\sum_{k=1}^{N-1}\sum_{\ell,\ell'=1}^2 \left({\psi_{k}^{\ell}}^\dagger U_{k}^{\ell\ell'}\psi_{k+1}^{\ell'}+\text{H.c.}\right)\\ &
+ m\sum_{k=1}^N\sum_{\ell=1}^2 (-1)^k{\psi_{k}^{\ell}}^\dagger\psi_{k}^{\ell} +\frac{ag^2}{2}\sum_{k=1}^{N-1} \mathbf{J}_{k}^2.
\label{hamiltonian}
\end{aligned}
\end{align}
In the expression above, ${\psi_{k}^\ell}^\dagger$ is a single-component fermionic field creating a fermion of color $\ell$ on site $k$,  $U_{k}^{\ell\ell'}$ acts on the gauge link between sites $k$ and $k+1$, and $\mathbf{J}^2_k$ gives the color-electric energy on the link. The parameter $g$ is the coupling constant, $a$ the lattice spacing, and $m$ the bare fermion mass.

The operators $U^{\ell\ell'}_k$ are SU(2) matrices in the fundamental representation and can be interpreted as rotation matrices. Hence, the Hilbert space for each gauge link is analogous to a quantum rigid rotor with total angular momentum $j$, which can be described in two reference frames, the body-fixed system and the space-fixed (inertial) frame of reference~\cite{Kogut1975}. Consequently, the links can be labeled by the angular momentum $z$ components of the rotor $\ell$, $\ell'$, one corresponding to the body-fixed coordinate system and one corresponding to the space-fixed coordinate system and the total angular momentum $j$ (the same in both reference frames). The angular momentum operators $L^\tau$, $\tau\in\{x,y,z\}$ (for the body-fixed reference frame) and $R^\tau$ (for the space-fixed reference frame) can be interpreted as the left and right electric field on a link and they are related to the color-electric flux energy as $\mathbf{J}_{k}^2 = \sum_\tau L^\tau_kL^\tau_k = \sum_\tau R^\tau_kR^\tau_k$. Hence, the operator for the color-electric flux in this basis is simply a total angular momentum operator with eigenvalues $j(j+1)$, $j=0,1/2,1,\dots$. From the considerations above, a suitable basis for addressing the Hamiltonian Eq. \eqref{hamiltonian} is given by $|n^1,n^2\rangle\otimes |j \ell \ell'\rangle\otimes|n^1,n^2\rangle\otimes\dots$ where $n^{\ell}$ is the fermionic occupation number for color $\ell$. 

The physical states $|\phi\rangle$ of the system have to satisfy the Gauss law $G^\tau_k|\phi\rangle=0$, $\forall k, \tau$, where 
\begin{align}
G^\tau_k =L^\tau_k-R^\tau_{k-1}-Q^\tau_k,
\label{gauss_law}
\end{align}
are the generators for gauge transformations. In the formula above, $Q_k^\tau=\sum_{\ell=1}^2\frac{1}{2}{\psi_k^\ell}^\dagger\sigma^\tau_{\ell \ell'}\psi_k^{\ell''}$ are the components of the non-Abelian charge at site $k$ and $\sigma^\tau$ are the usual Pauli matrices. 

For the following spectral calculations it is convenient to use a dimensionless formulation of Hamiltonian Eq. \eqref{hamiltonian}, $W= xV + W_0$, where 
\begin{align}
V=\sum_{k=1}^{N-1}\sum_{\ell, \ell'=1}^2 \left({\psi_{k}^{\ell}}^\dagger U_{k}^{\ell \ell'}\psi_{k+1}^{\ell'}+\text{H.c.}\right),  \label{hopping_operator} \\
W_0= \mu\sum_{k=1}^N\sum_{\ell=1}^2 (-1)^k{\psi_{k}^{\ell}}^\dagger\psi_{k}^{\ell} +\sum_{k=1}^{N-1} \mathbf{J}_{k}^2.\label{mass_flux_operator}
\end{align}
The adimensional parameters of the problem in units of the coupling $g$ are $x=1/(ag)^2$ and $\mu = 2\sqrt{x} m/g$.  In the strong coupling limit, the hopping term can be neglected and the Hamiltonian can be solved analytically. The gauge-invariant ground state is then simply given by the lattice analog of the Dirac sea corresponding to odd sites occupied by a fermion of each color, empty even sites, and vanishing color flux on the links~\cite{Hamer1982a}:
\begin{align*}
 |\phi_\mathrm{SC}\rangle = |\mathbf{1},\mathbf{1}\rangle\otimes|000\rangle\otimes|\mathbf{0},\mathbf{0}\rangle\otimes|000\rangle\cdots.
\end{align*}
In the formula above, the numbers in bold face represent the fermionic occupation numbers while $|000\rangle$ represents a link carrying no flux.

The basis considered in this paragraph still contains all the information about the color d.o.f.\ and, in particular, states which are not color singlets. However, the eigenstates for any physical observable consist of color-neutral superpositions of basis states satisfying the Gauss law. As we show in the next paragraph, restricting oneself to the physically relevant subspace of these color-neutral superpositions allows for significantly reducing these superfluous d.o.f.

\section{Integrating out the gauge field\label{sec:basis_no_links}}
\subsection{Color-neutral basis}
\begin{figure}[htp!]
\centering
\includegraphics[width=0.47\textwidth]{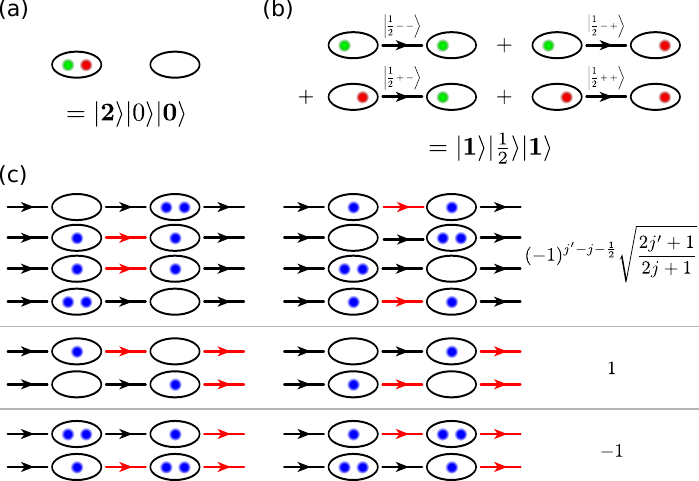}
\caption{(a) Strong coupling configuration with an odd site filled with two fermions, one of every color, and its neighboring empty even site. (b) Resulting color-neutral superposition of four states after applying the operator $V$. Each of the four states has a single fermion per site and a color-electric flux of $j=1/2$ on the intermediate link, with a different combination of $z$ components. The corresponding state in the color-neutral basis for those two cases is written below. (c) Transitions induced by the operator $V$ in the color-neutral basis. The left block represents the possible gauge-invariant starting configurations $|\phi_i\rangle$, the right block the final states $|\phi_f\rangle$ after application of the operator $V$. The arrows show the gauge links, where the black arrows indicate a color-electric flux of $j$ and the red arrows a value of $j'=j\pm 1/2$. The sites are represented by ovals, where the small blue dots indicate the number of fermions sitting on the site. The numbers to the right show the matrix element $\langle \phi_f|V|\phi_i\rangle$.}
\label{fig:sc_ex}
\end{figure}

A first step towards a physical basis was made by Hamer in the context of a strong coupling expansion of the model~\cite{Hamer1982a}. Here, we briefly review the basis formulation developed there. As shown in Refs.~\cite{Hamer1977,Hamer1982a}, the physically relevant states can be generated by applying the operator $V$ from Eq. \eqref{hopping_operator} repeatedly to a certain color-neutral initial state having the desired quantum numbers. This operator has no uncontracted color indices, thus, it can locally only generate or annihilate excitations consisting of color-neutral superpositions of quark-antiquark (antiquark-quark) pairs connected by a color-flux string, as illustrated in Figs. \ref{fig:sc_ex}(a) and \ref{fig:sc_ex}(b). The resulting superposition has a well-defined value of $j$ on the links and fermionic occupation number $n_k=n^1_k+n^2_k$. In particular, applying $V$ to such a color singlet characterized by $n_k$, $n_{k+1}$, and $j_k$ results, in general, again in a superposition of different color singlets with $n_k'$, $n_{k+1}'$ and $j_k'$. In Ref.~\cite{Hamer1982a}, all possible transitions were worked out and the matrix elements of the operator $V$ for each of those vertices are shown in Fig. \ref{fig:sc_ex}(c).

Looking at Eq. \eqref{mass_flux_operator}, one can easily see that the states generated in that manner are eigenstates of the mass term, as it depends only on the total occupation number, and of the color-electric energy, as it only depends on $j$. Thus, $W_0$ acts identically on all those states. Consequently, instead of working in the basis containing the full color information, we can restrict ourselves to a basis formed by those color-singlet states, characterized by the fermionic occupation number of each site and by the color-electric flux $j$ carried by a link,
\begin{align*}
 |\phi\rangle = |\mathbf{n}\rangle\otimes|j\rangle\otimes|\mathbf{n}\rangle\otimes|j\rangle\cdots.
\end{align*}
Here, $\mathbf{n}\in\{0,1,2\}$, as we are not distinguishing between fermions of different colors anymore. Gauss law in this color-singlet basis simply reduces to the fact that the electric flux $j_{k+1}$ can differ only from the one on the previous link by one quantum, if the site is occupied by a single fermion:
\begin{align}
 j_{k+1} = \begin{cases} j_k &\mbox{if } n_{k+1}=0,2 \\ j_k\pm \frac{1}{2} &\mbox{if } n_{k+1}=1. \end{cases} \label{gauss_law_reduced_basis}
\end{align}

In addition to reducing the d.o.f.\ significantly compared to the full basis, the color-singlet basis also offers the possibility to trivially truncate the color-electric flux at a certain value of $\jmax$ in a gauge-invariant manner. Taking into account only states with $j\leq \jmax$ results in a truncated model with Hilbert spaces of dimension $\dlink = 2\jmax + 1$ for the gauge links. Compared to the full basis, where for $\jmax=1/2$, $1$, $3/2$, $2$ one would have link Hilbert spaces of dimension $5$, $14$, $30$, $55$, one only has to deal with spaces of dimension $\dlink = 2$, $3$, $4$, $5$.

\subsection{Removing the gauge fields}
The color-singlet basis reviewed in the previous paragraph still contains redundant information which can be (partially) removed. While the discussion above still applies to both open and periodic boundary conditions, we restrict ourselves from now on to OBC. 

Realizing that due to Eq. \eqref{gauss_law_reduced_basis} the flux $j_{k+1}$ depends only on $j_k$ and $n_{k+1}$, one can reconstruct the color-electric flux at every link from the value $j_0$ at the left boundary, by recursive application of the Gauss law.  However, the case of a site occupied by a single fermion is ambiguous, as the SU(2) non-Abelian Gauss law allows a change of the color-electric flux by $\pm 1/2$. To lift this ambiguity, we use two states $|1_+\rangle$, $|1_-\rangle$ to describe the singly occupied site, which encode if the electric flux is increasing ($|1_+\rangle$) or decreasing ($|1_-\rangle$) with respect to the link to the left. As a result, the basis for a single fermionic site is again four dimensional and consists of the set of states $\{|0\rangle, |1_-\rangle, |1_+\rangle, |2\rangle\}$. The expense of increasing the basis for the fermionic sites by one allows us to integrate out the gauge links from the Hamiltonian, as the $j_k$ can now be reconstructed solely from the fermionic content via
\begin{align}
 \hat{\jmath}_k = j_0 + \sum_{p=1}^k \frac{1}{2}\bigl(|1_+\rangle\langle 1_+|_p - |1_-\rangle\langle 1_-|_p\bigr),
 \label{jk_recursion}
\end{align}
where $j_0$ is the color-flux value at the left boundary. Hence, a suitable basis for a system with $N$ sites is given by
\begin{align}
 |\phi\rangle = |\alpha_1\rangle\otimes|\alpha_2\rangle \otimes \dots \otimes |\alpha_N\rangle
 \label{eq:redbasis}
\end{align}
with $|\alpha_k\rangle\in \{|0\rangle, |1_-\rangle, |1_+\rangle, |2\rangle\}$. Similar to the (Abelian) Schwinger model, in this basis the configuration of the sites uniquely determines the content of the gauge links, thus, effectively \emph{Abelianizing} the model. Additionally, one can immediately see that this construction leads to long-range interactions in Hamer's color-electric energy term. Moreover, as some of the matrix elements for the hopping term in Fig. \ref{fig:sc_ex}(c) also depend on the color-electric flux, the hopping term becomes nonlocal, too (details about the Hamiltonian in this basis are given in Appendix \ref{app:hamiltonian}).

It is instructive to study the dimension of the physical subspace in our basis. Without further constraint, it still contains unphysical states implying negative values of $j_k$; e.g., $|\phi\rangle = |1_-\rangle...$  implies a value of $j_1=-1/2$~\footnote{Notice that the values of $j$ represent the total angular momentum corresponding to the quantum rigid rotor on that link and, thus, are positive.}. For vanishing background field, $j_0=0$, the case on which we focus in our numerical calculations, physical basis states are characterized by a simple condition: the number $l_{k,-}$ of sites with $|1_-\rangle$ up to a site $k$ can never exceed the corresponding number $l_{k,+}$ of sites with $|1_+\rangle$, $l_{k,+}\geq l_{k,-}$ $\forall k=1,\dots,N$. The dimension of the physically relevant subspace fulfilling this condition is given by $4^N(1-\sum_{k=1}^N C_k/ 4^k)$, where $C_k=(2k)!/(k+1)!k!$ is the Catalan number (for details, see Appendix \ref{app:phys_states}). Compared to formulations for the physical subspace for the U(1) case with dynamical fermions~\cite{Hamer1997,Silvi2014}, we observe that the number of basis states in our formulation is exactly the square.

\begin{figure}[htp!]
\centering
\includegraphics[width=0.42\textwidth]{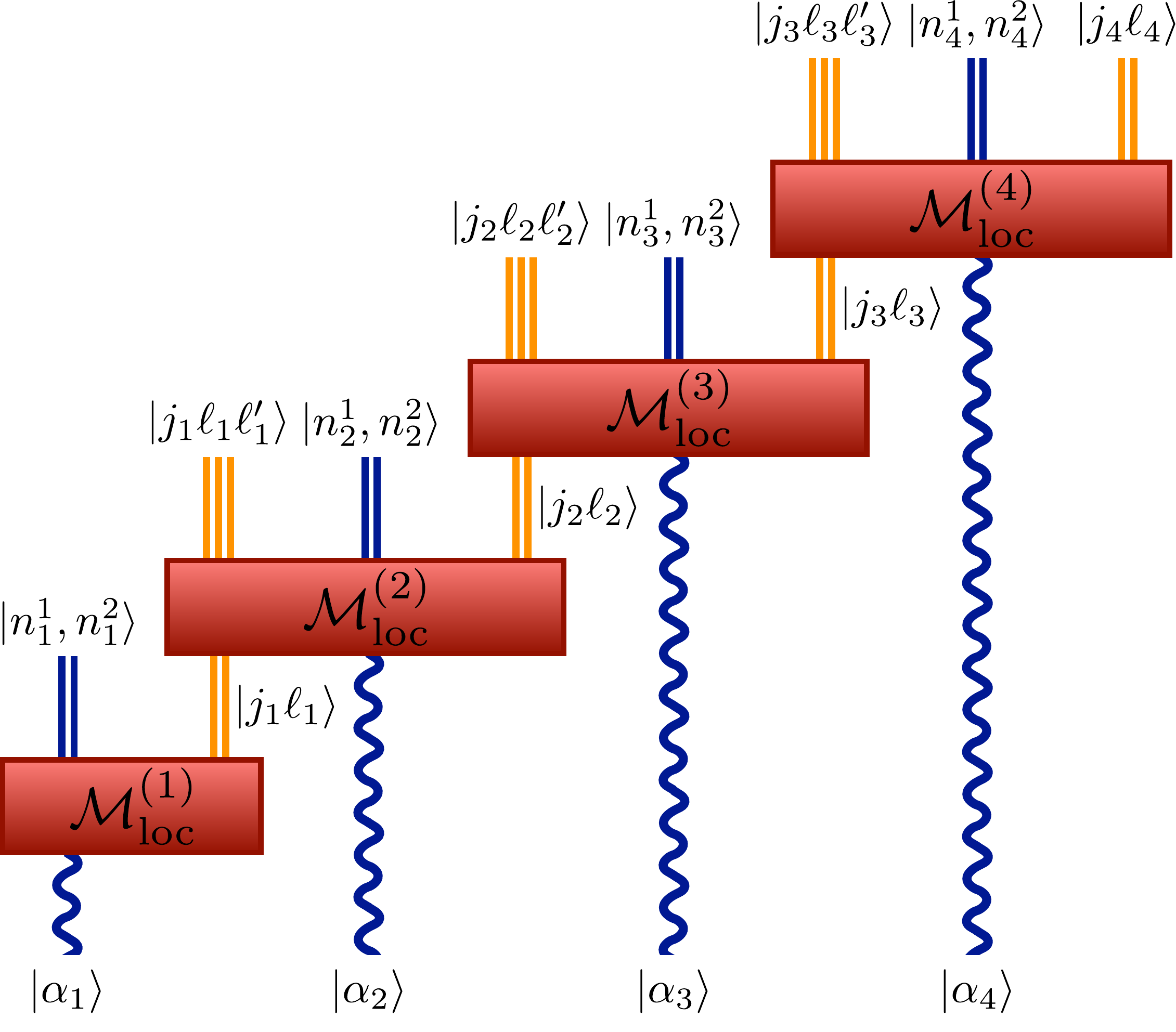}
\caption{Quantum circuit recovering the state in the full basis from our formulation for a system of four sites.}
\label{fig:circuit}
\end{figure}

A simple isometry maps the states of the reduced basis from Eq. \eqref{eq:redbasis} to the full one. This transformation, explicitly shown in Appendix \ref{app:entanglement}, sequentially reconstructs the color flux on each link from the fermion content, and prepares a suitable combination of states in the full basis such that the state is a color singlet. Thus, the map can be written as a sequence of isometries $\Mloc$, schematically shown in  Fig. \ref{fig:circuit}, that act from left to right and take as input one fermionic site and the corresponding incoming flux link, and expand the basis to include the outgoing link, too (see Appendix \ref{app:entanglement} for more details). From a quantum information point of view, this is simply a quantum circuit of depth equal to the system size~\cite{Nielsen2004}. 

Notice that the reduced basis formulation is completely general and contains the entire information about the physical subspace. Hence, it lends itself to any analytical or numerical method. Moreover, one could, in principle, treat arbitrary gauge groups SU($N_c$) with $N_c\geq 2$ in a similar fashion. After obtaining the corresponding matrix elements for the vertices, the gauge field for color-singlet states can likewise be encoded in the fermionic sites.

\section{Results\label{sec:results}}
While the formulation derived in Sec.~\ref{sec:model} can be approached with any numerical method, it is particularly suited for MPS. In order to demonstrate that, we compute the ground state and the vector state of the theory at vanishing background field, $j_0=0$, in the sector of vanishing total charge, $j_N=0$. Moreover, in contrast to Monte Carlo methods, the MPS approach also allows for access to the entanglement entropy, and we can also study the scaling of the von Neumann entanglement entropy in the ground state while approaching the continuum limit.

\subsection{Numerical methods}
For our numerical simulations we use the MPS \emph{Ansatz} with OBC~\cite{Perez-Garcia2007} which is given by
\begin{align*}
 |\phi\rangle = \sum_{i_1,i_2,\dots i_N} A^{i_1}_1A^{i_2}_2 \dots A^{i_N}_N|i_1\rangle \otimes \dots\otimes |i_N\rangle
\end{align*}
for a system of $N$ sites. In the formula above the states $\{|i_k\rangle_{k=1}^d\}$ form a basis for the $d$-dimensional Hilbert space on site $k$, $A^{i_k}_k\in\mathds{C}^{D\times D}$ for $1<k<N$ and $A^{i_1}_1\in\mathds{C}^{1\times D}$ ($A^{i_N}_N\in\mathds{C}^{D\times 1}$).  The bond dimension of the MPS, $D$, determines the number of variational parameters in the \emph{Ansatz} and limits the amount of entanglement that can be present in the state (for detailed reviews about MPS, see, e.g., Refs.~\cite{Verstraete2008,Schollwoeck2011,Orus2014a}).

Using standard methods~\cite{Verstraete2004}, we can determine MPS approximations for the ground-state as well as for low-lying excitations. The ground-state approximation is obtained by variationally minimizing the energy with respect to the tensor $A^{i_k}_k$, while keeping the others fixed and iterating these updates from left to right and back, until the relative change of the energy per sweep is below a certain tolerance $\varepsilon$. The optimal tensor in every step is found by solving an eigenvalue problem~\cite{Stathopoulos2010,Wu2016} for an effective Hamiltonian describing the interaction of site $k$ with its environment. Excited states can be computed by projecting the Hamiltonian onto a subspace orthogonal to each of the previously computed states and then using the same variational method with the projected Hamiltonian~\cite{Banuls2013}. In the continuum, the vector candidate is the lowest-lying zero momentum mesonic excitation of the ground state with charge conjugation quantum number $-1$ and parity $-1$. On the finite lattices with OBC we are working with, however, charge conjugation as well as the momentum are no longer good quantum numbers due to the broken translational invariance. Nonetheless, the remnants of these symmetries allow us to properly identify the vector state (see Appendix~\ref{app:dist_states} for details). 

In addition, to be able to address the problem with MPS, we have to express the Hamiltonian as a matrix product operator (MPO)~\cite{McCulloch2007}. This can be done exactly with the bond dimension of the MPO representation growing linearly with the maximum color-electric flux $\jmax$ present in the system (for details, see Appendix \ref{app:hamiltonian}). In the sector $j_0=0=j_N$, on which we focus in our calculations, $\jmax$ is upper bounded by $N/2 \times 1/2$ for a system with $N$ sites. For the system sizes we are interested in, this would lead to a very large computational effort and, hence, we truncate $\jmax$ to smaller values. In particular, this allows us to explore the effects of truncating the gauge d.o.f.\ to a finite dimension as might be necessary for a potential future quantum simulator~\cite{Zohar2013a,Zohar2013,Zohar2015,Zohar2015a}. Taking advantage of the fact that our basis formulation allows for an efficient truncation, we can easily reach maximum values for the color-electric flux far beyond $\jmax=1/2$ amenable in previous numerical studies with TNs~\cite{Kuehn2015,Silvi2016}.

In order to avoid any influence of the unphysical states, one could in principle implement the symmetries directly at the level of the tensors~\cite{Buyens2013,Silvi2014}. In our calculations, we are targeting only the low-lying spectrum; hence, we choose a simpler approach and remove the unphysical states by adding an energy penalty to the Hamiltonian. Moreover, as we are only interested in the vector excitations, we also remove possible baryonic states from the low-lying spectrum with an additional penalty term. A third penalty ensures that we are in the sector of vanishing total charge (see Appendix \ref{app:penalties} for more details on the penalty terms).

\subsection{Low-lying spectrum}
In order to demonstrate the power of the basis developed in Sec. \ref{sec:model} for TN calculations, we compute the ground state and the vector state of the model for a range of masses, $m/g\in[0.1,1.6]$. To probe for truncation effects, we explore a family of models with maximum color-electric flux $\jmax=1/2,\,1,\,3/2,\,2$. Moreover, we consider for each combination of $(m/g,\jmax)$ several system sizes $N\in[100,200]$ with lattice spacings corresponding to $x\in[50,150]$ to be able to extrapolate to the continuum. Compared to a conventional lattice calculation, we have an additional source of error due to the limited bond dimension that can be reached in the numerical simulations. To control this error, we repeat the calculation for each combination of parameters $(m/g,\jmax,x,N)$ for several bond dimensions $D\in[50,200]$. To estimate the exact ground-state energy values, $E_0(N,x)$, and vector mass gap values, $\Delta_\mathrm{vec}(N,x) = E_{1}(N,x) - E_{0}(N,x)$, we first extrapolate our data to the limit $1/D\to 0$, as illustrated in Figs. \ref{fig:extD_extN}(a) and \ref{fig:extD_extN}(b) (details about the extrapolation procedure are given in Appendix \ref{app:error_estimation}). Subsequently, we can proceed in a standard manner and estimate the continuum values by first extrapolating to the thermodynamic limit and then to the limit of vanishing lattice spacing.

Figures \ref{fig:extD_extN}(c) and \ref{fig:extD_extN}(d) show examples for the extrapolation to the thermodynamic limit for $\jmax=2$. Even for this case with the largest color-electric flux, for which we expect the error due to the finite bond dimension to be most pronounced, the error bars resulting from the extrapolation in $D$ are small and we can obtain precise estimates for the ground-state energy density and vector mass gap in the thermodynamic limit.
\begin{figure}[htp!]
\centering
\includegraphics[width=0.48\textwidth]{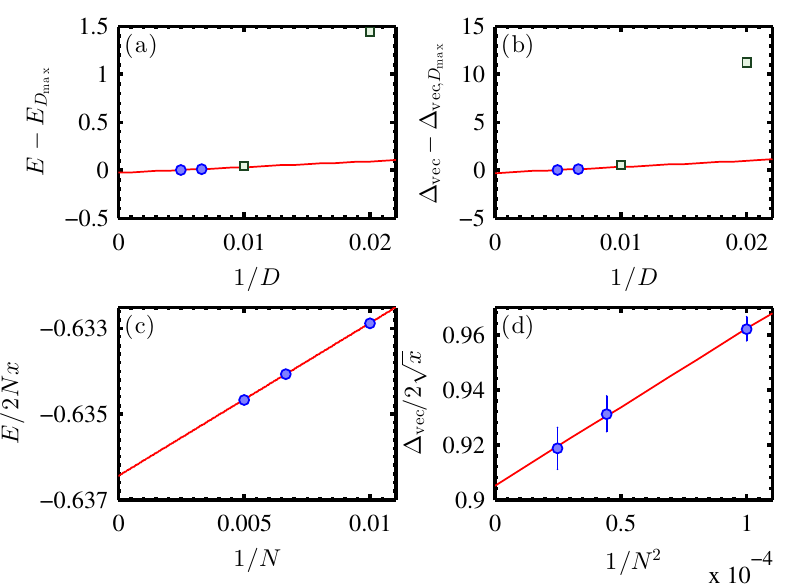}
\caption{Upper row: Extrapolation in bond dimension for the ground-state energy (a) and the vector mass gap (b) for $m/g=0.3$, $\jmax=2$, $N=150$, and $x=150$. The central value is determined with a linear fit through the largest two bond dimensions represented by the blue dots. Lower row: Extrapolation to the thermodynamic limit for the ground-state energy density (c) and the vector mass gap (d) for $\jmax=2$ and $x=150$.}
\label{fig:extD_extN}
\end{figure}

\begin{figure}[htp!]
\centering
\includegraphics[width=0.48\textwidth]{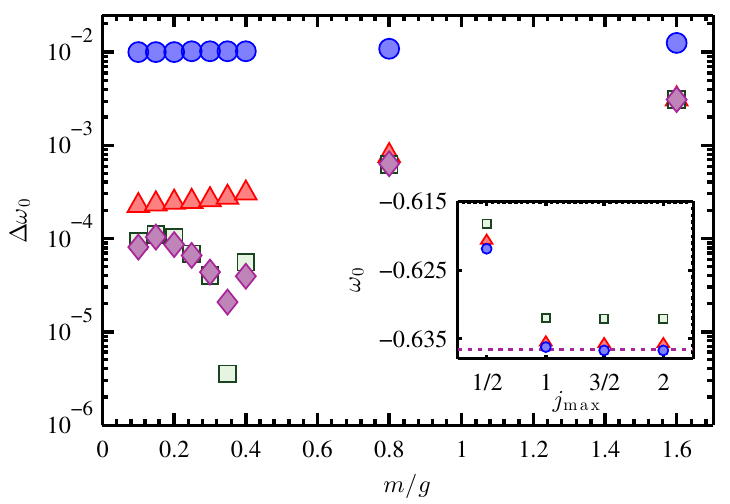}
\caption{Relative deviation of the ground-state energy density with respect to the continuum solution of the full theory $-2/\pi$, $\Delta \omega_0$, as a function of $m/g$. The markers indicate different values for $\jmax$, where blue circles represent $\jmax=1/2$, red triangles $\jmax=1$, green squares $\jmax=3/2$, and magenta diamonds $\jmax=2$. Inset: Ground-state energy density as a function of $\jmax$ for $m/g=0.1$ (blue circles), $m/g=0.8$ (red triangles), and $m/g=1.6$ (green squares). The horizontal dashed line indicates the analytic solution for the ground-state energy, $-2/\pi$, for the full lattice Hamiltonian without truncation in the limit $ag=1/\sqrt{x}\to 0$~\cite{Hamer1982a}. The error bars are smaller than the markers.}
\label{fig:GS_bond}
\end{figure}

In the final step, we extrapolate the data obtained in the previous step to the continuum limit $ ag = 1/\sqrt{x} \to 0$ by fitting a polynomial in $ag$. To estimate our systematic error, we compare different fits up to quadratic order using different ranges of $ag$ (details about the error estimation procedure are given in Appendix \ref{app:error_estimation}).
\begin{figure}[htp!]
\centering
\includegraphics[width=0.48\textwidth]{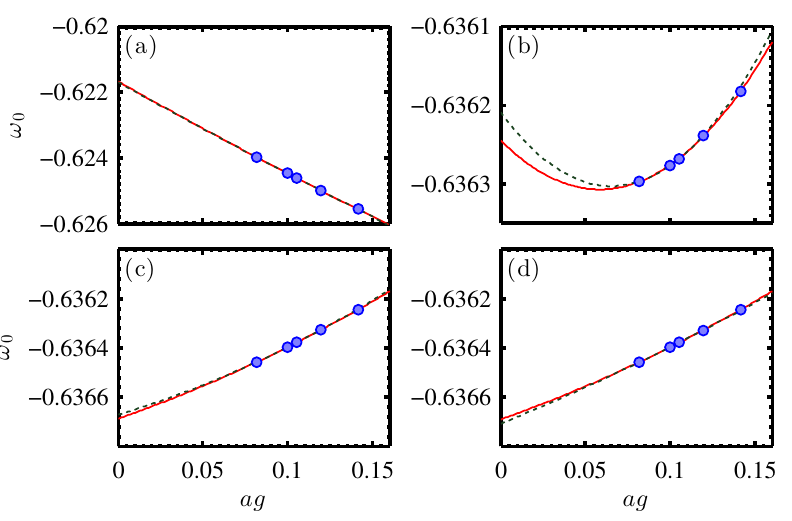}
\caption{Continuum extrapolation for the ground-state energy density for $m/g=0.3$ and $\jmax=1/2$ (a), $\jmax=1$ (b), $\jmax=3/2$ (c), and $\jmax=2$ (d). The red line shows the quadratic fit through all data points used to extract the central value. The dashed green line shows a quadratic fit omitting the largest lattice spacing to estimate the systematic error of the central value.}
\label{fig:extX_gs}
\end{figure}
For the ground state, we observe that, in general, lattice effects are well under control, independently of the truncation, and we can reliably extrapolate to the continuum limit, as can be seen in Fig. \ref{fig:extX_gs}. The values obtained for the ground-state energy density, shown in Fig. \ref{fig:GS_bond}, deviate at most at the percentage level from the result for the continuum theory, even for the simplest nontrivial truncation $\jmax=1/2$. For larger $\jmax$ our data are closer to the analytic solution for the untruncated lattice Hamiltonian in the limit $ag=1\sqrt{x}\to 0$, especially for smaller masses. In particular, the data obtained for the largest two values of $\jmax$ show hardly any difference. The dip around $m/g=0.35$ for $\jmax=3/2,\,2$ is due to the fact that for $m/g<0.4$ our estimates for the ground-state energy are lower than the exact results, whereas for $m/g\geq 0.4$ we obtain values slightly above the analytical prediction. For $\jmax=1/2,\,1$ the values are consistently larger than the exact continuum solution; hence, in these cases there is no dip.

\begin{figure}[htp!]
\centering
\includegraphics[width=0.48\textwidth]{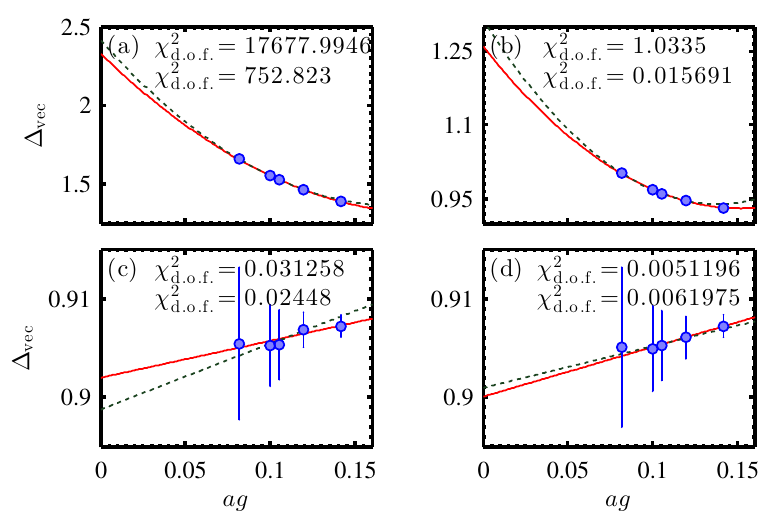}
\caption{Continuum extrapolation for the vector mass gap for $m/g=0.3$ and $\jmax=1/2$ (a), $\jmax=1$ (b), $\jmax=3/2$ (c), and $\jmax=2$ (d). The red line shows the fit used to extract the central value. The dashed green line shows the same fit omitting the largest lattice spacing to estimate the systematic error. The values indicate the $\chi^2_\text{d.o.f.}$ of the two fits, where the upper one corresponds to the red solid line and the lower one to the green dashed line.  Notice the different scales of the $y$ axis between panels (a), (b) and (c), (d), showing that systematic errors are much larger for $\jmax=1/2$ and $1$.}
\label{fig:extX_vec}
\end{figure}

\begin{figure}[htp!]
\centering
\includegraphics[width=0.48\textwidth]{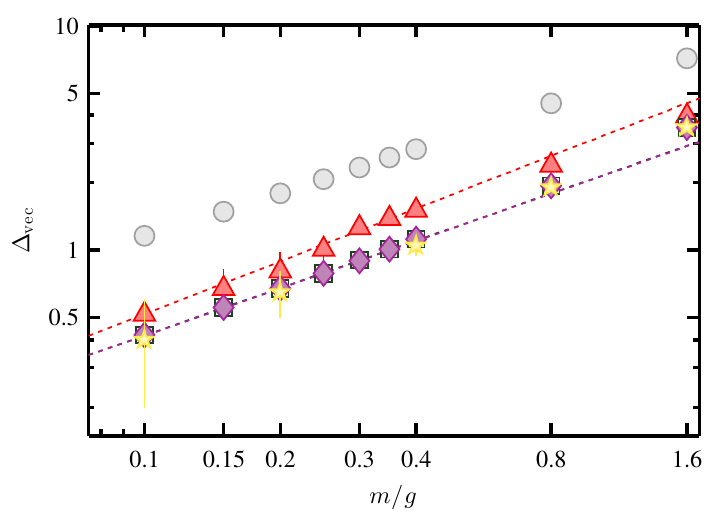}
\caption{Vector state mass gap as a function of $m/g$ for $\jmax=1$ (red triangles), $\jmax=3/2$ (green squares), and $\jmax=2$ (magenta diamonds) on double logarithmic scale. The yellow stars represent the numerical values obtained from the strong coupling expansion~\cite{Hamer1982a}. The dotted lines represent the best fit of the form $\gamma (m/g)^\nu$ to the data obtained on the interval $[0.1;(m/g)_\mathrm{max}]$ with  $0.25\leq (m/g)_\mathrm{max}\leq 0.4$. For completeness, we also show the data for $\jmax=1/2$ (light gray circles), although in this case our lattice spacings do not allow for a reliable estimate.}
\label{fig:VEC_bond}
\end{figure}

Looking at the continuum extrapolation for the vector mass gaps, we observe a noticeably different behavior. While the ground-state energy densities do not show any significantly different behavior for small $\jmax$, the vector mass gaps do, as Fig. \ref{fig:extX_vec} reveals. In particular, for $\jmax=1/2$ our data suggest that higher than quadratic order corrections in $ag$ are still relevant, which results in large $\chi^2_\text{d.o.f.}$ in our fits. With the range of lattice spacings available, we do not seem to control lattice effects well enough to give a reliable error estimate for that case. For $\jmax=1$, our data are reasonably well described with a quadratic function in $ag$ ($\chi^2_\text{d.o.f.}$ around $1$); nevertheless, the error of the continuum estimate is still large. In contrast, for $\jmax=3/2$ and $2$, quadratic corrections in $ag$ are irrelevant and our data are well described by a linear fit in the range of lattice spacings we study (for details, see Appendix \ref{app:error_estimation}). 
The final results obtained after the extrapolation for various masses and truncations are shown in Fig. \ref{fig:VEC_bond}. As the figure reveals, there is a considerable difference between the values obtained for $\jmax=1/2,1,3/2$. Only for values of $\jmax\geq 3/2$ our data agree well with the numerical results obtained by a strong coupling expansion in Ref.~\cite{Hamer1982a}. In particular, for the largest mass, $m/g=1.6$, the data for $\jmax = 3/2,2$ and from Ref.~\cite{Hamer1982a} are already close to the nonrelativistic limit, $m/g\to\infty$, for which the vector mass gap is given by $\Delta_\mathrm{vec}=2m/g$. On the contrary, the values obtained for $\jmax=1/2,\,1$ severely differ from the nonrelativistic prediction.

The fact that we do not recover the continuum limit for the full theory for $\jmax=1/2,\,1$ might be due to several reasons. On the one hand, the truncation to a small value of $\jmax$ might lead to enhanced lattice effects. While the extrapolations to the bulk limit are in general unproblematic for all truncations we study, for the continuum extrapolations we observe that higher than linear order corrections are relevant for small $\jmax$, whereas this is not the case for $\jmax=3/2,\,2$. This could indicate that one would need smaller lattice spacings for truncations to a small color-electric flux, to control lattice effects properly. On the other hand, this might be a hint that the continuum limit for small $\jmax$ does not exist, similar to quantum link models~\cite{Chandrasekharan1997,Brower1999}. These different types of truncated gauge models, in which the gauge links are replaced by spins, are known to approach the continuum limit by dimensional reduction of an extra dimension. 

Contrary to the Abelian Schwinger model, in the SU(2) case the chiral symmetry is restored in the limit $m/g\to0$ and, hence, the vector mass gap goes to zero and the theory becomes critical. From our data we can extract the critical exponent for the vector mass gap, too. In order to obtain the critical exponent, we fit our data to a power law $\gamma(m/g)^\nu$ in the region of small masses $m/g\leq 0.4$, for which the model is still far away from the nonrelativistic limit. The final results obtained for the critical exponents are shown in Table \ref{tab:final_result_exponent}.
\begin{table}[htp!]
\centering
\begin{tabular}{|c|c|}
\hline
$\jmax$ & $\nu$ \\ \hline \hline
$1$    &     $0.781(93)_\text{stat}(65)_\text{syst}$\\ \hline
$3/2$  &     $0.700(29)_\text{stat}(11)_\text{syst}$\\ \hline
$2$    &     $0.700(29)_\text{stat}(12)_\text{syst}$\\ \hline
\end{tabular}
\caption{Critical exponent for various values of $\jmax$. The first error is the fitting error with respect to a $1\sigma$ confidence interval, the second one the systematic error (for more details, see Appendix \ref{app:error_estimation}).}
\label{tab:final_result_exponent}
\end{table}
For $\jmax=3/2,\,2$ our estimates for the critical exponents essentially agree, within error bars, with $2/3$, obtained for the large $N_c$ limit of the model~\cite{Steinhardt1980}. The central value for $\jmax=1$ is not so close to $2/3$; nevertheless, within the relatively large error bars it is still compatible. In the case of the simplest nontrivial truncation, $\jmax=1/2$, a fit to our data yields $0.639$ for the critical exponent. However, due to the large lattice effects in the vector mass gaps, the value is not trustworthy and we cannot reliably estimate the uncertainty, which we expect to be large, too.

These observations also have important implications for a potential future quantum simulator. Our data show that the ground-state energy densities obtained from our family of truncated models already give a good estimate for the one of the full model in the continuum, even with the simplest nontrivial truncation. In contrast, the vector state is much more sensitive to truncation effects. Although with our data it is not possible to fully rule out that for small $\jmax$ one suffers from enhanced lattice effects, they might indicate that the model does not have a proper continuum limit in those cases.

\subsection{Entanglement entropy}
There is renewed interest in understanding the structure of entanglement in the gauge-invariant scenario, motivated in part by a deep connection between entanglement and space-time geometry that has been suggested in the context of the gauge-gravity duality~\cite{Ryu2006,VanRaamsdonk2010}. Our reduced gauge-invariant formulation, together with TN techniques, allows us not only to determine the mass spectrum of the theory, but also to compute the entanglement entropy of the corresponding ground state and to analyze the behavior of this quantity as we approach the continuum limit, as well as to tell to what extent these features are sensitive to a truncation in the gauge d.o.f.

The definition of entanglement entropy for the vacuum of a gauge theory entails some subtleties~\cite{Casini2014,Ghosh2015,Soni2016,VanAcoleyen2016,Aoki2017}. Recently it has been shown that for gauge theories the reduced density matrix (RDM) for a subsystem can be written as a direct sum of terms supported on sectors corresponding to different flux configuration of the boundary links~\cite{Ghosh2015,Soni2016,VanAcoleyen2016}. Specifically for the (1+1)-dimensional case, we can decompose the RDM for the leftmost $L$ sites and links as $\rho=\oplus_j\hat{\rho}_j$, where $j$ labels the flux on the $L$th link. Hence, the von Neumann entropy can be written as
\begin{align*}
S(\rho)&=-\tr\left[\rho\log_2(\rho)\right]\\
       &=-\sum_j p_j \log_2 (p_j) + \sum_j p_j S(\rho_j).
\end{align*}
Here, $\rho_j$ is the (normalized) RDM corresponding to sector $j$ and $p_j=\tr (\hat{\rho}_j)$. For non-Abelian theories, the second term can be further simplified.  For a given sector $j$, the Gauss law fixes the sum of the charge and the incoming flux in the last vertex. As a consequence, $\rho_j$ has a block diagonal structure $\rho_j=\bar{\rho}_j\otimes\mathds{1}_j$, where $\mathds{1}_j$ is the identity on the subspace corresponding to $j$ for the combined incoming flux plus vertex charge. Specifically for SU(2) the identity for the sector $j$ is $(2j+1)$ dimensional, which finally yields
\begin{align}
\begin{aligned}
S(\rho)=&-\sum_j p_j \log_2 (p_j) + \sum_j p_j \log_2(2 j+1)\\
        &+\sum_j p_j S(\bar{\rho}_j).
\end{aligned}
\label{eq:Sfull}
\end{align}

Looking at Eq. \eqref{eq:Sfull}, we can identify three contributions to the entanglement entropy. The last part, $\Sdist:=\sum_j p_j S(\bar{\rho}_j)$, represents the physical entropy which can be distilled from the system by means of local operations and classical communication (LOCC). The first two terms, $\Sclass:=-\sum_j p_j \log_2 p_j$ and $\Srep:=\sum_j p_j \log_2(2 j+1)$, which we, respectively, call the classical and the representation part, have their origin in the Gauss law, implying that the physical subspace is not a direct product of the Hilbert spaces for the links and the sites. They cannot be extracted with LOCC and appear due to the embedding of the physical subspace in the larger space spanned by the basis states discussed in Sec. \ref{sec:model} supporting a tensor product structure~\cite{Ghosh2015,Soni2016,VanAcoleyen2016}. 

With the MPS approach, it is straightforward to access the different contributions to the entropy. Looking at a state in the full basis, we can compute the RDM for the $L$ leftmost sites in the sector $j$ by simply applying a local projector $\Pi_j^{(L)}$ on link $L$ projecting it onto flux $j$. In our reduced formulation the gauge d.o.f.\ are integrated out, but the value of $j$ can be readout from the fermionic content, and the RDM $\tilde{\rho}$ for the corresponding state in our basis is still block diagonal. 
Thus, we can write
\begin{align*}
S(\tilde{\rho})=-\sum_j \tilde{p}_j \log_2(\tilde{p}_j) + \sum_j \tilde{p}_j S(\tilde{\rho_j}),
\end{align*}
where $\tilde{p}_j=\tr(\tilde{\rho}_j)$. Similarly to the full basis, we can obtain $\tilde{\rho}_j$ by applying the corresponding projector $\tilde{\Pi}_j$. In our formulation the model is effectively Abelianized; hence, compared to the full basis, the last term cannot be further simplified and does not give rise to a representation contribution. Since the quantum circuit which takes a state from our basis to the full one does not change $j$, the weights for the different sectors of the reduced density matrices are equal in both bases, $p_j=\tilde{p}_j$. Moreover, since $S(\bar{\rho}_j)=S(\tilde{\rho}_j)$, the distillable entropy is also equal in both bases (see Appendix \ref{app:entanglement} for the formal argument). Thus, we can directly compute the different contributions $\Sdist$, $\Sclass$, and $\Srep$ and therefore the total entropy in the full basis from our formulation. Notice, however, that the calculation in the reduced basis is much more efficient, because of the smaller physical dimensions we need to manipulate.

In (1+1) dimensions, a massive relativistic quantum field theory corresponds to a spin model off criticality in the scaling limit, for which the correlation length in lattice units $\hat{\xi}$ is large. For such a system, the entanglement entropy for the RDM describing half of the system is given by $S\propto (c/6)\log_2(\hat{\xi})$~\cite{Calabrese2004}, where the parameter $c$ is the central charge of the underlying conformal field theory describing the system at the critical point. Taking the continuum limit of the lattice formulation, $ag=1/\sqrt{x}\to 0$, corresponds to approaching the limit of diverging correlation length in lattice units~\cite{Rothe2006}. Consequently, for the full theory without truncation, we expect the entropy for the RDM for half of the system to be logarithmically UV divergent as
\begin{align}
 S = -\frac{c}{6}\log_2\left(ag\right) + c_2\times ag+ c_3 + \mathcal{O}\left((ag)^2\right),
 \label{eq:ccharge}
\end{align}
where $c_2$, $c_3$ are constants and we take into account finite lattice corrections as in Ref.~\cite{Buyens2015}.

\begin{figure}[htp!]
\centering
\includegraphics[width=0.48\textwidth]{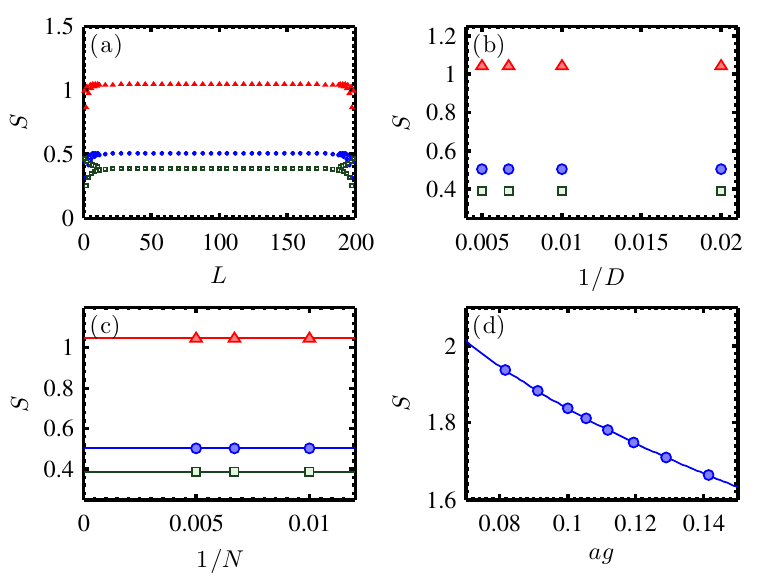}
\caption{(a) The different contributions to the entanglement entropy, $\Sdist$ (blue circles), $\Sclass$ (red triangles), and $\Srep$ (green squares), for the RDM of the leftmost $L$ sites for $N=200$, $D=200$, $m/g=0.8$ and $\jmax=2$. (b) Entropy contributions averaged over four bipartitions close to the center as a function of bond dimension. (c) Averaged entropy contributions for $D=200$, $m/g=0.8$, and $\jmax=2$ as a function of system size. (d) Continuum extrapolation for the total entropy. In panels (b)--(d) the error bars of the data points are smaller than the markers.}
\label{fig:entanglement_entropy}
\end{figure}

\begin{figure}[htp!]
\centering
\includegraphics[width=0.48\textwidth]{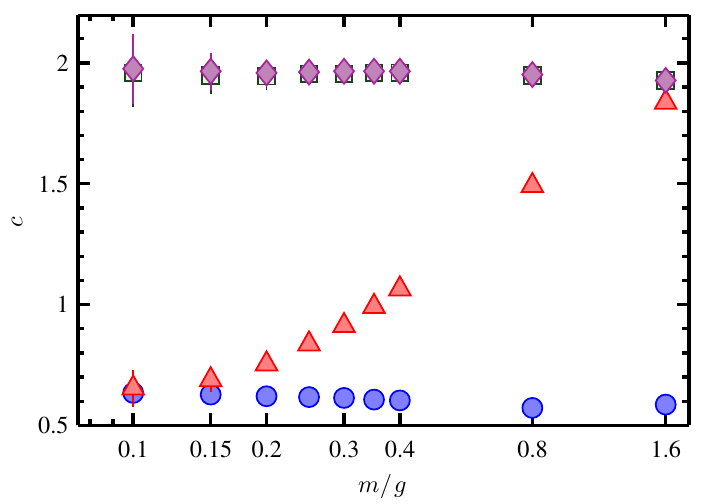}
\caption{Central charges extracted from the scaling of the entanglement entropy as a function of $m/g$ for $\jmax=1/2$ (blue circles), $\jmax=1$ (red triangles), $\jmax=3/2$ (green squares), and  $\jmax=2$ (magenta diamonds). }
\label{fig:central_charges}
\end{figure}

With our numerical data, we can check if the entropy in the ground state for our family of truncated models diverges, too. To this end, we look at the different contributions to the entanglement entropy for a cut along the center of the system in the same range of values for $(\jmax,m/g,x,N,D)$ as in the previous sections and study the scaling of $S$ for $ag=1/\sqrt{x}\to 0$. In general, we observe that none of the different contributions to the entropy show strong finite-size effects for bipartitions that are far away from the boundaries [see Fig. \ref{fig:entanglement_entropy}(a) for an example]. Nevertheless, we may expect an oscillating contribution to the entropy that becomes smaller as the system size increases~\cite{Laflorencie2006,Calabrese2010}. To minimize these effects, we average over the values obtained for $4$ bipartitions around the center to estimate the different entropy contributions for the half-chain. As Figs. \ref{fig:entanglement_entropy}(b) and \ref{fig:entanglement_entropy}(c) indicate, these averaged values are essentially converged in bond dimension and their dependence on the system size is negligible. Hence, we simply take the values obtained for $D=200$ as the central value for every combination of $(\jmax,m/g,x,N)$ and estimate our error as the difference with respect to the value obtained for $D=150$. Additionally, we take into account a systematic error due to the finite precision in our simulations (see Appendix \ref{app:error_estimation} for details). To compensate for residual finite-size effects, we take the weighted average for every $(\jmax,m/g,x)$ for the largest two system sizes available. In a final step, we extrapolate the total entropies, obtained from the sum of the different contributions, to the limit $ag=1/\sqrt{x}\to 0$ by fitting our data to Eq. \eqref{eq:ccharge}. Figure \ref{fig:entanglement_entropy}(d) shows an example for the continuum extrapolation.  
We clearly observe a curvature in the data, thus indicating that the logarithmic term contributes and the entropy is indeed UV divergent. The final results for $c$ for different truncations as a function of the bare fermion mass are shown in  Fig. \ref{fig:central_charges}.
Notice that for the full theory, i.e., without a $\jmax$ truncation, we expect a central charge $c=2$, corresponding to the two Dirac fermions that constitute the independent d.o.f.\ of the theory. We see that our numerical results for $\jmax=3/2,\,2$ follow Eq. \eqref{eq:ccharge} well ($\chi^2_\text{d.o.f.}\ll 1$ in all our fits), and the values for $c$ are close to the one for the full theory (see Fig. \ref{fig:central_charges}). Again, there is hardly any difference between the data for $\jmax=3/2$ and $\jmax=2$. For the smallest two truncations $\jmax=1/2,\,1$, instead, the picture is significantly different. In these cases our data are not very well compatible with a logarithmic divergence for $m/g\geq 0.2$ (resulting in $\chi^2_\text{d.o.f.}\gg 1$ in our fits). In the region of small $m/g$, for which our data follow Eq. \eqref{eq:ccharge} reasonably well, the central charges we obtain differ noticeably from $2$. Hence, although the ground-state energy densities are rather insensitive to the truncation, the entanglement entropies of the same states show another sign that we do not recover the proper continuum limit for a small $\jmax$ truncation. 

\section{Conclusion\label{sec:conclusion}}
We introduce an efficient color-neutral basis for a (1+1)-dimensional SU(2) lattice gauge theory on a finite lattice with OBC. Building on a color-neutral basis used for the strong coupling expansion of the model~\cite{Hamer1977,Hamer1982a}, we show how to remove the gauge d.o.f. Moreover, our formulation allows us to truncate the maximum color-electric flux at a finite value $\jmax$ in a gauge-invariant manner, yielding a family of SU(2) gauge models with finite-dimensional Hilbert spaces, that coincide with a SU(2) lattice gauge theory in the limit $\jmax\to\infty$. While general methods exist to truncate arbitrary gauge models with discrete finite or continuous compact Lie groups to a finite dimension~\cite{Zohar2015}, the truncation achieved for this particular case is a lot more efficient.

The basis we develop is completely general and can in principle be used with any numerical technique. Here, we combine the use of MPS with an efficient truncation for the color-electric flux to explore different limits. Because of the reduced number of d.o.f., we are able to reach values for $\jmax$ far beyond the ones reached in previous numerical work with TNs~\cite{Kuehn2015,Silvi2016}. To systematically study truncation effects, we compute the ground-state energy density, the entanglement entropy in the ground state, the vector mass gap and its critical exponent for a family of truncated SU(2) models with a maximum color-electric flux of $\jmax=1/2,1,3/2,2$. 

We observe that the continuum estimates for the ground-state energy density are rather insensitive to the truncation. Even for the simplest nontrivial truncation, the deviation between the values obtained from our family of truncated model with respect to the continuum result of the full theory is only at the percentage level. Moreover, the results converge quickly with increasing $\jmax$ such that between results for $\jmax=3/2$ and $2$ we observe hardly any difference. 

In contrast, the vector mass gap is a lot more sensitive to a truncation of the maximum color-electric flux. For the simplest nontrivial truncation, $\jmax=1/2$, we cannot control lattice effects in the extrapolations well and reliably estimate the errors. The final value obtained for the mass gap in this case differs significantly from previous numerical results. For $\jmax=1$, lattice effects are becoming smaller, thus allowing for a reliable error estimate. Nevertheless, they are still pronounced, and again the continuum estimate for the vector mass gap is not compliant with previous numerical results within error bars. On the contrary, for $\jmax=3/2,\,2$ the continuum extrapolations are unproblematic and we obtain precise values for the vector mass gap which agree with the ones from Ref.~\cite{Hamer1982a}. Although our data for $\jmax=1/2,\,1$ do not allow us to rule out with certainty that for finer lattices the results would approach the continuum result of the full model, the pronounced lattice effects in those cases might indicate that the continuum limit for these truncated models does not exist,  as it is the case for quantum link models~\cite{Brower1999}. Our findings for the critical exponents for the vector mass gap are essentially in agreement with a calculation in the large $N_c$ limit~\cite{Steinhardt1980}. 

Looking at the scaling of the bipartite entanglement entropy in the ground state towards the continuum limit, we observe similar effects as for the vector mass gap. The central charges for the two simplest nontrivial truncations differ noticeably from the expected value of $2$ for two Dirac fermions and, in particular, for large bare fermion masses our data do not follow the expected logarithmic UV divergence well. On the contrary, for $\jmax=3/2,\,2$ our numerical values show a clear indication of a logarithmic divergence and we find values close to $2$ throughout our entire regime of bare fermion masses we study. Thus, although the ground-state energy densities extracted for $\jmax=1/2,\,1$ are close to the values for the continuum model, this is giving a further indication that for these truncations we do not recover the continuum theory in the limit of vanishing lattice spacing.

In general, our findings for the ground state in the SU(2) case are consistent with those recently reported for the Schwinger model with truncated gauge links~\cite{Buyens2017}. There it was also observed that truncating the maximum electric field to a modest value yields a ground state close to the one of the full model in a wide range of bare fermion masses and lattice spacings. 

In our calculations we target only the vector state besides the ground state. Other masses in the theory, such as scalar mass gap or baryon masses, can be computed in a similar fashion with the basis we develop. Moreover, our formulation is not restricted to static problems and can be used to compute time evolution, thus also giving access to dynamical properties. 

Additionally, the formulation is also potentially suitable for designing future quantum simulators~\cite{Wiese2014,Zohar2015a,Dalmonte2016}. As the number of basis states is drastically reduced with respect to the full basis, and a truncation at a maximum value of $\jmax$ is straightforward, this could allow for simpler experimental realizations compared to previous proposals. Our results also show that in such a simulator one would be able to obtain good estimates for the ground-state energy for the full theory in the continuum, even with the simplest nontrivial truncation for the color-electric flux. However, other quantities as, for example, low-lying excitations or the scaling of entanglement entropy, seem to be more delicate and we only recover the values for the full theory in the continuum, if $\jmax$ is chosen large enough.

In this work, we focus on the (1+1)-dimensional case, for which it is possible to remove the gauge d.o.f.\ completely due the the absence of transversal directions. Although in higher dimensions a complete elimination is not possible, it is feasible to construct formulations that reduce the number of d.o.f.\ as much as possible~\cite{Ligterink2000}. In principle, the physical basis we present here could be modified to realize such a maximal gauge fixing. This would be useful for both the numerical simulations with TN in higher dimensions and the potential quantum simulation of the models, since in both cases the (computational or physical) resources needed would be largely reduced.

\begin{acknowledgments}
We thank E. L\'{o}pez and U.-J. Wiese for helpful discussions. This work was partially funded by SIQS Grant No. FP7 600645. K. C.\ was supported by the Deutsche Forschungsgemeinschaft (DFG), Project No. CI 236/1-1 (Sachbeihilfe). 
\end{acknowledgments}

\appendix

\section{Dimension of the physical subspace\label{app:phys_states}}
Here, we compute the dimension of the physical subspace contained in the basis developed in the main text. For all the following, we focus on a system with $N$ sites and the case of vanishing background field, $j_0=0$. 

\begin{figure}[htp!]
\centering
\includegraphics[width=0.48\textwidth]{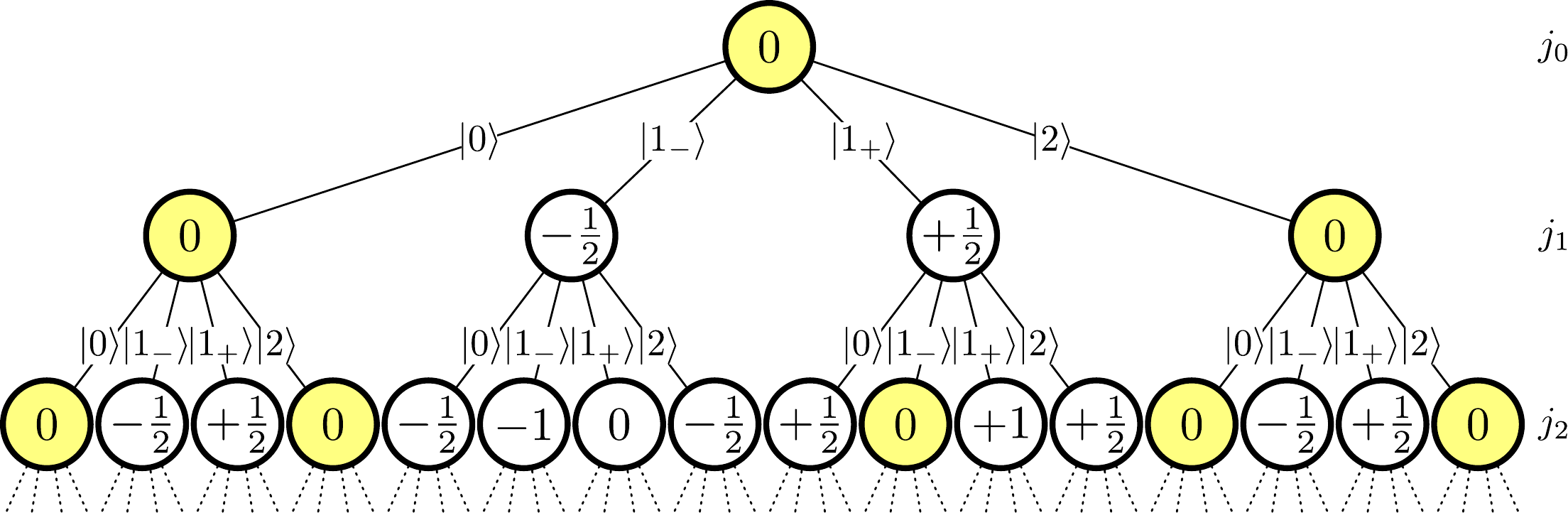}
\caption{The first three levels of the quaternary perfect tree representing the basis states. The vertices represent the color-electric flux indicated by the fermionic states on the edges along the path to each vertex using Eq. \eqref{jk_recursion}. The yellow filled circles represent those vertices, for which one encounters for the first time a negative value, if the path leading to the vertex is continued along the edge carrying $|1_-\rangle$.}
\label{fig:tree}
\end{figure}

As shown in the main text, an arbitrary basis state can be expressed as $|\phi\rangle = \otimes_{k=1}^N |\alpha_k\rangle$, $|\alpha_k\rangle\in\{|0\rangle,|1_-\rangle,|1_+\rangle,|2\rangle\}$. To calculate the dimension of the physical subspace, it is convenient to represent the $4^N$ basis states as directed paths from the root $r$ to one of the leaves in a perfect quaternary tree of depth $N$. The vertices at level $k$ are labeled with the color-electric flux $j_k$ at link $k$, implied by the fermionic states sitting at the edges along the path from the root to the vertex due to Eq. \eqref{jk_recursion} (cf. Fig. \ref{fig:tree}).
Unphysical states now correspond to directed paths from the root to one of the leaves that contain at least one vertex labeled by a negative number. %
Looking at a path starting from the root along vertices with non-negative labels to the vertex $v$ at level $k-1$, a vertex $v'$ with negative label in level $k$ can arise if and only if (i) $v$ is labeled by a $0$ and (ii) from $v$ the path is continued along the edge corresponding to $|1_-\rangle$, thus ending up in a vertex $v'$ labeled by $-1/2$. Hence, the first vertex $v'$ with negative label along a path corresponding to an unphysical state is always one carrying a $-1/2$. Moreover, all paths containing $v'$ necessarily correspond to unphysical states. The number of paths containing $v'$ is simply the number of paths through the perfect quaternary subtree of depths $N-k$ rooted by $v'$, $4^{N-k}$. Consequently, the number of unphysical states is given by $\sum_{k=1}^{N}\bar{t}_k4^{N-k}=\sum_{k=0}^{N-1}\bar{t}_{k+1}4^{N-(k+1)}$. Here, $\bar{t}_k$ is the number of vertices carrying label $-1/2$ at level $k$, for which the path starting from $r$ does not yet pass any other vertex with a negative label. Because of observations (i) and (ii) $\bar{t}_k$ is equivalent to the number of paths $t_{k-1}$ starting from $r$ to a vertex at level $k-1$ with label $0$ that did not yet pass any vertex with a negative label. As we show in the following, $t_{k-1}$ is exactly given by the Catalan number $C_k=(2k)!/(k+1)!k!$. As a result, the number of unphysical states is $\sum_{k=0}^{N-1} t_k 4^{N-(k+1)} = \sum_{k=0}^{N-1} C_{k+1} 4^{N-(k+1)} = \sum_{k=1}^{N} C_k 4^{N-k} $.

To compute the number of paths starting from the root to a vertex at level $k$ with label $0$ that did not yet pass any vertex with negative label, we use to following observations.
\begin{enumerate}
 \item As explained in the main text, the number of edges $l_{k,-}$ passed with $|1_-\rangle$  at any level $k'<k$ must not exceed the ones with $|1_+\rangle$, $l_{k,+}$, to avoid encountering any negative vertices along the path.
 \item Looking at a path from the root with $j_0=0$, to any vertex labeled by $0$ at level $k$, we immediately see that the condition $l_{k,+}=l_{k,-}$ has to be fulfilled in order to compensate for the flux changes induced by $|1_-\rangle$ and $|1_+\rangle$. In particular, this implies that $2l_{k,+}\leq k$ or, equivalently, $l_{k,+}\leq \llcorner k/2\lrcorner$. The other $k-2l_{k,+}$ edges along the path have to carry $|0\rangle$ or $|2\rangle$, as those states do not lead to a flux change while going from one layer to the other. 
\end{enumerate}
The number of paths of length $2l_{k,+}$ which contain at any point at least as many $|1_-\rangle$ as $|1_+\rangle$ is exactly the number of Dyck paths and given by the Catalan number $C_{l_{k,+}}$~\cite{Deutsch1999}. Hence, the number of paths fulfilling conditions 1 and 2 at level $k$ is given by 
\begin{align*}                                                                                                                                                                                                                                                                        
t_k=\sum_{l_{k,+}=0}^{\llcorner k/2 \lrcorner} C_{l_{k,+}}\begin{pmatrix}k\\2l_{k,+}\end{pmatrix}2^{k-2l_{k,+}}=C_{k+1},
\end{align*}
where the factor $\begin{pmatrix}k\\2l_{k,+}\end{pmatrix}$ takes into account the number of ways that the $2l_{k,+}$ symbols $|1_+\rangle$ and $|1_-\rangle$ can be distributed among the $k$ levels and $2^{k-2l_{k,+}}$ the possible ways of filling the remaining edges with $|0\rangle$ and $|2\rangle$. In the last step, we use an identity for the Catalan numbers.

Thus, the dimension of the physical subspace is given by
\begin{align}
 d_{N,\mathrm{phys}}=4^N\left(1-\sum_{k=1}^N\frac{C_k}{4^k}\right).
 \label{phys_subspace}
\end{align}

\begin{figure}[htp!]
\centering
\includegraphics[width=0.48\textwidth]{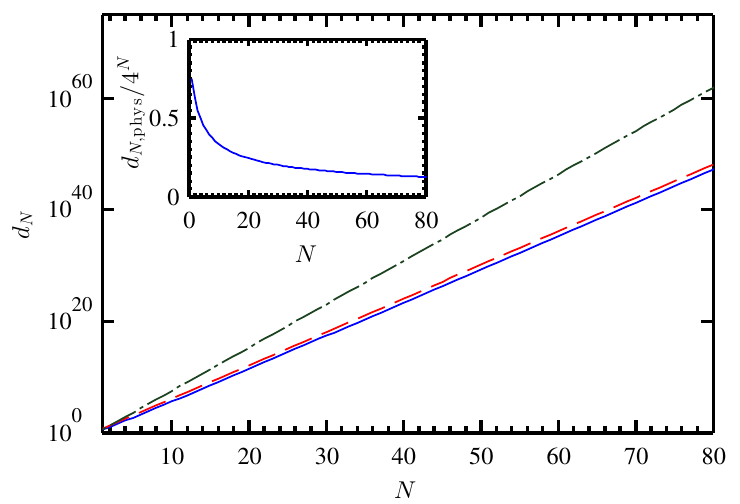}
\caption{Dimension of the physical subspace $d_{N,\mathrm{phys}}$ (blue solid line), the total number of basis states $4^N$ in our formulation (red dashed line) and the dimension of the basis from Ref.~\cite{Hamer1982a}, $3^N(2\jmax+1)^{N-1}$, for the simplest nontrivial truncation $\jmax = 1/2$ (green dash-dotted line) as a function of system size. Inset: Fraction of the physical subspace with respect to the total amount of basis states.}
\label{fig:basis_scaling}
\end{figure}

In Fig. \ref{fig:basis_scaling}, we show a comparison between the scaling of our basis and the basis from Ref.~\cite{Hamer1982a}, which has dimension $d_{N,\mathrm{Hamer}} = 3^N(2\jmax+1)^{N-1}$. 
As the figure reveals, for systems with OBC our basis offers a vast improvement over the color-neutral basis from Ref.~\cite{Hamer1982a} already in the case of the simplest nontrivial truncation $\jmax=1/2$. Even though the fraction of physical states in our basis, $d_{N,\mathrm{phys}}/4^N$, quickly decreases with system size, the total number of states is still exponentially smaller.

\section{Hamiltonian\label{app:hamiltonian}}
In this appendix we show how the terms of the Hamiltonian given in Eqs. \eqref{hopping_operator} and \eqref{mass_flux_operator} can be formulated in the basis presented in Sec. \ref{sec:basis_no_links}. As explained in the main text, the mass term is straightforward as even in the original formulation it depends only on the fermionic occupation number. The color-electric energy term can also readily be formulated in the new basis using Eq. \eqref{jk_recursion}. Hence, $W_0$ is given by
\begin{align}
\begin{aligned}
  W_0 &=\mu\sum_{k=1}^N(-1)^k \hat{n}_k \\
   &+ \sum_{k=1}^{N-1} \left(j_0 + \sum_{l=1}^k \hat{q}_l\right) \left(j_0+ \sum_{l=1}^k \hat{q}_l + 1\right),
\end{aligned}
\label{mass_flux_operator_dimless_newbasis}
\end{align}
where the single-site operators are
\begin{align}
\hat{n}_k &= |1_-\rangle\langle 1_-| + |1_+\rangle\langle 1_+| + 2\,|2\rangle\langle 2|,\label{eq:nk} \\ 
\hat{q}_k &=\frac{1}{2}\left(|1_+\rangle\langle 1_+| - |1_-\rangle\langle 1_-|\right). \label{eq:qk}
\end{align}
From Eq. \eqref{mass_flux_operator_dimless_newbasis} one can see explicitly that integrating out the gauge field leads to nonlocal interactions in the color-electric energy term.

The hopping term can be obtained by translating the possible hopping processes shown in Fig. \ref{fig:sc_ex}(c) in the new basis. The possible hopping processes in the new basis are listed in Table \ref{tab:transitions_no_gauge}. 
\begin{table}[htp!]
\centering
 \begin{tabular}{|ccc|c|}
 \hline
 Initial state & & Final state & Matrix element \\ \hline 
 $|0\rangle\otimes|2\rangle$ & $\to$ & $|1_-\rangle\otimes|1_+\rangle$ &\multirow{8}{*}{$\displaystyle(-1)^{j_k-j_{k-1}-1/2}\sqrt{\frac{2j_k+1}{2j_{k-1}+1}}$} \\
 $|0\rangle\otimes|2\rangle$ & $\to$ &  $|1_+\rangle\otimes|1_-\rangle$ & \\
 
 $|1_-\rangle\otimes|1_+\rangle$ & $\to$ &  $|0 \rangle\otimes|2\rangle$& \\
 $|1_+\rangle\otimes|1_-\rangle$ & $\to$ &  $|0\rangle\otimes|2\rangle$& \\
 
 $|1_-\rangle\otimes|1_+\rangle$ & $\to$ &  $|2\rangle\otimes|0\rangle$& \\
 $|1_+\rangle\otimes|1_-\rangle$ & $\to$ &  $|2\rangle\otimes|0\rangle$& \\
 
 $|2\rangle\otimes|0\rangle$ & $\to$ &  $|1_-\rangle\otimes|1_+\rangle$& \\
 $|2\rangle\otimes|0\rangle$ & $\to$ &  $|1_+\rangle\otimes|1_-\rangle$& \\ \hline 
 
 $|1_- \rangle\otimes|0\rangle$ & $\to$ &  $|0\rangle\otimes|1_-\rangle$& \multirow{4}{*}{$\displaystyle +1$} \\
 $|1_+ \rangle\otimes|0\rangle$ & $\to$ &  $|0\rangle\otimes|1_+\rangle$& \\
 
 $|0 \rangle\otimes|1_- \rangle$ & $\to$ &  $|1_-\rangle\otimes|0\rangle$& \\
 $|0 \rangle\otimes|1_+ \rangle$ & $\to$ &  $|1_+\rangle\otimes|0\rangle$& \\ \hline
 
 $|2\rangle\otimes|1_-\rangle$ & $\to$ &  $|1_-\rangle\otimes|2\rangle$& \multirow{4}{*}{$\displaystyle -1$} \\
 $|2\rangle\otimes|1_+\rangle$ & $\to$ &  $|1_+\rangle\otimes|2\rangle$& \\
 
 $|1_-\rangle\otimes|2\rangle$ & $\to$ &  $|2\rangle\otimes|1_-\rangle$& \\
 $|1_+\rangle\otimes|2\rangle$ & $\to$ &  $|2\rangle\otimes|1_+\rangle$& \\ \hline
 \end{tabular}
\caption{Gauge-invariant transitions induced by the hopping term from Fig. \ref{fig:sc_ex}(c) expressed in the new basis. The value on the right-hand side shows the corresponding matrix elements for the hopping operator.}
\label{tab:transitions_no_gauge}
\end{table}
As the table reveals, the matrix elements for certain transitions depend on the color-electric flux, thus also leading to long-range interactions in the hopping term. The hopping term $V$ can then be expressed in the new basis by defining the operators $O_{i,k}$
\begin{align*}
O_{1,k} &= |0\rangle\langle 1_-|_k,  &O_{2,k} &= |0\rangle\langle 1_+|_k,\\ 
O_{3,k} &= |1_-\rangle\langle 2|_k, &O_{4,k} &= |1_+\rangle\langle 2|_k,
\end{align*}
and translating the possible transitions from Table \ref{tab:transitions_no_gauge} in operator form:
\begin{align}
\begin{alignedat}{10}
 V &= \sum_{k=1}^N && \bigl(&& c_{k-1}^-O_{1,k}^\dagger  O_{4,k+1} +  c_{k-1}^+O_{2,k}^\dagger O_{3,k+1}\bigr.  \\
                     & && +&&c_{k-1}^-O_{3,k}^\dagger O_{2,k+1} +  c_{k-1}^+O_{4,k}^\dagger O_{1,k+1} \\
                     & && + &&O_{1,k}^\dagger O_{1,k+1} +  O_{2,k}^\dagger O_{2,k+1} \\
                      & && -&&\bigl. O_{3,k}^\dagger O_{3,k+1} -  O_{4,k}^\dagger O_{4,k+1} + \text{H.c.} \bigr),\\
\end{alignedat}
\label{hopping_dimless_newbasis}
\end{align}
where $\text{H.c.}$ refers to the Hermitian conjugates of all terms appearing in the formula above. The color-flux-dependent constants $c_k^\pm$ are given by 
\begin{align*}
c_k^+ = \sqrt{\frac{2j_k+2}{2j_k+1}},\quad\quad c_k^- = -\sqrt{\frac{2j_k}{2j_k+1}},
\end{align*}
and are nothing but the matrix elements shown in Table \ref{tab:transitions_no_gauge}. In order to compute these constants, the value for $j_k$ has to be reconstructed from the fermionic occupation number via Eq. \eqref{jk_recursion}.

\section{Penalty terms\label{app:penalties}}
As mentioned in Sec.~\ref{sec:model} and further discussed in Appedix \ref{app:phys_states}, the basis would in principle allow for unphysical states implying negative values for $j$. For our numerical calculations with MPS, we choose to remove those states with appropriate penalty terms shifting unphysical states high enough in the spectrum such that they do not interfere with the low-lying spectrum we are targeting. More specifically, we add the following penalty to the Hamiltonian terms from Eqs. \eqref{mass_flux_operator_dimless_newbasis} and \eqref{hopping_dimless_newbasis} which has a nonvanishing contribution at sites with negative value of $j_k$:
\begin{align*}
 P_1 = \lambda_1 \sum_{k=1}^{N}\Theta\left(-j_0-\sum_{l=1}^{k}\hat{q}_l\right),
\end{align*}
where $\Theta(x)$ is the Heaviside step function. The constant $\lambda_1$ has to be chosen large enough to shift states with a negative value for $j$ high enough in the spectrum such that they do not mix into the low-lying spectrum we are interested in.

Moreover, in our calculations we focus on the vector meson states. To avoid any baryon states, we restrict the total fermion number to the number of sites in the system. This can be easily achieved by adding another penalty term,
\begin{align*}
P_2 = \lambda_2\left(\sum_{k=1}^N \hat{n}_k - N \right)^2,
\end{align*}
to the Hamiltonian, where $\lambda_2$ again has to be chosen large enough to shift the states high enough in the spectrum to prevent them from mixing with the low-lying ones we are interested in.

In addition to that, we are interested in the subspace with $j_0=0=j_N$. Another energy penalty of the form 
\begin{align*}
 P_3=\lambda_3 \left(\sum_{k=1}^N \hat{q}_k\right)^2
\end{align*}
makes it possible to restrict the calculations to that sector, where $\lambda_3$ is again a constant that has to be chosen large enough to penalize unwanted states sufficiently.

For our calculations presented in the main text, we have checked the expectation values for all three penalties and found that they are negligible for $\lambda_i=1000$, $i=1,2,3$.

\section{Distinguishing vector and scalar states\label{app:dist_states}}
\begin{figure}[htp!]
\centering
\includegraphics[width=0.48\textwidth]{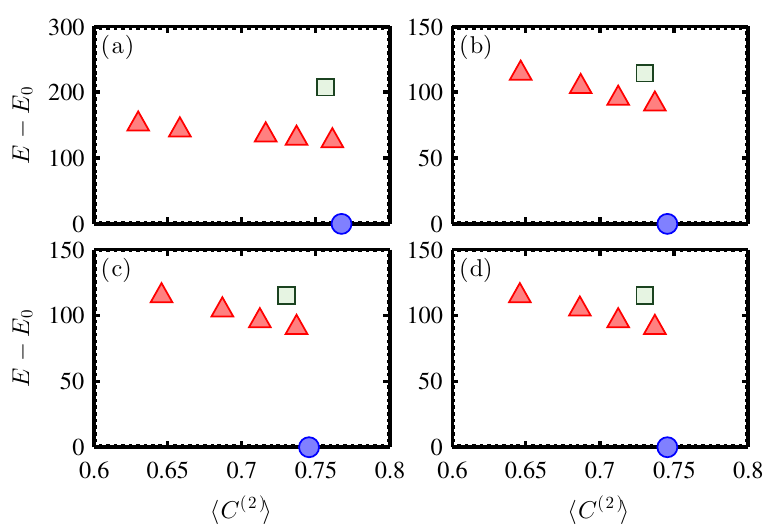}
\caption{Dispersion relation for $m/g=1.6$, $N=50$, $x=150$, $D=50$, $\jmax=1/2$ (a), $\jmax=1$ (b), $\jmax=3/2$ (c), and $\jmax=2$ (d). The blue dot indicates the ground state, the red triangles the vector states, and the green square the scalar candidate.}
\label{fig:dispersion}
\end{figure}

Because of the fact that we are working with finite lattices with OBC, the symmetries which allow for distinguishing between the different meson states are no longer preserved. Nevertheless, following the ideas from Ref.~\cite{Banuls2013}, the remnants make it possible to separate the different type of states. However, in the basis formulation presented in the main text, it is not straightforward to write down a pseudomomentum operator as was done in Ref.~\cite{Banuls2013}. Thus, to identify the zero momentum excitations of the ground state, we use a simpler approach. On a lattice with periodic boundary conditions, the zero momentum states correspond exactly to translational invariant states. For our finite lattice this should still be approximately fulfilled as long as the system size is large enough. Because of the staggered formulation a translational-invariant state should be invariant under a cyclic shift by two lattice sites. To assign a pseudomomentum to our states, we compute the expectation value of the operator $C^{(2)}$, where $C^{(k)}$ describes a cyclic shift by $k$ lattice sites to the right. Moreover, to probe for the charge conjugation number, we proceed again similar to Ref.~\cite{Banuls2013}, and apply a cyclic shift followed by exchanging the two states $|0\rangle \leftrightarrow |2\rangle$, $\sum_{k=1}^N \left(|0\rangle \langle 2|_k + |2\rangle \langle 0|_k\right)C^{(1)}$. While this lattice analog of charge conjugation is not a good quantum number in the case of OBC, the phase of this operator allows for distinguishing between vector candidates (charge conjugation number $-1$) and scalar candidates (charge conjugation number $+1$). For states with charge conjugation number $+1$, we observe phases close to $0$, whereas for states with charge conjugation number $-1$, the observed phase is close to $\pi$. Together with the dispersion obtained from the pseudomomentum operator this allows us to identify the different states as shown in Fig. \ref{fig:dispersion}.
Above the ground state we observe a vector candidate with $\langle C^{(2)}\rangle\approx 1$. Subsequently we discover the momentum excitations of the vector state which are characterized by decreasing $\langle C^{(2)}\rangle$ before we finally obtain a scalar candidate with $\langle C^{(2)}\rangle\approx 1$ again.

\section{Extrapolation procedure and error estimation\label{app:error_estimation}}
Here, we give more details on how we extrapolate our data in bond dimension, system size, and lattice spacing to reach the continuum limit, as well as for the extraction of the critical exponents presented in Table \ref{tab:final_result_exponent} and the central charges.

In a first step, we estimate the exact values for the ground-state energy and the vector mass gap by extrapolating for each combination of $(\jmax,m/g,x,N)$ to the limit $D\to \infty$. As the examples in Figs. \ref{fig:extD_extN}(a) and \ref{fig:extD_extN}(b) show, we plot our data versus $1/D$ and fit a line through the two data points with the largest bond dimension. As an estimate for the exact energy value (vector mass gap), we take the mean value between our value computed with maximum bond dimension, $E_{0,D_\mathrm{max}}(N,x)$, [$\Delta_{\mathrm{vec},D_\mathrm{max}}(N,x)$], and the extrapolated value with infinite bond dimension, $E_{0,D_\infty}(N,x)$ [$\Delta_{\mathrm{vec},D_\infty}(N,x)$]. We estimate the error of the central value in a standard manner by taking half of the difference between these two values $\delta E_\mathrm{fit} = (E_{0,D_\mathrm{max}}(N,x)- E_{0,D_\infty}(N,x))/2$. Additionally to the extrapolation error, the central value obtained also has a systematic error due to the convergence tolerance of $\varepsilon=10^{-6}$ set in the simulations, $\delta E_\mathrm{syst} = \varepsilon E_{0,D_\infty}(N,x)$. Taking into account this error, too, the final error for the ground-state energy (vector mass gap) is given by $\delta E_\mathrm{tot} = \sqrt{\left(\delta E_\mathrm{fit}\right)^2 + \left(\delta E_\mathrm{syst}\right)^2}$.

With the estimates for the exact ground-state energies and the vector mass gaps, we extrapolate to the thermodynamic limit for each combination of $(\jmax,m/g,x)$, where we use the asymptotic behavior up to linear order~\cite{Hamer1997},
\begin{align*}
 \frac{E_0}{2Nx} &\approx \omega_0 + \frac{a}{N} + \mathcal{O}\left(\frac{1}{N^2}\right),\\
 \frac{\Delta_\mathrm{vec}}{2\sqrt{x}} &\approx \omega_1 + \frac{a}{N^2} + \mathcal{O}\left(\frac{1}{N^3}\right),
\end{align*}
and propagate our errors from the previous extrapolation in $D$. As an estimate for the error we take the fitting error with a $1\sigma$ confidence interval.

In a final step, we extrapolate to the continuum limit $ag = 1/\sqrt{x}\to 0$ by fitting a polynomial in $1/\sqrt{x}$ up to second order. In general, we take the value obtained by the lowest order statistically significant fit, which achieves $\chi^2_\text{d.o.f.} < 1$ as the central value. Statistically significant means that the errors for the fit coefficients are smaller than the actual value of the coefficient. In addition to the statistical error of the fit, we estimate the systematic error as the difference between our central value and the next statistically significant fit of the next highest order and/or omitting the largest lattice spacing. For the ground-state energy density this is for all cases a quadratic fit. As an estimate for the systematic error, we take the difference with respect to the value obtained by a quadratic fit omitting the largest lattice spacing, meaning in the region $x\in[70,150]$. 

For the vector mass gap we observe largely enhanced lattice effects for small values of $\jmax$. In particular, for $\jmax=1/2$ quadratic fits have high values for $\chi^2_\text{d.o.f.}$. However, as we have only five different lattice spacings, we cannot take higher-order corrections into account. Thus, we determine the central value with a quadratic fit taking into account all lattice spacings and again estimate the systematic error as the difference with respect to the value obtained by a quadratic fit omitting the largest lattice spacing. Consequently, in this case the error might be heavily underestimated as we are neglecting higher order corrections. For $\jmax=1$, we find that for all $m/g\geq 0.3$ quadratic corrections are sufficient and proceed the same way for estimating the central value and its systematic error as for $\jmax=1/2$. For $m/g=0.25$, we estimate the central value via a linear fit taking into account lattice spacings corresponding to $x\in[90,150]$. The systematic error in this case is estimated as the difference with respect to a quadratic fit in the region $[70,150]$. In the region of smaller masses $m/g\leq 0.2$, we find that both linear and quadratic fits are statistically significant. Hence, we estimate our central value with a linear fit through all available lattice spacings and the systematic error as the difference with respect to a quadratic fit in the same region. For the largest two truncations, $\jmax=3/2,\,2$, the quadratic correction loses significance and thus we estimate our central value in those cases with a linear fit. The systematic error is then determined as the difference with respect to a linear fit discarding the largest lattice spacing, corresponding to $x=50$. The final results obtained for the ground-state energy densities and the vector mass gaps following the procedure described above are listed in Table \ref{tab:final_result}.

\begin{table*}[htp!]
\centering
\begin{tabular}{|c||c|c|c|c||c|c|c|c|}
\hline
{} &  \multicolumn{4}{c||}{Ground state energy density} & \multicolumn{4}{c|}{Vector mass gap} \\    
$m/g$   & $\jmax=1/2$  & $\jmax=1$ & $\jmax=3/2$ & $\jmax=2$ & $\jmax=1/2$  & $\jmax=1$ & $\jmax=3/2$ & $\jmax=2$ \\ \hline
$0.10$ & $-0.621933(20)$ & $-0.636285(28)$ & $-0.636758(31)$ & $-0.636740(84)$ & $1.154(43)$ & $0.521(81)$ & $0.418(20)$ & $0.418(20)$ \\
  $0.15$ & $-0.621878(17)$ & $-0.636275(36)$ & $-0.636782(53)$ & $-0.636773(24)$ & $1.489(46)$ & $0.67(14)$ & $0.557(15)$ & $0.555(15)$ \\
  $0.20$ & $-0.621815(18)$ & $-0.636258(56)$ & $-0.636771(28)$ & $-0.636745(33)$ & $1.788(63)$ & $0.81(16)$ & $0.678(12)$ & $0.675(13)$ \\
  $0.25$ & $-0.621743(25)$ & $-0.636255(45)$ & $-0.636721(18)$ & $-0.636716(22)$ & $2.067(74)$ & $1.01(13)$ & $0.790(10)$ & $0.788(10)$ \\
  $0.30$ & $-0.621668(28)$ & $-0.636239(45)$ & $-0.636679(20)$ & $-0.636684(39)$ & $2.329(85)$ & $1.260(57)$ & $0.9020(85)$ & $0.9001(84)$ \\
  $0.35$ & $-0.621582(31)$ & $-0.636210(40)$ & $-0.636625(17)$ & $-0.636650(22)$ & $2.576(97)$ & $1.389(58)$ & $1.0108(71)$ & $1.0100(79)$ \\
  $0.40$ & $-0.621499(47)$ & $-0.636170(19)$ & $-0.636538(50)$ & $-0.636562(25)$ & $2.82(10)$ & $1.508(59)$ & $1.1180(56)$ & $1.1168(76)$ \\
  $0.80$ & $-0.62070(11)$ & $-0.63548(10)$ & $-0.63570(11)$ & $-0.63569(13)$ & $4.51(11)$ & $2.391(88)$ & $1.9329(28)$ & $1.9322(28)$ \\ 
  $1.60$ & $-0.61823(40)$ & $-0.63200(63)$ & $-0.63205(65)$ & $-0.63205(65)$ & $7.14(17)$ & $3.954(97)$ & $3.5196(11)$ & $3.5191(10)$ \\ \hline 
\end{tabular}
\caption{Ground-state energy densities and vector mass gaps obtained for various values for $m/g$ and $\jmax$. The errors represent the sum in quadrature of the fitting uncertainty with a $1\sigma$ confidence interval and the systematic error.}
\label{tab:final_result}
\end{table*}

The values for the critical exponents for the vector mass gaps are estimated in a similar fashion. Our data for $\jmax=3/2,\,2$, which are close to the ones from Ref.~\cite{Hamer1982a}, reveal that for our largest fermion mass $m/g=1.6$ we are already relatively close to the nonrelativistic limit. Hence, we restrict ourselves to data for small fermion mass to estimate the critical exponent. For each value of $\jmax$, we fit our data to a power law, $\gamma(m/g)^\nu$, for every interval $[0.1,(m/g)_\mathrm{max}]$ with $0.25\leq (m/g)_\mathrm{max}\leq 0.4$. As the central value, we take the fit with the smallest $\chi^2_\mathrm{d.o.f.}$. To estimate our systematic error we take the difference between our central value and the fit giving the most outlying value. The statistical error is again given by $1\sigma$ error bar for the fitting error.

\begin{figure}[htp!]
\centering
\includegraphics[width=0.48\textwidth]{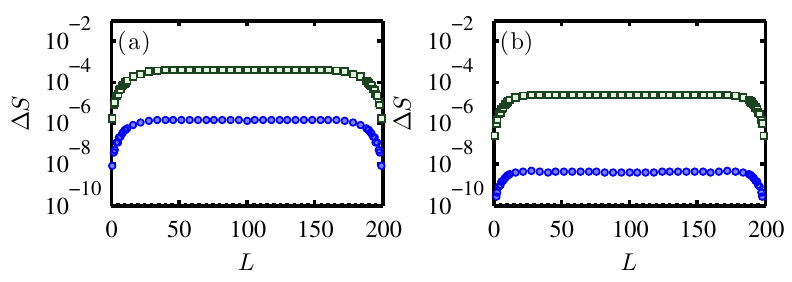}
\caption{Difference in the total entropy for the RDM for the leftmost $L$ sites between simulation results obtained with $\varepsilon=10^{-10}$ and $\varepsilon=10^{-6}$. The panels show the results for $N=200$, $x=50$, $m/g=0.1$ (a) and $m/g=0.3$ (b). The blue dots indicate $\jmax=1/2$, the green squares $\jmax=2$.}
\label{fig:entropy_precision}
\end{figure}

To study the scaling of the entropy towards the continuum limit and obtain an estimate for the central charges, we proceed as described in the main text. First, for $D=200$ and every combination of $(j_\mathrm{max},m/g,x,N)$, we average over the values obtained for 4 bipartitions close to the center for each of the different entropy contributions. To estimate our systematic error, we take the difference with respect to the values obtained with $D=150$. Additionally, our data have another systematic error due to the finite precision in our simulations, which has to be added on top. For the entropies we cannot give the same precise estimates for this systematic error as for the energies. To get, nevertheless, a rough idea of the order of magnitude, we compare results with convergence tolerance $\varepsilon=10^{-6}$ and $\varepsilon=10^{-10}$. Figure \ref{fig:entropy_precision} reveals that even for the largest value of $\jmax$, where we expect the largest differences between these results, it is around $10^{-5}$. Hence, we simply assume a systematic error of $10^{-5}$ due to the finite precision of our simulations in all cases.

\section{Entanglement entropy in the full basis and our formulation\label{app:entanglement}}
In this appendix, we discuss the relation between the different contributions to the von Neumann entropy in our reduced basis formulation and the full basis. In a first step, we show that  one can recover the full basis state from the reduced one by means of an isometry which can be written as a sequence of $N$ local isometries, and thus corresponds to a quantum circuit of depth equal to the system size, $N$. Afterwards, we formally argue why the weights of different sectors, $p_j$, are the same in both bases, and show the explicit relation between the entanglement entropies computed in each basis.

\subsection{Mapping to the full basis}
\begin{figure}[htp!]
\centering
\includegraphics[width=0.43\textwidth]{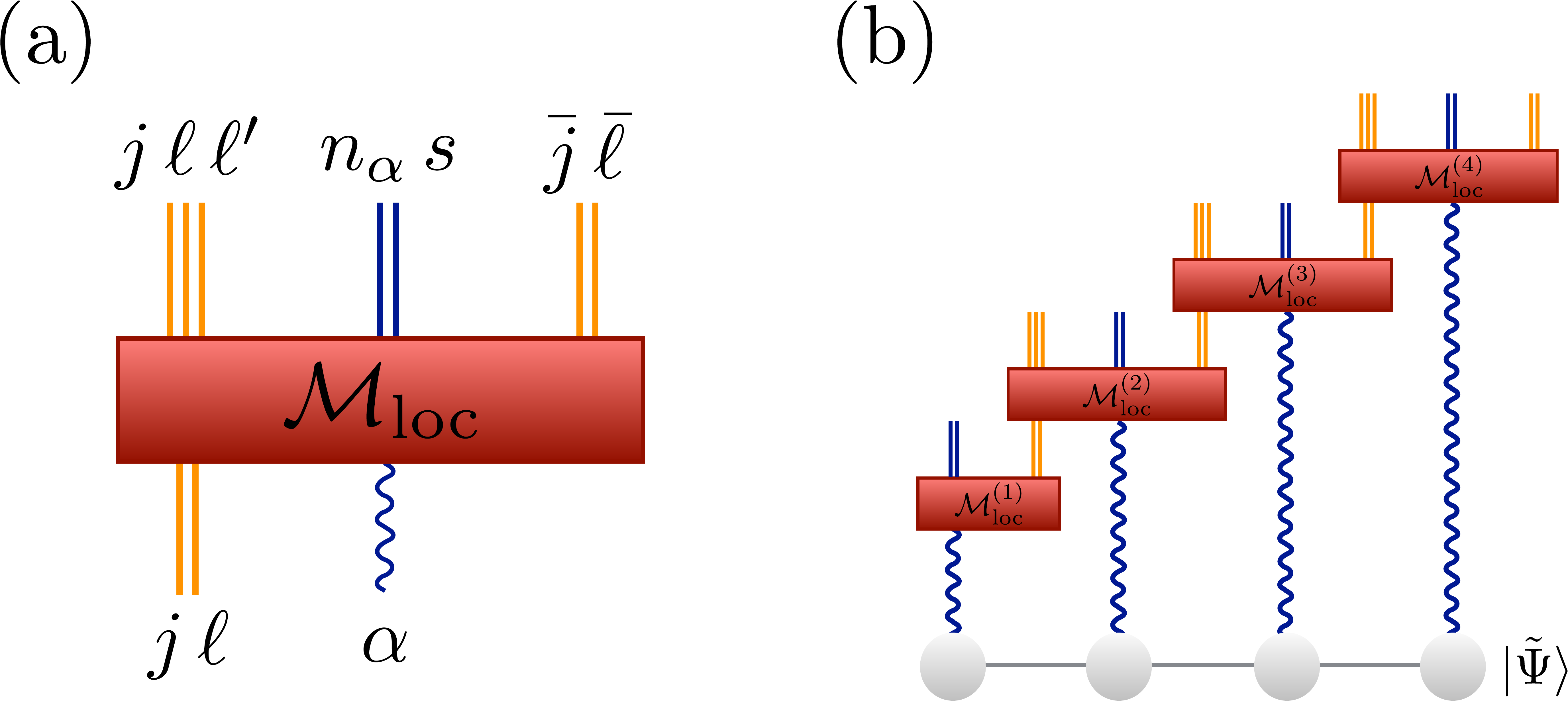}
\caption{(a) Schematic representation of $\Mloc$ that locally maps the reduced basis to the full one. Different line styles are used to indicate the different spaces where $\bar{j} = j+q_{\alpha}$ and $\bar{\ell}=\ell'+s$. (b) Mapping a MPS in the reduced basis to the full one by applying $\mathcal{M}=\Mloc^{(1)}\Mloc^{(2)}\dots \Mloc^{(N)}$.}
\label{fig:Vloc}
\end{figure}

Here we show how the full basis state can be recovered from our reduced basis formulation by sequentially applying a local, isometric map. The map is given by
\begin{align}
\begin{aligned}
 \Mloc = \sum_{j=0,1/2,\dots}^{\jmax}&\sum_{\alpha} \sum_{\ell,\ell'=-j}^j \sum_{s=-|q_{\alpha}|}^{|q_{\alpha}|} \frac{\cg{|q_{\alpha}|}{s}{j}{\ell'}{j+q_{\alpha}}{\ell'+s}}{\sqrt{2(j+q_{\alpha})+1}} \\
 &\times\ket{j\ell\ell';n_{\alpha} s;j+q_{\alpha}, \ell'+s}\bra{j\ell;\alpha}.
\end{aligned}
\label{eq:Vloc}
\end{align}
where $\cg{j_1}{\ell_1}{j_2}{\ell_2}{J}{M} = \langle J,M|j_1,\ell_1;j_2,\ell_2\rangle$ are the usual Clebsch-Gordan coefficients for coupling two angular momenta $j_1$, $j_2$ with $z$ components $\ell_1$, $\ell_2$ to a total angular momentum $J$ with $z$ component $M$. The symbol $\alpha\in\{0,1_-,1_+,2\}$ labels the decorated fermionic occupation. The state $|\alpha\rangle$ is an eigenstate of the operators  $\hat{n}$ from Eq.  \eqref{eq:nk} and $\hat{q}$ from Eq.  \eqref{eq:qk}, with respective eigenvalues $n_{\alpha}$ and $q_{\alpha}$. The states $|n_{\alpha}s\rangle$ correspond to a relabeling of the full basis $|n^1,n^2\rangle$. Different from the main text, we label them with the total occupation number and the $z$ component of the related angular momentum, $n_{\alpha}=n^1+n^2$, $s=(n^1-n^2)/2$, to make the dependence explicit. The effect of the map is to introduce extra d.o.f., $\ell'$ for the incoming link and $j,\ \ell$ on the outgoing link, by simultaneously respecting the proper SU(2) composition rules, which is ensured by the Clebsch-Gordan coefficients [see Fig. \ref{fig:Vloc}(a)].
Note, that for empty or doubly occupied sites, $q_{\alpha}=0$ and the Clebsch-Gordan coefficients are trivial. In the case of singly occupied sites, $|q_{\alpha}|=1/2$ and the spin $1/2$ of the single fermion couples to the angular momentum of the previous link to ensure a color-neutral superposition. The prefactors $1/\sqrt{2(j+q_{\alpha})+1}$ ensure proper normalization of the resulting state and have to be chosen such that $\Mloc^\dagger\Mloc$ is the projector on the physical subspace, as we show in the following. A straightforward calculation yields
\begin{align}
\begin{aligned}
\Mloc^\dagger\Mloc = \sum_{j,\alpha,\ell}&\frac{1}{2(j+q_{\alpha})+1}\left(\sum_{\ell',s}\cg{|q_{\alpha}|}{s}{j}{\ell'}{j+q_{\alpha}}{\ell'+s}^2\right) \\
&\ket{j\ell;\alpha}\bra{j\ell;\alpha}.
\end{aligned}
\label{mmdag}
\end{align}
The sum inside the bracket over the squares of the Clebsch-Gordan coefficients can be simplified as follows:
\begin{align*}
&\sum_{\ell'=-j}^j\sum_{s=-|q_{\alpha}|}^{|q_{\alpha}|}\left(\cg{|q_{\alpha}|}{s}{j}{\ell'}{j+q_{\alpha}}{\ell'+s}\right)^2 \\
&=\sum_{\ell'=-j}^j\sum_{s=-|q_{\alpha}|}^{|q_{\alpha}|}\sum_{t=-(j+q_{\alpha})}^{j+q_{\alpha}}\left(\cg{|q_{\alpha}|}{s}{j}{\ell'}{j+q_{\alpha}}{t}\right)^2\\
&=\sum_{t=-(j+q_{\alpha})}^{j+q_{\alpha}}\left(\sum_{\ell'=-j}^j\sum_{s=-1/2}^{1/2}\left(\cg{|q_{\alpha}|}{s}{j}{\ell'}{j+q_{\alpha}}{t}\right)^2\right)\\
&=\sum_{t=-(j+q_{\alpha})}^{j+q_{\alpha}}1 = 2\left(j+q_{\alpha}\right)+1,
\end{align*}
where in the step from the second to the third line we use that the Clebsch-Gordan coefficients vanish, if the resulting $z$ component differs from the sum of the individual $z$ components and, hence, we can sum over $t$. To arrive at the last line we use the orthogonality relations. Thus, we see that Eq. \eqref{mmdag} is the identity on the physical subspace and $\Mloc$ is indeed an isometry.

Considering a system with $N$ sites, we can recover the full state after fixing the link on the left boundary, $|j_0 \ell_0\rangle$, via a sequential application of $\Mloc$, $\mathcal{M} = \Mloc^{(1)}\Mloc^{(2)}\dots \Mloc^{(N)}$. As sketched in Fig. \ref{fig:Vloc}(b), the sequential application of the map corresponds to a quantum circuit of depth $N$. In all our calculations, we work in the sector $j_0=0$; hence the left electric field necessarily has to vanish and the input left link is $|0 0\rangle$ (thus, not explicitly shown in the figure).

\subsection{Classical part of the entropy}
Here, we show that the weights of sectors with a particular value of $j$ on a certain link $p_j$ are identical in both the full basis and our formulation. For all the following, we assume a system of $N$ sites in a physical state $|\tilde\Psi\rangle$ in the reduced basis, corresponding, in the full one, to $|\Psi\rangle=\mathcal{M}|\tilde{\Psi}\rangle$. We consider the bipartition of the system obtained by cutting at the $L$th gauge link.

\begin{figure}[htp!]
\centering
\includegraphics[width=0.4\textwidth]{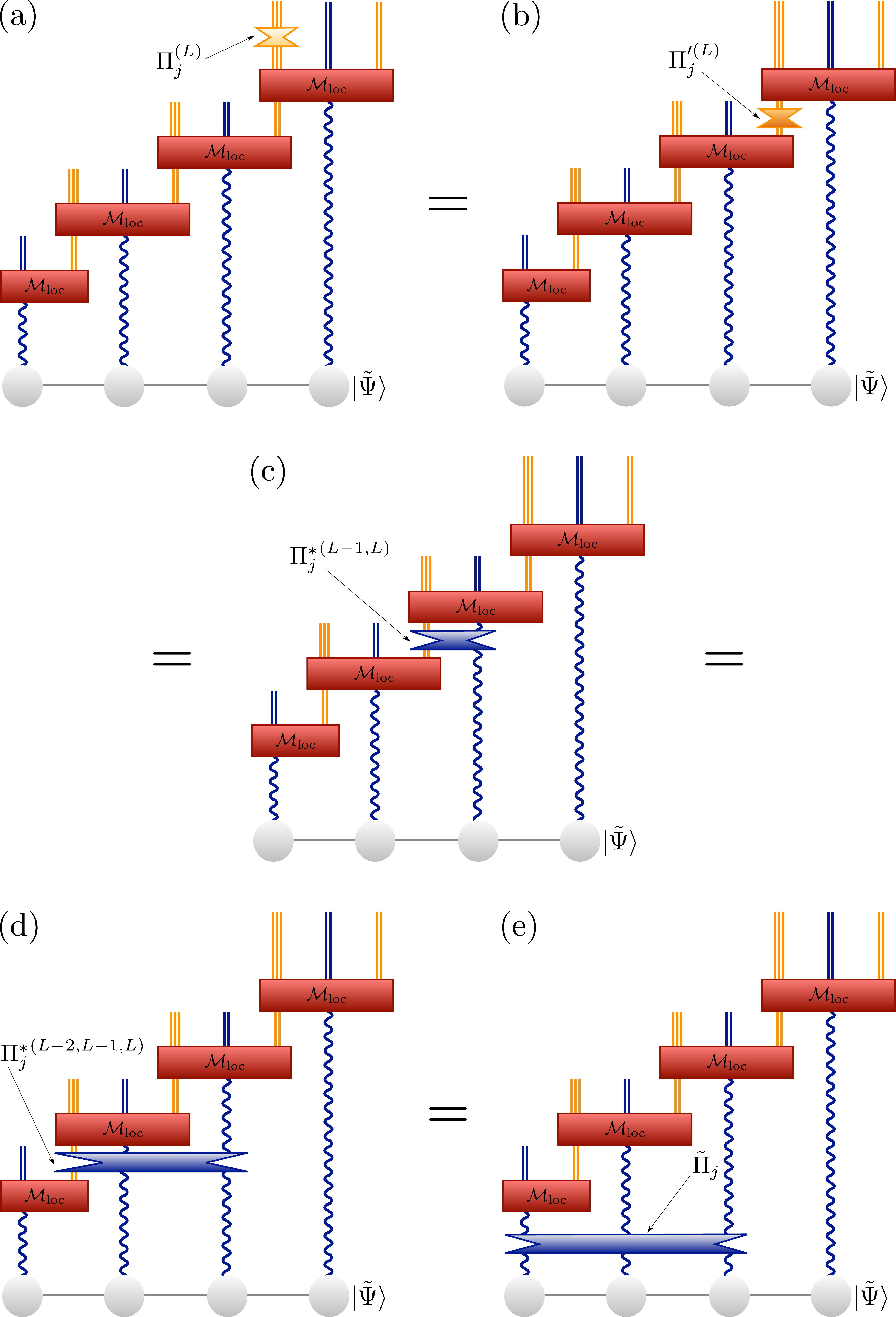}
\caption{(a) The action of a projector onto a given sector of flux $j$, for an intermediate link on the full basis. For physical states, it can be 
pulled through the basis changing isometries (b)-(d), and expressed in the reduced basis as a projector onto the corresponding sum of $q_{\alpha_k}$ values for the vertices to the left of the target link (e).}
\label{fig:const_pj}
\end{figure}

In the full basis, the RDM for the leftmost $L$ sites has block diagonal structure thanks to the gauge constraints, and we can write 
\begin{align}
\begin{aligned}
\rho_j &= \Pi_j \tr_{L+1,\ldots, N}(\ket{\Psi}\bra{\Psi})\Pi_j =\tr_{L+1,\ldots, N}(\Pi_j \ket{\Psi}\bra{\Psi}\Pi_j ) \\
       &=\tr_{L+1,\ldots, N}\left(\Pi_j \mathcal{M}\ket{\tilde{\Psi}}\bra{\tilde{\Psi}}\mathcal{M}^\dagger\Pi_j\right),
\end{aligned}
\label{eq:rdmFull}
\end{align}
where $\Pi_j$ is the projector on total flux $j$ for the $L$th link. In the full basis, this projector acts locally on the link and thus can be written $\Pi_j=\Id_{\mathrm{in}} \otimes \Pi_j^{(L)}\otimes \Id$, where the left factor is the identity on the inner part, i.e., the part where $\rho$ is defined [see Fig. \ref{fig:const_pj}(a)].

The corresponding projection in our basis formulation is given by
\begin{align*}
\tilde{\Pi}_j=\sum_{q_{\alpha_1}+\cdots+q_{\alpha_L}=j}\ket{\alpha_1\ldots\alpha_N}\bra{\alpha_1\ldots\alpha_N},
\end{align*}
where $\sum_{q_{\alpha_1}+\cdots+q_{\alpha_L}=j}$ takes into account all basis states, for which the sum of the eigenvalues $q_{\alpha_k}$, $k=1,\dots,L$, for the single-site operators from Eq. \eqref{eq:qk} is equal to $j$. The corresponding RDM thus reads
\begin{align*}
\tilde{\rho_j} &= \tilde{\Pi}_j \tr_{L+1,\ldots, N}(\ket{\tilde{\Psi}}\bra{\tilde{\Psi}})\tilde{\Pi}_j =\tr_{L+1,\ldots, N}(\tilde{\Pi}_j \ket{\tilde{\Psi}}\bra{\tilde{\Psi}}\tilde{\Pi}_j )
\end{align*}

It turns out, as we show next, that the action of the projector $\Pi_j$ on a certain value of the flux link \emph{commutes} with the isometry that changes the basis, namely, $\Pi_j\mathcal{M}\ket{\tilde{\Psi}} = \mathcal{M}\tilde{\Pi}_j\ket{\tilde{\Psi}}$. This implies the following:
\begin{align*}
p_j&=\tr(\rho_j)=\tr\left ( \Pi_j \mathcal{M}\ket{\tilde{\Psi}} \bra{\tilde{\Psi}}\mathcal{M}^{\dagger} \Pi_j\right)\\
&=\tr\left (\mathcal{M} \tilde{\Pi}_j \ket{\tilde{\Psi}} \bra{\tilde{\Psi}} \tilde{\Pi}_j \mathcal{M}^{\dagger}\right)=\tr\left (\tilde{\Pi}_j \ket{\tilde{\Psi}} \bra{\tilde{\Psi}}\tilde{\Pi}_j\right)=\tilde{p}_j.
\label{eq:pj}
\end{align*}

To prove the statement $\Pi_j\mathcal{M}\ket{\tilde{\Psi}} = \mathcal{M}\tilde{\Pi}_j\ket{\tilde{\Psi}}$, we proceed as sketched in Fig. \ref{fig:const_pj}. The individual steps are justified as follows.
\begin{itemize}
\item{(a)=(b):}
In the full basis, the projector onto a flux value of $j$  for link $L$ is the identity everywhere, except for the local basis of the link, where it acts as $\Pi_j^{(L)}=\sum_{\ell,\ell'=-j}^j\ket{j\, \ell\, \ell'}\bra{j\, \ell\, \ell'}.$ Looking at Eq. \eqref{eq:Vloc}, it is clear that its action on $\Mloc$ just fixes the value of $j$, and thus 
\begin{align*}
(\Pi_j^{(L)}\otimes \Id_{n_{\alpha}s,j\ell}) \Mloc^{(L)}= \Mloc^{(L)} (\Pi_j'^{(L)}\otimes \Id_{\alpha}),
\end{align*}
where $\Pi_j'^{(L)}$ acts to the same effect on the link variables before the isometry, as $\Pi_j'^{(L)}=\sum_{\ell=-j}^j\ket{j\, \ell}\bra{j\, \ell}.$ 
\item{(b)=(c):}
The second step is guaranteed by gauge invariance, in particular, by the form of $\Mloc$ in Eq. \eqref{eq:Vloc}. It is immediate to see that $(\Id\otimes\Pi_j'^{(L)})\Mloc^{(L-1)}= \Mloc^{(L-1)} {\Pi_j^*}^{(L-1,L)},$ where the projector ${\Pi_j^*}^{(L-1,L)}$ acts on the d.o.f.\ $j,\,\ell$ of the $(L-1)$th link and the \emph{decorated} fermion occupation number $\alpha$ of the $L$th vertex, as  
\begin{align*}
{\Pi_j^*}^{(L-1,L)}=\sum_{j'}\sum_{\alpha}\sum_{\ell=-j'}^{j'} \delta_{j'+q_{\alpha},\,j}\ket{j'\,\ell;\,\alpha}\bra{j'\,\ell;\,\alpha}.
\end{align*}
\item{(c)=(d):}
The third equivalence can be formally expressed as $ (\Id \otimes {\Pi_j^*}^{(L-1,L)}) (\Mloc^{(L-1)} \otimes \Id_{\alpha}^{(L)} )=\left(\Mloc^{(L-1)} \otimes \Id_{\alpha}^{(L)}\right ){\Pi_j^*}^{(L-2,L-1,L)}$, with 
\begin{align*}
&{\Pi_j^*}^{(L-2,L-1,L)}= \\
&=\sum_{j' \alpha \alpha'}\sum_{\ell=-j'}^{j'} \delta_{j'+q_{\alpha}+q_{\alpha'},\,j}\ket{j'\,\ell;\,\alpha;\alpha'}\bra{j'\,\ell;\,\alpha;\,\alpha'}.
\end{align*}
On the left-hand side, the only effect of the projector for a sector $\bar{j}$ which acts on $\Mloc\otimes \Id_{\alpha}^{(L)}$ is to restrict the sum over $j$ in Eq. \eqref{eq:Vloc} to values such that $j+q_{\alpha}+q_{\alpha}'=\bar{j}$, which is precisely the effect of the projector on the right-hand side.
\item{(d)=(e):}
Iterating the step above, we can pull the projector through very $\Mloc$ block, until the edge of the chain, where the input $j$ is fixed to $0$ and can be ignored.
\end{itemize}

\subsection{Distillable part of the entropy}
We assume that for a physical state, in the reduced basis, the unnormalized RDM for the $L$ leftmost sites for the sector with (outgoing) flux $j$ has spectrum ${\lambda_{\sigma}}$, i.e.,
\begin{align*}
\tilde{\rho}_j=\sum_{\sigma} \lambda_{\sigma}\ket{\sigma} \bra{\sigma},\quad\quad \tilde{p}_j=\tr(\tilde{\rho}_j)=\sum_\sigma \lambda_\sigma,
\end{align*}
with the eigenvectors $\ket{\sigma}$ supported on the reduced basis for the $L$ leftmost sites.

\begin{figure}[htp!]
\centering
\includegraphics[width=0.46\textwidth]{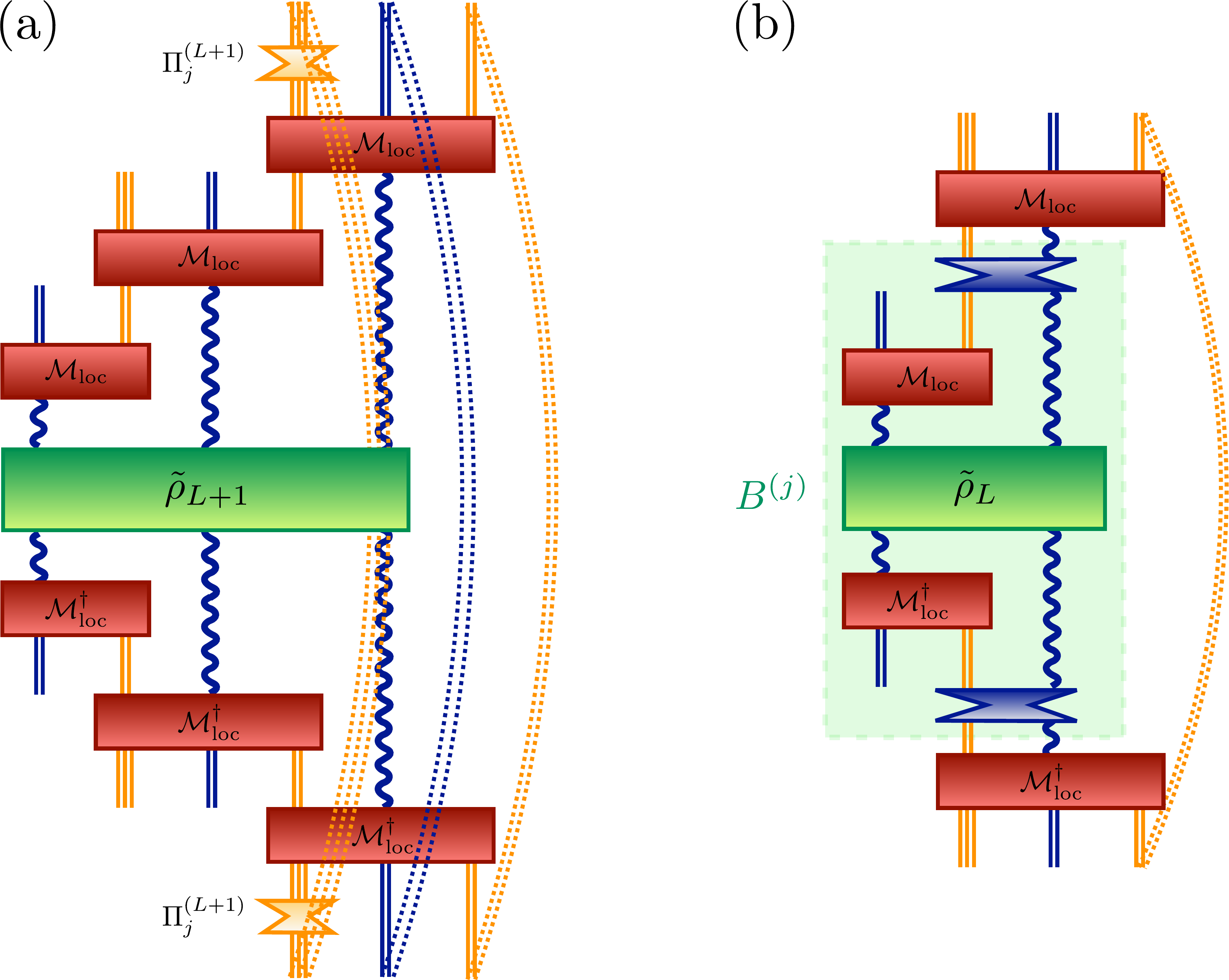} 
\caption{RDM corresponding to a fixed flux sector, $j$, computed in the full basis. (a) $\rho_L^j$ in the full basis. After pushing the projectors through as explained in the text and using that $\Mloc^{(L+1)}$ is an isometry one obtains the equivalent form depicted in panel (b).}
\label{fig:rhoj}
\end{figure}

We can use the relations discussed above to relate the density operators in the full and reduced basis. The RDM in the full basis, given by \eqref{eq:rdmFull}, can be computed from the RDM for $L+1$ sites in the reduced basis, as sketched in Fig. \ref{fig:rhoj}(a), since the isometries acting to the right of the projector cancel out in the trace. Thus, the RDM in the full basis is obtained by first applying the isometry $\mathcal{M}_{L+1}=\Mloc^{(L+1)} \Mloc^{(L)} \cdots \Mloc^{(1)}$ to the RDM for $L+1$ sites in the reduced basis, $\tilde{\rho}_{L+1}$, then projecting onto the sector $j$, 
and finally tracing out the $(L+1)$th site and the gauge d.o.f.\ introduced by $\Mloc^{(L)}$ and $\Mloc^{(L+1)}$. Following the above discussion, the projector can be pushed through the local isometries [see Fig. \ref{fig:rhoj}(b)] so that $\Mloc^{(L+1)}$ cancels and we finally obtain $\rho_j=\tr_{(j\,\ell)_L}(\Mloc^{(L)} B^{(j)} {\Mloc^{(L)}}^{\dagger})$ where $B^{(j)}:=\mathcal{M}_{L-1} \tilde{\rho}_j \mathcal{M}_{L-1}^\dagger$. Since $B^{(j)}$ is simply an isometric transformation of $\tilde{\rho}_j$, it has the same spectrum. We can write
\begin{align*}
B^{(j)}=\sum_{k_\mathrm{in}\ell\alpha,k_\mathrm{in}'\ell'\alpha'} &B_{k_\mathrm{in}\ell\alpha,\,k_\mathrm{in}'\ell'\alpha'} \times\\
        &\ket{k_\mathrm{in};j-q_{\alpha},\,\ell;\alpha}\bra{k_\mathrm{in}';j-q_{\alpha'},\,\ell';\alpha'},
\end{align*}
where we make the d.o.f.\ of the $(L-1)$th link and the $L$th vertex explicit, and represent all the others for the inner part with global indices $k_\mathrm{in}$, $k_\mathrm{in}'$. Then,
\begin{align*}
B_{k_\mathrm{in}\ell\alpha,\,k_\mathrm{in}'\ell'\alpha'} =\sum_{\sigma} \lambda_{\sigma}\,\mathcal{U}_{k_\mathrm{in}\ell\alpha,\sigma} \mathcal{U}_{\sigma,k_\mathrm{in}'\ell'\alpha'}^{\dagger},
\end{align*}
where $\mathcal{U}_{k_\mathrm{in}\ell\alpha,\sigma}=\bra{k_\mathrm{in};j-q_{\alpha},\,\ell;\alpha}\mathcal{M}_{L-1}\ket{\sigma}$.

Applying the local isometry $\Mloc^{(L)}$ and tracing out the $L$th link, we get
\begin{align*}
&\rho_j=\frac{1}{2j+1} \sum_{k_\mathrm{in}\ell\alpha,k_\mathrm{in}'\ell'\alpha'} B_{k_\mathrm{in}\ell\alpha,\,k_\mathrm{in}'\ell'\alpha'}\times\\
&\sum_{s=-|q_{\alpha}|}^{|q_{\alpha}|}\sum_{s'=-|q_{\alpha'}|}^{|q_{\alpha'}|}\sum_{r=-j}^{j} \cg{|q_{\alpha}|}{s}{j-q_{\alpha}}{r-s}{j}{r}\, \cg{|q_{\alpha'}|}{s'}{j-q_{\alpha'}}{r-s'}{j}{r} \times \\
&\quad\ket{k_\mathrm{in};j-q_{\alpha},\,\ell,\,r-s;n_{\alpha}\,s} \bra{k_\mathrm{in}';j-q_{\alpha'},\,\ell',\,r-s';n_{\alpha'}\,s'}.
\end{align*}

We observe that defining
\begin{align*}
\ket{v(k_\mathrm{in};jr;\alpha \ell)}=\sum_{s=-|q_{\alpha}|}^{|q_{\alpha}|} &\cg{|q_{\alpha}|}{s}{j-q_{\alpha}}{r-s}{j}{r} \\
&\ket{k_\mathrm{in};j-q_{\alpha},\,\ell,\,r-s;n_{\alpha}\,s},
\end{align*}
we obtain a set of orthogonal vectors (with respect to all the labels), corresponding to changing the basis of the $(L-1)$th link and the $L$th vertex to a basis of total angular momentum. As a result, if we use the unitary $\mathcal{U}_{k_\mathrm{in}\ell \alpha,\sigma}$ to express the internal d.o.f.\ in the diagonal basis, 
\begin{align*}
\ket{w(\sigma;jr)}=\sum_{k_\mathrm{in} \ell q}\mathcal{U}_{k_\mathrm{in} \ell \alpha,\sigma}^* \ket{v(k_\mathrm{in};jr;\alpha \ell)},
\end{align*}
the resulting vectors are also orthogonal, and
\begin{align}
\rho_j=&\frac{1}{2j+1} \sum_{\sigma} \sum_{r=-j}^j \lambda_{\sigma} \ket{w(\sigma;jr)} \bra{w(\sigma;jr)}.
\end{align}
Thus $\rho_j$ consists of $2j+1$ blocks, with identical spectrum $\{\lambda_{\sigma}/(2j+1)\}$. Consequently, the entropy is given by
\begin{align*}
S(\rho_j)&=-\sum_{r=-j}^j\sum_{\sigma}\frac{\lambda_{\sigma}}{2j+1} \log_2 \frac{\lambda_{\sigma}}{2j+1}  \\
&=-\sum_{\sigma}\lambda_{\sigma} \log_2 \frac{\lambda_{\sigma}}{2j+1} \\
&=\tilde{p}_j\log_2(2j+1) -\sum_{\sigma} \lambda_{\sigma} \log_2 \lambda_{\sigma} \\
&=\tilde{p}_j\log_2(2j+1) +S(\tilde{\rho}_j).
\end{align*}

Putting this result together with the fact that each $j$ sector has the same weight in the reduced and the full representations, $p_j=\tilde{p}_j$, we find that the relation between the entropies is
\begin{align}
S(\rho)&=-\sum_j p_j \log_2 (p_j) + \sum_j p_j S(\rho_j) \nonumber \\
&=-\sum_j p_j \log_2 (p_j) + \sum_j p_j  (\log_2(2j+1) +S(\tilde{\rho}_j)) \nonumber \\
&= \sum_j p_j  \log_2(2j+1) + S(\tilde{\rho}).
\end{align}

\bibliography{Papers_MPQ_converted}

\begin{thebibliography}{83}%
\makeatletter
\providecommand \@ifxundefined [1]{%
 \@ifx{#1\undefined}
}%
\providecommand \@ifnum [1]{%
 \ifnum #1\expandafter \@firstoftwo
 \else \expandafter \@secondoftwo
 \fi
}%
\providecommand \@ifx [1]{%
 \ifx #1\expandafter \@firstoftwo
 \else \expandafter \@secondoftwo
 \fi
}%
\providecommand \natexlab [1]{#1}%
\providecommand \enquote  [1]{``#1''}%
\providecommand \bibnamefont  [1]{#1}%
\providecommand \bibfnamefont [1]{#1}%
\providecommand \citenamefont [1]{#1}%
\providecommand \href@noop [0]{\@secondoftwo}%
\providecommand \href [0]{\begingroup \@sanitize@url \@href}%
\providecommand \@href[1]{\@@startlink{#1}\@@href}%
\providecommand \@@href[1]{\endgroup#1\@@endlink}%
\providecommand \@sanitize@url [0]{\catcode `\\12\catcode `\$12\catcode
  `\&12\catcode `\#12\catcode `\^12\catcode `\_12\catcode `\%12\relax}%
\providecommand \@@startlink[1]{}%
\providecommand \@@endlink[0]{}%
\providecommand \url  [0]{\begingroup\@sanitize@url \@url }%
\providecommand \@url [1]{\endgroup\@href {#1}{\urlprefix }}%
\providecommand \urlprefix  [0]{URL }%
\providecommand \Eprint [0]{\href }%
\providecommand \doibase [0]{http://dx.doi.org/}%
\providecommand \selectlanguage [0]{\@gobble}%
\providecommand \bibinfo  [0]{\@secondoftwo}%
\providecommand \bibfield  [0]{\@secondoftwo}%
\providecommand \translation [1]{[#1]}%
\providecommand \BibitemOpen [0]{}%
\providecommand \bibitemStop [0]{}%
\providecommand \bibitemNoStop [0]{.\EOS\space}%
\providecommand \EOS [0]{\spacefactor3000\relax}%
\providecommand \BibitemShut  [1]{\csname bibitem#1\endcsname}%
\let\auto@bib@innerbib\@empty
\bibitem [{\citenamefont {Coleman}(1976)}]{Coleman1976}%
  \BibitemOpen
  \bibfield  {author} {\bibinfo {author} {\bibfnamefont {S.}~\bibnamefont
  {Coleman}},\ }\href
  {http://www.sciencedirect.com/science/article/pii/0003491676902803}
  {\bibfield  {journal} {\bibinfo  {journal} {Ann. Phys. (N.Y.)}\ }\textbf
  {\bibinfo {volume} {101}},\ \bibinfo {pages} {239} (\bibinfo {year}
  {1976})}\BibitemShut {NoStop}%
\bibitem [{\citenamefont {Melnikov}\ and\ \citenamefont
  {Weinstein}(2000)}]{Melnikov2000}%
  \BibitemOpen
  \bibfield  {author} {\bibinfo {author} {\bibfnamefont {K.}~\bibnamefont
  {Melnikov}}\ and\ \bibinfo {author} {\bibfnamefont {M.}~\bibnamefont
  {Weinstein}},\ }\href {\doibase 10.1103/PhysRevD.62.094504} {\bibfield
  {journal} {\bibinfo  {journal} {Phys. Rev. D}\ }\textbf {\bibinfo {volume}
  {62}},\ \bibinfo {pages} {094504} (\bibinfo {year} {2000})}\BibitemShut
  {NoStop}%
\bibitem [{\citenamefont {Hamer}\ \emph {et~al.}(1997)\citenamefont {Hamer},
  \citenamefont {Weihong},\ and\ \citenamefont {Oitmaa}}]{Hamer1997}%
  \BibitemOpen
  \bibfield  {author} {\bibinfo {author} {\bibfnamefont {C.~J.}\ \bibnamefont
  {Hamer}}, \bibinfo {author} {\bibfnamefont {Z.}~\bibnamefont {Weihong}}, \
  and\ \bibinfo {author} {\bibfnamefont {J.}~\bibnamefont {Oitmaa}},\ }\href
  {\doibase 10.1103/PhysRevD.56.55} {\bibfield  {journal} {\bibinfo  {journal}
  {Phys. Rev. D}\ }\textbf {\bibinfo {volume} {56}},\ \bibinfo {pages} {55}
  (\bibinfo {year} {1997})}\BibitemShut {NoStop}%
\bibitem [{\citenamefont {Wilson}(1974)}]{Wilson1974}%
  \BibitemOpen
  \bibfield  {author} {\bibinfo {author} {\bibfnamefont {K.~G.}\ \bibnamefont
  {Wilson}},\ }\href {\doibase 10.1103/PhysRevD.10.2445} {\bibfield  {journal}
  {\bibinfo  {journal} {Phys. Rev. D}\ }\textbf {\bibinfo {volume} {10}},\
  \bibinfo {pages} {2445} (\bibinfo {year} {1974})}\BibitemShut {NoStop}%
\bibitem [{\citenamefont {Tagliacozzo}\ and\ \citenamefont
  {Vidal}(2011)}]{Tagliacozzo2011}%
  \BibitemOpen
  \bibfield  {author} {\bibinfo {author} {\bibfnamefont {L.}~\bibnamefont
  {Tagliacozzo}}\ and\ \bibinfo {author} {\bibfnamefont {G.}~\bibnamefont
  {Vidal}},\ }\href {\doibase 10.1103/PhysRevB.83.115127} {\bibfield  {journal}
  {\bibinfo  {journal} {Phys. Rev. B}\ }\textbf {\bibinfo {volume} {83}},\
  \bibinfo {pages} {115127} (\bibinfo {year} {2011})}\BibitemShut {NoStop}%
\bibitem [{\citenamefont {Rico}\ \emph {et~al.}(2014)\citenamefont {Rico},
  \citenamefont {Pichler}, \citenamefont {Dalmonte}, \citenamefont {Zoller},\
  and\ \citenamefont {Montangero}}]{Rico2013}%
  \BibitemOpen
  \bibfield  {author} {\bibinfo {author} {\bibfnamefont {E.}~\bibnamefont
  {Rico}}, \bibinfo {author} {\bibfnamefont {T.}~\bibnamefont {Pichler}},
  \bibinfo {author} {\bibfnamefont {M.}~\bibnamefont {Dalmonte}}, \bibinfo
  {author} {\bibfnamefont {P.}~\bibnamefont {Zoller}}, \ and\ \bibinfo {author}
  {\bibfnamefont {S.}~\bibnamefont {Montangero}},\ }\href {\doibase
  10.1103/PhysRevLett.112.201601} {\bibfield  {journal} {\bibinfo  {journal}
  {Phys. Rev. Lett.}\ }\textbf {\bibinfo {volume} {112}},\ \bibinfo {pages}
  {201601} (\bibinfo {year} {2014})}\BibitemShut {NoStop}%
\bibitem [{\citenamefont {Buyens}\ \emph
  {et~al.}(2014{\natexlab{a}})\citenamefont {Buyens}, \citenamefont {Haegeman},
  \citenamefont {Van~Acoleyen}, \citenamefont {Verschelde},\ and\ \citenamefont
  {Verstraete}}]{Buyens2013}%
  \BibitemOpen
  \bibfield  {author} {\bibinfo {author} {\bibfnamefont {B.}~\bibnamefont
  {Buyens}}, \bibinfo {author} {\bibfnamefont {J.}~\bibnamefont {Haegeman}},
  \bibinfo {author} {\bibfnamefont {K.}~\bibnamefont {Van~Acoleyen}}, \bibinfo
  {author} {\bibfnamefont {H.}~\bibnamefont {Verschelde}}, \ and\ \bibinfo
  {author} {\bibfnamefont {F.}~\bibnamefont {Verstraete}},\ }\href {\doibase
  10.1103/PhysRevLett.113.091601} {\bibfield  {journal} {\bibinfo  {journal}
  {Phys. Rev. Lett.}\ }\textbf {\bibinfo {volume} {113}},\ \bibinfo {pages}
  {091601} (\bibinfo {year} {2014}{\natexlab{a}})}\BibitemShut {NoStop}%
\bibitem [{\citenamefont {Silvi}\ \emph {et~al.}(2014)\citenamefont {Silvi},
  \citenamefont {Rico}, \citenamefont {Calarco},\ and\ \citenamefont
  {Montangero}}]{Silvi2014}%
  \BibitemOpen
  \bibfield  {author} {\bibinfo {author} {\bibfnamefont {P.}~\bibnamefont
  {Silvi}}, \bibinfo {author} {\bibfnamefont {E.}~\bibnamefont {Rico}},
  \bibinfo {author} {\bibfnamefont {T.}~\bibnamefont {Calarco}}, \ and\
  \bibinfo {author} {\bibfnamefont {S.}~\bibnamefont {Montangero}},\ }\href
  {\doibase 10.1088/1367-2630/16/10/103015} {\bibfield  {journal} {\bibinfo
  {journal} {New J. Phys.}\ }\textbf {\bibinfo {volume} {16}},\ \bibinfo
  {pages} {103015} (\bibinfo {year} {2014})}\BibitemShut {NoStop}%
\bibitem [{\citenamefont {Zohar}\ and\ \citenamefont
  {Burrello}(2015)}]{Zohar2015}%
  \BibitemOpen
  \bibfield  {author} {\bibinfo {author} {\bibfnamefont {E.}~\bibnamefont
  {Zohar}}\ and\ \bibinfo {author} {\bibfnamefont {M.}~\bibnamefont
  {Burrello}},\ }\href {\doibase 10.1103/PhysRevD.91.054506} {\bibfield
  {journal} {\bibinfo  {journal} {Phys. Rev. D}\ }\textbf {\bibinfo {volume}
  {91}},\ \bibinfo {pages} {054506} (\bibinfo {year} {2015})}\BibitemShut
  {NoStop}%
\bibitem [{\citenamefont {Zohar}\ and\ \citenamefont
  {Burrello}(2016)}]{Zohar2015b}%
  \BibitemOpen
  \bibfield  {author} {\bibinfo {author} {\bibfnamefont {E.}~\bibnamefont
  {Zohar}}\ and\ \bibinfo {author} {\bibfnamefont {M.}~\bibnamefont
  {Burrello}},\ }\href {\doibase 10.1088/1367-2630/18/4/043008} {\bibfield
  {journal} {\bibinfo  {journal} {New J. Phys.}\ }\textbf {\bibinfo {volume}
  {18}},\ \bibinfo {pages} {043008} (\bibinfo {year} {2016})}\BibitemShut
  {NoStop}%
\bibitem [{\citenamefont {Zohar}\ \emph {et~al.}(2015)\citenamefont {Zohar},
  \citenamefont {Burrello}, \citenamefont {Wahl},\ and\ \citenamefont
  {Cirac}}]{Zohar2015c}%
  \BibitemOpen
  \bibfield  {author} {\bibinfo {author} {\bibfnamefont {E.}~\bibnamefont
  {Zohar}}, \bibinfo {author} {\bibfnamefont {M.}~\bibnamefont {Burrello}},
  \bibinfo {author} {\bibfnamefont {T.~B.}\ \bibnamefont {Wahl}}, \ and\
  \bibinfo {author} {\bibfnamefont {J.~I.}\ \bibnamefont {Cirac}},\ }\href
  {\doibase 10.1016/j.aop.2015.10.009} {\bibfield  {journal} {\bibinfo
  {journal} {Ann. Phys. (Amsterdam)}\ }\textbf {\bibinfo {volume} {363}},\
  \bibinfo {pages} {385 } (\bibinfo {year} {2015})}\BibitemShut {NoStop}%
\bibitem [{\citenamefont {Tagliacozzo}\ \emph {et~al.}(2013)\citenamefont
  {Tagliacozzo}, \citenamefont {Celi}, \citenamefont {Zamora},\ and\
  \citenamefont {Lewenstein}}]{Tagliacozzo2013}%
  \BibitemOpen
  \bibfield  {author} {\bibinfo {author} {\bibfnamefont {L.}~\bibnamefont
  {Tagliacozzo}}, \bibinfo {author} {\bibfnamefont {A.}~\bibnamefont {Celi}},
  \bibinfo {author} {\bibfnamefont {A.}~\bibnamefont {Zamora}}, \ and\ \bibinfo
  {author} {\bibfnamefont {M.}~\bibnamefont {Lewenstein}},\ }\href {\doibase
  10.1016/j.aop.2012.11.009} {\bibfield  {journal} {\bibinfo  {journal} {Ann.
  Phys.}\ }\textbf {\bibinfo {volume} {330}},\ \bibinfo {pages} {160 }
  (\bibinfo {year} {2013})}\BibitemShut {NoStop}%
\bibitem [{\citenamefont {Tagliacozzo}\ \emph {et~al.}(2014)\citenamefont
  {Tagliacozzo}, \citenamefont {Celi},\ and\ \citenamefont
  {Lewenstein}}]{Tagliacozzo2014}%
  \BibitemOpen
  \bibfield  {author} {\bibinfo {author} {\bibfnamefont {L.}~\bibnamefont
  {Tagliacozzo}}, \bibinfo {author} {\bibfnamefont {A.}~\bibnamefont {Celi}}, \
  and\ \bibinfo {author} {\bibfnamefont {M.}~\bibnamefont {Lewenstein}},\
  }\href {\doibase 10.1103/PhysRevX.4.041024} {\bibfield  {journal} {\bibinfo
  {journal} {Phys. Rev. X}\ }\textbf {\bibinfo {volume} {4}},\ \bibinfo {pages}
  {041024} (\bibinfo {year} {2014})}\BibitemShut {NoStop}%
\bibitem [{\citenamefont {Zohar}\ \emph
  {et~al.}(2016{\natexlab{a}})\citenamefont {Zohar}, \citenamefont {Wahl},
  \citenamefont {Burrello},\ and\ \citenamefont {Cirac}}]{Zohar2016}%
  \BibitemOpen
  \bibfield  {author} {\bibinfo {author} {\bibfnamefont {E.}~\bibnamefont
  {Zohar}}, \bibinfo {author} {\bibfnamefont {T.~B.}\ \bibnamefont {Wahl}},
  \bibinfo {author} {\bibfnamefont {M.}~\bibnamefont {Burrello}}, \ and\
  \bibinfo {author} {\bibfnamefont {J.~I.}\ \bibnamefont {Cirac}},\ }\href
  {\doibase 10.1016/j.aop.2016.08.008} {\bibfield  {journal} {\bibinfo
  {journal} {Ann. Phys.}\ }\textbf {\bibinfo {volume} {374}},\ \bibinfo {pages}
  {84 } (\bibinfo {year} {2016}{\natexlab{a}})}\BibitemShut {NoStop}%
\bibitem [{\citenamefont {Ba\~{n}uls}\ \emph {et~al.}(2013)\citenamefont
  {Ba\~{n}uls}, \citenamefont {Cichy}, \citenamefont {Jansen},\ and\
  \citenamefont {Cirac}}]{Banuls2013}%
  \BibitemOpen
  \bibfield  {author} {\bibinfo {author} {\bibfnamefont {M.~C.}\ \bibnamefont
  {Ba\~{n}uls}}, \bibinfo {author} {\bibfnamefont {K.}~\bibnamefont {Cichy}},
  \bibinfo {author} {\bibfnamefont {K.}~\bibnamefont {Jansen}}, \ and\ \bibinfo
  {author} {\bibfnamefont {J.~I.}\ \bibnamefont {Cirac}},\ }\href {\doibase
  10.1007/JHEP11(2013)158} {\bibfield  {journal} {\bibinfo  {journal} {J. High
  Energy Phys.}\ }\textbf {\bibinfo {volume} {2013}},\ \bibinfo {pages} {158}
  (\bibinfo {year} {2013})}\BibitemShut {NoStop}%
\bibitem [{\citenamefont {Buyens}\ \emph
  {et~al.}(2014{\natexlab{b}})\citenamefont {Buyens}, \citenamefont {Haegeman},
  \citenamefont {Van~Acoleyen},\ and\ \citenamefont {Verstraete}}]{Buyens2014}%
  \BibitemOpen
  \bibfield  {author} {\bibinfo {author} {\bibfnamefont {B.}~\bibnamefont
  {Buyens}}, \bibinfo {author} {\bibfnamefont {J.}~\bibnamefont {Haegeman}},
  \bibinfo {author} {\bibfnamefont {K.}~\bibnamefont {Van~Acoleyen}}, \ and\
  \bibinfo {author} {\bibfnamefont {F.}~\bibnamefont {Verstraete}},\ }\href
  {http://arxiv.org/abs/1411.0020} {\bibfield  {journal} {\bibinfo  {journal}
  {PoS(LATTICE 2014)308}\ } (\bibinfo {year} {2014}{\natexlab{b}})}\BibitemShut
  {NoStop}%
\bibitem [{\citenamefont {Buyens}\ \emph {et~al.}(2015)\citenamefont {Buyens},
  \citenamefont {Haegeman}, \citenamefont {Verstraete},\ and\ \citenamefont
  {Van~Acoleyen}}]{Buyens2015a}%
  \BibitemOpen
  \bibfield  {author} {\bibinfo {author} {\bibfnamefont {B.}~\bibnamefont
  {Buyens}}, \bibinfo {author} {\bibfnamefont {J.}~\bibnamefont {Haegeman}},
  \bibinfo {author} {\bibfnamefont {F.}~\bibnamefont {Verstraete}}, \ and\
  \bibinfo {author} {\bibfnamefont {K.}~\bibnamefont {Van~Acoleyen}},\ }\href
  {http://arxiv.org/abs/1511.04288} {\bibfield  {journal} {\bibinfo  {journal}
  {PoS(LATTICE 2015)280}\ } (\bibinfo {year} {2015})}\BibitemShut {NoStop}%
\bibitem [{\citenamefont {Ba\~{n}uls}\ \emph {et~al.}(2017)\citenamefont
  {Ba\~{n}uls}, \citenamefont {Cichy}, \citenamefont {Cirac}, \citenamefont
  {Jansen}, \citenamefont {K\"uhn},\ and\ \citenamefont {Saito}}]{Banuls2016c}%
  \BibitemOpen
  \bibfield  {author} {\bibinfo {author} {\bibfnamefont {M.~C.}\ \bibnamefont
  {Ba\~{n}uls}}, \bibinfo {author} {\bibfnamefont {K.}~\bibnamefont {Cichy}},
  \bibinfo {author} {\bibfnamefont {J.~I.}\ \bibnamefont {Cirac}}, \bibinfo
  {author} {\bibfnamefont {K.}~\bibnamefont {Jansen}}, \bibinfo {author}
  {\bibfnamefont {S.}~\bibnamefont {K\"uhn}}, \ and\ \bibinfo {author}
  {\bibfnamefont {H.}~\bibnamefont {Saito}},\ }\href {\doibase
  10.1051/epjconf/201713704001} {\bibfield  {journal} {\bibinfo  {journal} {EPJ
  Web Conf.}\ }\textbf {\bibinfo {volume} {137}},\ \bibinfo {pages} {04001}
  (\bibinfo {year} {2017})}\BibitemShut {NoStop}%
\bibitem [{\citenamefont {Ba\~{n}uls}\ \emph {et~al.}(2015)\citenamefont
  {Ba\~{n}uls}, \citenamefont {Cichy}, \citenamefont {Cirac}, \citenamefont
  {Jansen},\ and\ \citenamefont {Saito}}]{Banuls2015}%
  \BibitemOpen
  \bibfield  {author} {\bibinfo {author} {\bibfnamefont {M.~C.}\ \bibnamefont
  {Ba\~{n}uls}}, \bibinfo {author} {\bibfnamefont {K.}~\bibnamefont {Cichy}},
  \bibinfo {author} {\bibfnamefont {J.~I.}\ \bibnamefont {Cirac}}, \bibinfo
  {author} {\bibfnamefont {K.}~\bibnamefont {Jansen}}, \ and\ \bibinfo {author}
  {\bibfnamefont {H.}~\bibnamefont {Saito}},\ }\href {\doibase
  10.1103/PhysRevD.92.034519} {\bibfield  {journal} {\bibinfo  {journal} {Phys.
  Rev. D}\ }\textbf {\bibinfo {volume} {92}},\ \bibinfo {pages} {034519}
  (\bibinfo {year} {2015})}\BibitemShut {NoStop}%
\bibitem [{\citenamefont {Ba\~{n}uls}\ \emph
  {et~al.}(2016{\natexlab{a}})\citenamefont {Ba\~{n}uls}, \citenamefont
  {Cichy}, \citenamefont {Jansen},\ and\ \citenamefont {Saito}}]{Banuls2016}%
  \BibitemOpen
  \bibfield  {author} {\bibinfo {author} {\bibfnamefont {M.~C.}\ \bibnamefont
  {Ba\~{n}uls}}, \bibinfo {author} {\bibfnamefont {K.}~\bibnamefont {Cichy}},
  \bibinfo {author} {\bibfnamefont {K.}~\bibnamefont {Jansen}}, \ and\ \bibinfo
  {author} {\bibfnamefont {H.}~\bibnamefont {Saito}},\ }\href {\doibase
  10.1103/PhysRevD.93.094512} {\bibfield  {journal} {\bibinfo  {journal} {Phys.
  Rev. D}\ }\textbf {\bibinfo {volume} {93}},\ \bibinfo {pages} {094512}
  (\bibinfo {year} {2016}{\natexlab{a}})}\BibitemShut {NoStop}%
\bibitem [{\citenamefont {Saito}\ \emph {et~al.}(2014)\citenamefont {Saito},
  \citenamefont {Ba\~{n}uls}, \citenamefont {Cichy}, \citenamefont {Cirac},\
  and\ \citenamefont {Jansen}}]{Saito2014}%
  \BibitemOpen
  \bibfield  {author} {\bibinfo {author} {\bibfnamefont {H.}~\bibnamefont
  {Saito}}, \bibinfo {author} {\bibfnamefont {M.~C.}\ \bibnamefont
  {Ba\~{n}uls}}, \bibinfo {author} {\bibfnamefont {K.}~\bibnamefont {Cichy}},
  \bibinfo {author} {\bibfnamefont {J.~I.}\ \bibnamefont {Cirac}}, \ and\
  \bibinfo {author} {\bibfnamefont {K.}~\bibnamefont {Jansen}},\ }\href
  {http://arxiv.org/abs/1412.0596} {\bibfield  {journal} {\bibinfo  {journal}
  {PoS(LATTICE 2014)302}\ } (\bibinfo {year} {2014})}\BibitemShut {NoStop}%
\bibitem [{\citenamefont {Saito}\ \emph {et~al.}(2015)\citenamefont {Saito},
  \citenamefont {Ba\~{n}uls}, \citenamefont {Cichy}, \citenamefont {Cirac},\
  and\ \citenamefont {Jansen}}]{Saito2015}%
  \BibitemOpen
  \bibfield  {author} {\bibinfo {author} {\bibfnamefont {H.}~\bibnamefont
  {Saito}}, \bibinfo {author} {\bibfnamefont {M.~C.}\ \bibnamefont
  {Ba\~{n}uls}}, \bibinfo {author} {\bibfnamefont {K.}~\bibnamefont {Cichy}},
  \bibinfo {author} {\bibfnamefont {J.~I.}\ \bibnamefont {Cirac}}, \ and\
  \bibinfo {author} {\bibfnamefont {K.}~\bibnamefont {Jansen}},\ }\href
  {http://arxiv.org/abs/1511.00794} {\bibfield  {journal} {\bibinfo  {journal}
  {PoS(LATTICE 2015)283}\ } (\bibinfo {year} {2015})}\BibitemShut {NoStop}%
\bibitem [{\citenamefont {Buyens}\ \emph
  {et~al.}(2016{\natexlab{a}})\citenamefont {Buyens}, \citenamefont
  {Verstraete},\ and\ \citenamefont {Van~Acoleyen}}]{Buyens2016}%
  \BibitemOpen
  \bibfield  {author} {\bibinfo {author} {\bibfnamefont {B.}~\bibnamefont
  {Buyens}}, \bibinfo {author} {\bibfnamefont {F.}~\bibnamefont {Verstraete}},
  \ and\ \bibinfo {author} {\bibfnamefont {K.}~\bibnamefont {Van~Acoleyen}},\
  }\href {\doibase 10.1103/PhysRevD.94.085018} {\bibfield  {journal} {\bibinfo
  {journal} {Phys. Rev. D}\ }\textbf {\bibinfo {volume} {94}},\ \bibinfo
  {pages} {085018} (\bibinfo {year} {2016}{\natexlab{a}})}\BibitemShut
  {NoStop}%
\bibitem [{\citenamefont {Troyer}\ and\ \citenamefont
  {Wiese}(2005)}]{Troyer2005}%
  \BibitemOpen
  \bibfield  {author} {\bibinfo {author} {\bibfnamefont {M.}~\bibnamefont
  {Troyer}}\ and\ \bibinfo {author} {\bibfnamefont {U.-J.}\ \bibnamefont
  {Wiese}},\ }\href {\doibase 10.1103/PhysRevLett.94.170201} {\bibfield
  {journal} {\bibinfo  {journal} {Phys. Rev. Lett.}\ }\textbf {\bibinfo
  {volume} {94}},\ \bibinfo {pages} {170201} (\bibinfo {year}
  {2005})}\BibitemShut {NoStop}%
\bibitem [{\citenamefont {K\"uhn}\ \emph {et~al.}(2015)\citenamefont {K\"uhn},
  \citenamefont {Zohar}, \citenamefont {Cirac},\ and\ \citenamefont
  {Ba\~{n}uls}}]{Kuehn2015}%
  \BibitemOpen
  \bibfield  {author} {\bibinfo {author} {\bibfnamefont {S.}~\bibnamefont
  {K\"uhn}}, \bibinfo {author} {\bibfnamefont {E.}~\bibnamefont {Zohar}},
  \bibinfo {author} {\bibfnamefont {J.}~\bibnamefont {Cirac}}, \ and\ \bibinfo
  {author} {\bibfnamefont {M.~C.}\ \bibnamefont {Ba\~{n}uls}},\ }\href
  {\doibase 10.1007/JHEP07(2015)130} {\bibfield  {journal} {\bibinfo  {journal}
  {J. High Energy Phys.}\ }\textbf {\bibinfo {volume} {2015}},\ \bibinfo
  {pages} {130} (\bibinfo {year} {2015})}\BibitemShut {NoStop}%
\bibitem [{\citenamefont {Pichler}\ \emph {et~al.}(2016)\citenamefont
  {Pichler}, \citenamefont {Dalmonte}, \citenamefont {Rico}, \citenamefont
  {Zoller},\ and\ \citenamefont {Montangero}}]{Pichler2015}%
  \BibitemOpen
  \bibfield  {author} {\bibinfo {author} {\bibfnamefont {T.}~\bibnamefont
  {Pichler}}, \bibinfo {author} {\bibfnamefont {M.}~\bibnamefont {Dalmonte}},
  \bibinfo {author} {\bibfnamefont {E.}~\bibnamefont {Rico}}, \bibinfo {author}
  {\bibfnamefont {P.}~\bibnamefont {Zoller}}, \ and\ \bibinfo {author}
  {\bibfnamefont {S.}~\bibnamefont {Montangero}},\ }\href {\doibase
  10.1103/PhysRevX.6.011023} {\bibfield  {journal} {\bibinfo  {journal} {Phys.
  Rev. X}\ }\textbf {\bibinfo {volume} {6}},\ \bibinfo {pages} {011023}
  (\bibinfo {year} {2016})}\BibitemShut {NoStop}%
\bibitem [{\citenamefont {Buyens}\ \emph
  {et~al.}(2016{\natexlab{b}})\citenamefont {Buyens}, \citenamefont {Haegeman},
  \citenamefont {Hebenstreit}, \citenamefont {Verstraete},\ and\ \citenamefont
  {Van~Acoleyen}}]{Buyens2016b}%
  \BibitemOpen
  \bibfield  {author} {\bibinfo {author} {\bibfnamefont {B.}~\bibnamefont
  {Buyens}}, \bibinfo {author} {\bibfnamefont {J.}~\bibnamefont {Haegeman}},
  \bibinfo {author} {\bibfnamefont {F.}~\bibnamefont {Hebenstreit}}, \bibinfo
  {author} {\bibfnamefont {F.}~\bibnamefont {Verstraete}}, \ and\ \bibinfo
  {author} {\bibfnamefont {K.}~\bibnamefont {Van~Acoleyen}},\ }\href
  {https://arxiv.org/abs/1612.00739} {\bibfield  {journal} {\bibinfo  {journal}
  {arXiv:1612.00739}\ } (\bibinfo {year} {2016}{\natexlab{b}})}\BibitemShut
  {NoStop}%
\bibitem [{\citenamefont {Silvi}\ \emph {et~al.}(2017)\citenamefont {Silvi},
  \citenamefont {Rico}, \citenamefont {Dalmonte}, \citenamefont {Tschirsich},\
  and\ \citenamefont {Montangero}}]{Silvi2016}%
  \BibitemOpen
  \bibfield  {author} {\bibinfo {author} {\bibfnamefont {P.}~\bibnamefont
  {Silvi}}, \bibinfo {author} {\bibfnamefont {E.}~\bibnamefont {Rico}},
  \bibinfo {author} {\bibfnamefont {M.}~\bibnamefont {Dalmonte}}, \bibinfo
  {author} {\bibfnamefont {F.}~\bibnamefont {Tschirsich}}, \ and\ \bibinfo
  {author} {\bibfnamefont {S.}~\bibnamefont {Montangero}},\ }\href {\doibase
  10.22331/q-2017-04-25-9} {\bibfield  {journal} {\bibinfo  {journal}
  {Quantum}\ }\textbf {\bibinfo {volume} {1}},\ \bibinfo {pages} {9} (\bibinfo
  {year} {2017})}\BibitemShut {NoStop}%
\bibitem [{\citenamefont {Ba\~nuls}\ \emph {et~al.}(2017)\citenamefont
  {Ba\~nuls}, \citenamefont {Cichy}, \citenamefont {Cirac}, \citenamefont
  {Jansen},\ and\ \citenamefont {K\"uhn}}]{Banuls2016a}%
  \BibitemOpen
  \bibfield  {author} {\bibinfo {author} {\bibfnamefont {M.~C.}\ \bibnamefont
  {Ba\~nuls}}, \bibinfo {author} {\bibfnamefont {K.}~\bibnamefont {Cichy}},
  \bibinfo {author} {\bibfnamefont {J.~I.}\ \bibnamefont {Cirac}}, \bibinfo
  {author} {\bibfnamefont {K.}~\bibnamefont {Jansen}}, \ and\ \bibinfo {author}
  {\bibfnamefont {S.}~\bibnamefont {K\"uhn}},\ }\href {\doibase
  10.1103/PhysRevLett.118.071601} {\bibfield  {journal} {\bibinfo  {journal}
  {Phys. Rev. Lett.}\ }\textbf {\bibinfo {volume} {118}},\ \bibinfo {pages}
  {071601} (\bibinfo {year} {2017})}\BibitemShut {NoStop}%
\bibitem [{\citenamefont {Ba\~{n}uls}\ \emph
  {et~al.}(2016{\natexlab{b}})\citenamefont {Ba\~{n}uls}, \citenamefont
  {Cichy}, \citenamefont {Cirac}, \citenamefont {Jansen}, \citenamefont
  {K\"uhn},\ and\ \citenamefont {Saito}}]{Banuls2016b}%
  \BibitemOpen
  \bibfield  {author} {\bibinfo {author} {\bibfnamefont {M.~C.}\ \bibnamefont
  {Ba\~{n}uls}}, \bibinfo {author} {\bibfnamefont {K.}~\bibnamefont {Cichy}},
  \bibinfo {author} {\bibfnamefont {J.~I.}\ \bibnamefont {Cirac}}, \bibinfo
  {author} {\bibfnamefont {K.}~\bibnamefont {Jansen}}, \bibinfo {author}
  {\bibfnamefont {S.}~\bibnamefont {K\"uhn}}, \ and\ \bibinfo {author}
  {\bibfnamefont {H.}~\bibnamefont {Saito}},\ }\href
  {https://arxiv.org/abs/1611.01458} {\bibfield  {journal} {\bibinfo  {journal}
  {PoS(LATTICE 2016)316}\ } (\bibinfo {year} {2016}{\natexlab{b}})}\BibitemShut
  {NoStop}%
\bibitem [{\citenamefont {Zapp}\ and\ \citenamefont {Or\'us}(2017)}]{Zapp2017}%
  \BibitemOpen
  \bibfield  {author} {\bibinfo {author} {\bibfnamefont {K.}~\bibnamefont
  {Zapp}}\ and\ \bibinfo {author} {\bibfnamefont {R.}~\bibnamefont {Or\'us}},\
  }\href {\doibase 10.1103/PhysRevD.95.114508} {\bibfield  {journal} {\bibinfo
  {journal} {Phys. Rev. D}\ }\textbf {\bibinfo {volume} {95}},\ \bibinfo
  {pages} {114508} (\bibinfo {year} {2017})}\BibitemShut {NoStop}%
\bibitem [{\citenamefont {Buyens}\ \emph
  {et~al.}(2016{\natexlab{c}})\citenamefont {Buyens}, \citenamefont {Haegeman},
  \citenamefont {Verschelde}, \citenamefont {Verstraete},\ and\ \citenamefont
  {Van~Acoleyen}}]{Buyens2015}%
  \BibitemOpen
  \bibfield  {author} {\bibinfo {author} {\bibfnamefont {B.}~\bibnamefont
  {Buyens}}, \bibinfo {author} {\bibfnamefont {J.}~\bibnamefont {Haegeman}},
  \bibinfo {author} {\bibfnamefont {H.}~\bibnamefont {Verschelde}}, \bibinfo
  {author} {\bibfnamefont {F.}~\bibnamefont {Verstraete}}, \ and\ \bibinfo
  {author} {\bibfnamefont {K.}~\bibnamefont {Van~Acoleyen}},\ }\href {\doibase
  10.1103/PhysRevX.6.041040} {\bibfield  {journal} {\bibinfo  {journal} {Phys.
  Rev. X}\ }\textbf {\bibinfo {volume} {6}},\ \bibinfo {pages} {041040}
  (\bibinfo {year} {2016}{\natexlab{c}})}\BibitemShut {NoStop}%
\bibitem [{\citenamefont {Zohar}\ and\ \citenamefont
  {Reznik}(2011)}]{Zohar2011}%
  \BibitemOpen
  \bibfield  {author} {\bibinfo {author} {\bibfnamefont {E.}~\bibnamefont
  {Zohar}}\ and\ \bibinfo {author} {\bibfnamefont {B.}~\bibnamefont {Reznik}},\
  }\href {\doibase 10.1103/PhysRevLett.107.275301} {\bibfield  {journal}
  {\bibinfo  {journal} {Phys. Rev. Lett.}\ }\textbf {\bibinfo {volume} {107}},\
  \bibinfo {pages} {275301} (\bibinfo {year} {2011})}\BibitemShut {NoStop}%
\bibitem [{\citenamefont {Banerjee}\ \emph {et~al.}(2012)\citenamefont
  {Banerjee}, \citenamefont {Dalmonte}, \citenamefont {M\"{u}ller},
  \citenamefont {Rico}, \citenamefont {Stebler}, \citenamefont {Wiese},\ and\
  \citenamefont {Zoller}}]{Banerjee2012}%
  \BibitemOpen
  \bibfield  {author} {\bibinfo {author} {\bibfnamefont {D.}~\bibnamefont
  {Banerjee}}, \bibinfo {author} {\bibfnamefont {M.}~\bibnamefont {Dalmonte}},
  \bibinfo {author} {\bibfnamefont {M.}~\bibnamefont {M\"{u}ller}}, \bibinfo
  {author} {\bibfnamefont {E.}~\bibnamefont {Rico}}, \bibinfo {author}
  {\bibfnamefont {P.}~\bibnamefont {Stebler}}, \bibinfo {author} {\bibfnamefont
  {U.-J.}\ \bibnamefont {Wiese}}, \ and\ \bibinfo {author} {\bibfnamefont
  {P.}~\bibnamefont {Zoller}},\ }\href {\doibase
  10.1103/PhysRevLett.109.175302} {\bibfield  {journal} {\bibinfo  {journal}
  {Phys. Rev. Lett.}\ }\textbf {\bibinfo {volume} {109}},\ \bibinfo {pages}
  {175302} (\bibinfo {year} {2012})}\BibitemShut {NoStop}%
\bibitem [{\citenamefont {Banerjee}\ \emph {et~al.}(2013)\citenamefont
  {Banerjee}, \citenamefont {B\"{o}gli}, \citenamefont {Dalmonte},
  \citenamefont {Rico}, \citenamefont {Stebler}, \citenamefont {Wiese},\ and\
  \citenamefont {Zoller}}]{Banerjee2013}%
  \BibitemOpen
  \bibfield  {author} {\bibinfo {author} {\bibfnamefont {D.}~\bibnamefont
  {Banerjee}}, \bibinfo {author} {\bibfnamefont {M.}~\bibnamefont {B\"{o}gli}},
  \bibinfo {author} {\bibfnamefont {M.}~\bibnamefont {Dalmonte}}, \bibinfo
  {author} {\bibfnamefont {E.}~\bibnamefont {Rico}}, \bibinfo {author}
  {\bibfnamefont {P.}~\bibnamefont {Stebler}}, \bibinfo {author} {\bibfnamefont
  {U.-J.}\ \bibnamefont {Wiese}}, \ and\ \bibinfo {author} {\bibfnamefont
  {P.}~\bibnamefont {Zoller}},\ }\href {\doibase
  10.1103/PhysRevLett.110.125303} {\bibfield  {journal} {\bibinfo  {journal}
  {Phys. Rev. Lett.}\ }\textbf {\bibinfo {volume} {110}},\ \bibinfo {pages}
  {125303} (\bibinfo {year} {2013})}\BibitemShut {NoStop}%
\bibitem [{\citenamefont {Zohar}\ \emph {et~al.}(2012)\citenamefont {Zohar},
  \citenamefont {Cirac},\ and\ \citenamefont {Reznik}}]{Zohar2012}%
  \BibitemOpen
  \bibfield  {author} {\bibinfo {author} {\bibfnamefont {E.}~\bibnamefont
  {Zohar}}, \bibinfo {author} {\bibfnamefont {J.~I.}\ \bibnamefont {Cirac}}, \
  and\ \bibinfo {author} {\bibfnamefont {B.}~\bibnamefont {Reznik}},\ }\href
  {\doibase 10.1103/PhysRevLett.109.125302} {\bibfield  {journal} {\bibinfo
  {journal} {Phys. Rev. Lett.}\ }\textbf {\bibinfo {volume} {109}},\ \bibinfo
  {pages} {125302} (\bibinfo {year} {2012})}\BibitemShut {NoStop}%
\bibitem [{\citenamefont {Zohar}\ \emph
  {et~al.}(2013{\natexlab{a}})\citenamefont {Zohar}, \citenamefont {Cirac},\
  and\ \citenamefont {Reznik}}]{Zohar2013c}%
  \BibitemOpen
  \bibfield  {author} {\bibinfo {author} {\bibfnamefont {E.}~\bibnamefont
  {Zohar}}, \bibinfo {author} {\bibfnamefont {J.~I.}\ \bibnamefont {Cirac}}, \
  and\ \bibinfo {author} {\bibfnamefont {B.}~\bibnamefont {Reznik}},\ }\href
  {\doibase 10.1103/PhysRevLett.110.055302} {\bibfield  {journal} {\bibinfo
  {journal} {Phys. Rev. Lett.}\ }\textbf {\bibinfo {volume} {110}},\ \bibinfo
  {pages} {055302} (\bibinfo {year} {2013}{\natexlab{a}})}\BibitemShut
  {NoStop}%
\bibitem [{\citenamefont {Zohar}\ \emph
  {et~al.}(2013{\natexlab{b}})\citenamefont {Zohar}, \citenamefont {Cirac},\
  and\ \citenamefont {Reznik}}]{Zohar2013a}%
  \BibitemOpen
  \bibfield  {author} {\bibinfo {author} {\bibfnamefont {E.}~\bibnamefont
  {Zohar}}, \bibinfo {author} {\bibfnamefont {J.~I.}\ \bibnamefont {Cirac}}, \
  and\ \bibinfo {author} {\bibfnamefont {B.}~\bibnamefont {Reznik}},\ }\href
  {\doibase 10.1103/PhysRevLett.110.125304} {\bibfield  {journal} {\bibinfo
  {journal} {Phys. Rev. Lett.}\ }\textbf {\bibinfo {volume} {110}},\ \bibinfo
  {pages} {125304} (\bibinfo {year} {2013}{\natexlab{b}})}\BibitemShut
  {NoStop}%
\bibitem [{\citenamefont {Zohar}\ \emph
  {et~al.}(2013{\natexlab{c}})\citenamefont {Zohar}, \citenamefont {Cirac},\
  and\ \citenamefont {Reznik}}]{Zohar2013}%
  \BibitemOpen
  \bibfield  {author} {\bibinfo {author} {\bibfnamefont {E.}~\bibnamefont
  {Zohar}}, \bibinfo {author} {\bibfnamefont {J.~I.}\ \bibnamefont {Cirac}}, \
  and\ \bibinfo {author} {\bibfnamefont {B.}~\bibnamefont {Reznik}},\ }\href
  {\doibase 10.1103/PhysRevA.88.023617} {\bibfield  {journal} {\bibinfo
  {journal} {Phys. Rev. A}\ }\textbf {\bibinfo {volume} {88}},\ \bibinfo
  {pages} {023617} (\bibinfo {year} {2013}{\natexlab{c}})}\BibitemShut
  {NoStop}%
\bibitem [{\citenamefont {Marcos}\ \emph {et~al.}(2014)\citenamefont {Marcos},
  \citenamefont {Widmer}, \citenamefont {Rico}, \citenamefont {Hafezi},
  \citenamefont {Rabl}, \citenamefont {Wiese},\ and\ \citenamefont
  {Zoller}}]{Marcos2014}%
  \BibitemOpen
  \bibfield  {author} {\bibinfo {author} {\bibfnamefont {D.}~\bibnamefont
  {Marcos}}, \bibinfo {author} {\bibfnamefont {P.}~\bibnamefont {Widmer}},
  \bibinfo {author} {\bibfnamefont {E.}~\bibnamefont {Rico}}, \bibinfo {author}
  {\bibfnamefont {M.}~\bibnamefont {Hafezi}}, \bibinfo {author} {\bibfnamefont
  {P.}~\bibnamefont {Rabl}}, \bibinfo {author} {\bibfnamefont {U.-J.}\
  \bibnamefont {Wiese}}, \ and\ \bibinfo {author} {\bibfnamefont
  {P.}~\bibnamefont {Zoller}},\ }\href {\doibase 10.1016/j.aop.2014.09.011}
  {\bibfield  {journal} {\bibinfo  {journal} {Ann. Phys.}\ }\textbf {\bibinfo
  {volume} {351}},\ \bibinfo {pages} {634 } (\bibinfo {year}
  {2014})}\BibitemShut {NoStop}%
\bibitem [{\citenamefont {Wiese}(2014)}]{Wiese2014}%
  \BibitemOpen
  \bibfield  {author} {\bibinfo {author} {\bibfnamefont {U.-J.}\ \bibnamefont
  {Wiese}},\ }\href {\doibase 10.1016/j.nuclphysa.2014.09.102} {\bibfield
  {journal} {\bibinfo  {journal} {Nucl. Phys. A}\ }\textbf {\bibinfo {volume}
  {931}},\ \bibinfo {pages} {246 } (\bibinfo {year} {2014})}\BibitemShut
  {NoStop}%
\bibitem [{\citenamefont {Zohar}\ \emph
  {et~al.}(2016{\natexlab{b}})\citenamefont {Zohar}, \citenamefont {Cirac},\
  and\ \citenamefont {Reznik}}]{Zohar2015a}%
  \BibitemOpen
  \bibfield  {author} {\bibinfo {author} {\bibfnamefont {E.}~\bibnamefont
  {Zohar}}, \bibinfo {author} {\bibfnamefont {J.~I.}\ \bibnamefont {Cirac}}, \
  and\ \bibinfo {author} {\bibfnamefont {B.}~\bibnamefont {Reznik}},\ }\href
  {\doibase 10.1088/0034-4885/79/1/014401} {\bibfield  {journal} {\bibinfo
  {journal} {Rep. Prog. Phys.}\ }\textbf {\bibinfo {volume} {79}},\ \bibinfo
  {pages} {014401} (\bibinfo {year} {2016}{\natexlab{b}})}\BibitemShut
  {NoStop}%
\bibitem [{\citenamefont {Mezzacapo}\ \emph {et~al.}(2015)\citenamefont
  {Mezzacapo}, \citenamefont {Rico}, \citenamefont {Sab\'{\i}n}, \citenamefont
  {Egusquiza}, \citenamefont {Lamata},\ and\ \citenamefont
  {Solano}}]{Mezzacapo2015}%
  \BibitemOpen
  \bibfield  {author} {\bibinfo {author} {\bibfnamefont {A.}~\bibnamefont
  {Mezzacapo}}, \bibinfo {author} {\bibfnamefont {E.}~\bibnamefont {Rico}},
  \bibinfo {author} {\bibfnamefont {C.}~\bibnamefont {Sab\'{\i}n}}, \bibinfo
  {author} {\bibfnamefont {I.~L.}\ \bibnamefont {Egusquiza}}, \bibinfo {author}
  {\bibfnamefont {L.}~\bibnamefont {Lamata}}, \ and\ \bibinfo {author}
  {\bibfnamefont {E.}~\bibnamefont {Solano}},\ }\href {\doibase
  10.1103/PhysRevLett.115.240502} {\bibfield  {journal} {\bibinfo  {journal}
  {Phys. Rev. Lett.}\ }\textbf {\bibinfo {volume} {115}},\ \bibinfo {pages}
  {240502} (\bibinfo {year} {2015})}\BibitemShut {NoStop}%
\bibitem [{\citenamefont {Zohar}\ \emph
  {et~al.}(2017{\natexlab{a}})\citenamefont {Zohar}, \citenamefont {Farace},
  \citenamefont {Reznik},\ and\ \citenamefont {Cirac}}]{Zohar2016b}%
  \BibitemOpen
  \bibfield  {author} {\bibinfo {author} {\bibfnamefont {E.}~\bibnamefont
  {Zohar}}, \bibinfo {author} {\bibfnamefont {A.}~\bibnamefont {Farace}},
  \bibinfo {author} {\bibfnamefont {B.}~\bibnamefont {Reznik}}, \ and\ \bibinfo
  {author} {\bibfnamefont {J.~I.}\ \bibnamefont {Cirac}},\ }\href {\doibase
  10.1103/PhysRevA.95.023604} {\bibfield  {journal} {\bibinfo  {journal} {Phys.
  Rev. A}\ }\textbf {\bibinfo {volume} {95}},\ \bibinfo {pages} {023604}
  (\bibinfo {year} {2017}{\natexlab{a}})}\BibitemShut {NoStop}%
\bibitem [{\citenamefont {Zohar}\ \emph
  {et~al.}(2017{\natexlab{b}})\citenamefont {Zohar}, \citenamefont {Farace},
  \citenamefont {Reznik},\ and\ \citenamefont {Cirac}}]{Zohar2016c}%
  \BibitemOpen
  \bibfield  {author} {\bibinfo {author} {\bibfnamefont {E.}~\bibnamefont
  {Zohar}}, \bibinfo {author} {\bibfnamefont {A.}~\bibnamefont {Farace}},
  \bibinfo {author} {\bibfnamefont {B.}~\bibnamefont {Reznik}}, \ and\ \bibinfo
  {author} {\bibfnamefont {J.~I.}\ \bibnamefont {Cirac}},\ }\href {\doibase
  10.1103/PhysRevLett.118.070501} {\bibfield  {journal} {\bibinfo  {journal}
  {Phys. Rev. Lett.}\ }\textbf {\bibinfo {volume} {118}},\ \bibinfo {pages}
  {070501} (\bibinfo {year} {2017}{\natexlab{b}})}\BibitemShut {NoStop}%
\bibitem [{\citenamefont {Brennen}\ \emph {et~al.}(2016)\citenamefont
  {Brennen}, \citenamefont {Pupillo}, \citenamefont {Rico}, \citenamefont
  {Stace},\ and\ \citenamefont {Vodola}}]{Brennen2016}%
  \BibitemOpen
  \bibfield  {author} {\bibinfo {author} {\bibfnamefont {G.~K.}\ \bibnamefont
  {Brennen}}, \bibinfo {author} {\bibfnamefont {G.}~\bibnamefont {Pupillo}},
  \bibinfo {author} {\bibfnamefont {E.}~\bibnamefont {Rico}}, \bibinfo {author}
  {\bibfnamefont {T.~M.}\ \bibnamefont {Stace}}, \ and\ \bibinfo {author}
  {\bibfnamefont {D.}~\bibnamefont {Vodola}},\ }\href {\doibase
  10.1103/PhysRevLett.117.240504} {\bibfield  {journal} {\bibinfo  {journal}
  {Phys. Rev. Lett.}\ }\textbf {\bibinfo {volume} {117}},\ \bibinfo {pages}
  {240504} (\bibinfo {year} {2016})}\BibitemShut {NoStop}%
\bibitem [{\citenamefont {Muschik}\ \emph {et~al.}(2017)\citenamefont
  {Muschik}, \citenamefont {Heyl}, \citenamefont {Martinez}, \citenamefont
  {Monz}, \citenamefont {Schindler}, \citenamefont {Vogell}, \citenamefont
  {Dalmonte}, \citenamefont {Hauke}, \citenamefont {Blatt},\ and\ \citenamefont
  {Zoller}}]{Muschik2016}%
  \BibitemOpen
  \bibfield  {author} {\bibinfo {author} {\bibfnamefont {C.}~\bibnamefont
  {Muschik}}, \bibinfo {author} {\bibfnamefont {M.}~\bibnamefont {Heyl}},
  \bibinfo {author} {\bibfnamefont {E.}~\bibnamefont {Martinez}}, \bibinfo
  {author} {\bibfnamefont {T.}~\bibnamefont {Monz}}, \bibinfo {author}
  {\bibfnamefont {P.}~\bibnamefont {Schindler}}, \bibinfo {author}
  {\bibfnamefont {B.}~\bibnamefont {Vogell}}, \bibinfo {author} {\bibfnamefont
  {M.}~\bibnamefont {Dalmonte}}, \bibinfo {author} {\bibfnamefont
  {P.}~\bibnamefont {Hauke}}, \bibinfo {author} {\bibfnamefont
  {R.}~\bibnamefont {Blatt}}, \ and\ \bibinfo {author} {\bibfnamefont
  {P.}~\bibnamefont {Zoller}},\ }\href {\doibase 10.1088/1367-2630/aa89ab}
  {\bibfield  {journal} {\bibinfo  {journal} {New J. Phys.}\ }\textbf {\bibinfo
  {volume} {19}},\ \bibinfo {pages} {103020} (\bibinfo {year}
  {2017})}\BibitemShut {NoStop}%
\bibitem [{\citenamefont {Gonz\'{a}lez-Cuadra}\ \emph
  {et~al.}(2017)\citenamefont {Gonz\'{a}lez-Cuadra}, \citenamefont {Zohar},\
  and\ \citenamefont {Cirac}}]{Gonzalez-Cuadra2017}%
  \BibitemOpen
  \bibfield  {author} {\bibinfo {author} {\bibfnamefont {D.}~\bibnamefont
  {Gonz\'{a}lez-Cuadra}}, \bibinfo {author} {\bibfnamefont {E.}~\bibnamefont
  {Zohar}}, \ and\ \bibinfo {author} {\bibfnamefont {J.~I.}\ \bibnamefont
  {Cirac}},\ }\href {\doibase 10.1088/1367-2630/aa6f37} {\bibfield  {journal}
  {\bibinfo  {journal} {New J. Phys.}\ }\textbf {\bibinfo {volume} {19}},\
  \bibinfo {pages} {063038} (\bibinfo {year} {2017})}\BibitemShut {NoStop}%
\bibitem [{\citenamefont {Martinez}\ \emph {et~al.}(2016)\citenamefont
  {Martinez}, \citenamefont {Muschik}, \citenamefont {Schindler}, \citenamefont
  {Nigg}, \citenamefont {Erhard}, \citenamefont {Heyl}, \citenamefont {Hauke},
  \citenamefont {Dalmonte}, \citenamefont {Monz}, \citenamefont {Zoller},\ and\
  \citenamefont {Blatt}}]{Martinez2016}%
  \BibitemOpen
  \bibfield  {author} {\bibinfo {author} {\bibfnamefont {E.~A.}\ \bibnamefont
  {Martinez}}, \bibinfo {author} {\bibfnamefont {C.~A.}\ \bibnamefont
  {Muschik}}, \bibinfo {author} {\bibfnamefont {P.}~\bibnamefont {Schindler}},
  \bibinfo {author} {\bibfnamefont {D.}~\bibnamefont {Nigg}}, \bibinfo {author}
  {\bibfnamefont {A.}~\bibnamefont {Erhard}}, \bibinfo {author} {\bibfnamefont
  {M.}~\bibnamefont {Heyl}}, \bibinfo {author} {\bibfnamefont {P.}~\bibnamefont
  {Hauke}}, \bibinfo {author} {\bibfnamefont {M.}~\bibnamefont {Dalmonte}},
  \bibinfo {author} {\bibfnamefont {T.}~\bibnamefont {Monz}}, \bibinfo {author}
  {\bibfnamefont {P.}~\bibnamefont {Zoller}}, \ and\ \bibinfo {author}
  {\bibfnamefont {R.}~\bibnamefont {Blatt}},\ }\href {\doibase
  10.1038/nature18318} {\bibfield  {journal} {\bibinfo  {journal} {Nature}\
  }\textbf {\bibinfo {volume} {534}},\ \bibinfo {pages} {516–519} (\bibinfo
  {year} {2016})}\BibitemShut {NoStop}%
\bibitem [{\citenamefont {Milsted}(2016)}]{Milsted2016}%
  \BibitemOpen
  \bibfield  {author} {\bibinfo {author} {\bibfnamefont {A.}~\bibnamefont
  {Milsted}},\ }\href {\doibase 10.1103/PhysRevD.93.085012} {\bibfield
  {journal} {\bibinfo  {journal} {Phys. Rev. D}\ }\textbf {\bibinfo {volume}
  {93}},\ \bibinfo {pages} {085012} (\bibinfo {year} {2016})}\BibitemShut
  {NoStop}%
\bibitem [{\citenamefont {Horn}(1981)}]{Horn1981}%
  \BibitemOpen
  \bibfield  {author} {\bibinfo {author} {\bibfnamefont {D.}~\bibnamefont
  {Horn}},\ }\href {\doibase 10.1016/0370-2693(81)90763-2} {\bibfield
  {journal} {\bibinfo  {journal} {Phys. Lett. B}\ }\textbf {\bibinfo {volume}
  {100}},\ \bibinfo {pages} {149 } (\bibinfo {year} {1981})}\BibitemShut
  {NoStop}%
\bibitem [{\citenamefont {Orland}\ and\ \citenamefont
  {Rohrlich}(1990)}]{Orland1990}%
  \BibitemOpen
  \bibfield  {author} {\bibinfo {author} {\bibfnamefont {P.}~\bibnamefont
  {Orland}}\ and\ \bibinfo {author} {\bibfnamefont {D.}~\bibnamefont
  {Rohrlich}},\ }\href {\doibase 10.1016/0550-3213(90)90646-U} {\bibfield
  {journal} {\bibinfo  {journal} {Nucl. Phys. B}\ }\textbf {\bibinfo {volume}
  {338}},\ \bibinfo {pages} {647 } (\bibinfo {year} {1990})}\BibitemShut
  {NoStop}%
\bibitem [{\citenamefont {Chandrasekharan}\ and\ \citenamefont
  {Wiese}(1997)}]{Chandrasekharan1997}%
  \BibitemOpen
  \bibfield  {author} {\bibinfo {author} {\bibfnamefont {S.}~\bibnamefont
  {Chandrasekharan}}\ and\ \bibinfo {author} {\bibfnamefont {U.-J.}\
  \bibnamefont {Wiese}},\ }\href {\doibase 10.1016/S0550-3213(97)80041-7}
  {\bibfield  {journal} {\bibinfo  {journal} {Nucl. Phys. B}\ }\textbf
  {\bibinfo {volume} {492}},\ \bibinfo {pages} {455 } (\bibinfo {year}
  {1997})}\BibitemShut {NoStop}%
\bibitem [{\citenamefont {Brower}\ \emph {et~al.}(1999)\citenamefont {Brower},
  \citenamefont {Chandrasekharan},\ and\ \citenamefont {Wiese}}]{Brower1999}%
  \BibitemOpen
  \bibfield  {author} {\bibinfo {author} {\bibfnamefont {R.}~\bibnamefont
  {Brower}}, \bibinfo {author} {\bibfnamefont {S.}~\bibnamefont
  {Chandrasekharan}}, \ and\ \bibinfo {author} {\bibfnamefont {U.-J.}\
  \bibnamefont {Wiese}},\ }\href {\doibase 10.1103/PhysRevD.60.094502}
  {\bibfield  {journal} {\bibinfo  {journal} {Phys. Rev. D}\ }\textbf {\bibinfo
  {volume} {60}},\ \bibinfo {pages} {094502} (\bibinfo {year}
  {1999})}\BibitemShut {NoStop}%
\bibitem [{\citenamefont {Hamer}(1982)}]{Hamer1982a}%
  \BibitemOpen
  \bibfield  {author} {\bibinfo {author} {\bibfnamefont {C.}~\bibnamefont
  {Hamer}},\ }\href {\doibase 10.1016/0550-3213(82)90009-8} {\bibfield
  {journal} {\bibinfo  {journal} {Nucl. Phys. B}\ }\textbf {\bibinfo {volume}
  {195}},\ \bibinfo {pages} {503 } (\bibinfo {year} {1982})}\BibitemShut
  {NoStop}%
\bibitem [{\citenamefont {Ryu}\ and\ \citenamefont
  {Takayanagi}(2006)}]{Ryu2006}%
  \BibitemOpen
  \bibfield  {author} {\bibinfo {author} {\bibfnamefont {S.}~\bibnamefont
  {Ryu}}\ and\ \bibinfo {author} {\bibfnamefont {T.}~\bibnamefont
  {Takayanagi}},\ }\href {\doibase 10.1103/PhysRevLett.96.181602} {\bibfield
  {journal} {\bibinfo  {journal} {Phys. Rev. Lett.}\ }\textbf {\bibinfo
  {volume} {96}},\ \bibinfo {pages} {181602} (\bibinfo {year}
  {2006})}\BibitemShut {NoStop}%
\bibitem [{\citenamefont {Van~Raamsdonk}(2010)}]{VanRaamsdonk2010}%
  \BibitemOpen
  \bibfield  {author} {\bibinfo {author} {\bibfnamefont {M.}~\bibnamefont
  {Van~Raamsdonk}},\ }\href {\doibase 10.1007/s10714-010-1034-0} {\bibfield
  {journal} {\bibinfo  {journal} {Gen. Relativ. Gravit.}\ }\textbf {\bibinfo
  {volume} {42}},\ \bibinfo {pages} {2323} (\bibinfo {year}
  {2010})}\BibitemShut {NoStop}%
\bibitem [{\citenamefont {Casini}\ \emph {et~al.}(2014)\citenamefont {Casini},
  \citenamefont {Huerta},\ and\ \citenamefont {Rosabal}}]{Casini2014}%
  \BibitemOpen
  \bibfield  {author} {\bibinfo {author} {\bibfnamefont {H.}~\bibnamefont
  {Casini}}, \bibinfo {author} {\bibfnamefont {M.}~\bibnamefont {Huerta}}, \
  and\ \bibinfo {author} {\bibfnamefont {J.~A.}\ \bibnamefont {Rosabal}},\
  }\href {\doibase 10.1103/PhysRevD.89.085012} {\bibfield  {journal} {\bibinfo
  {journal} {Phys. Rev. D}\ }\textbf {\bibinfo {volume} {89}},\ \bibinfo
  {pages} {085012} (\bibinfo {year} {2014})}\BibitemShut {NoStop}%
\bibitem [{\citenamefont {Ghosh}\ \emph {et~al.}(2015)\citenamefont {Ghosh},
  \citenamefont {Soni},\ and\ \citenamefont {Trivedi}}]{Ghosh2015}%
  \BibitemOpen
  \bibfield  {author} {\bibinfo {author} {\bibfnamefont {S.}~\bibnamefont
  {Ghosh}}, \bibinfo {author} {\bibfnamefont {R.~M.}\ \bibnamefont {Soni}}, \
  and\ \bibinfo {author} {\bibfnamefont {S.~P.}\ \bibnamefont {Trivedi}},\
  }\href {\doibase 10.1007/JHEP09(2015)069} {\bibfield  {journal} {\bibinfo
  {journal} {J. High Energy Phys.}\ }\textbf {\bibinfo {volume} {2015}},\
  \bibinfo {pages} {69} (\bibinfo {year} {2015})}\BibitemShut {NoStop}%
\bibitem [{\citenamefont {Soni}\ and\ \citenamefont
  {Trivedi}(2016)}]{Soni2016}%
  \BibitemOpen
  \bibfield  {author} {\bibinfo {author} {\bibfnamefont {R.~M.}\ \bibnamefont
  {Soni}}\ and\ \bibinfo {author} {\bibfnamefont {S.~P.}\ \bibnamefont
  {Trivedi}},\ }\href {\doibase 10.1007/JHEP01(2016)136} {\bibfield  {journal}
  {\bibinfo  {journal} {J. High Energy Phys.}\ }\textbf {\bibinfo {volume}
  {2016}},\ \bibinfo {pages} {136} (\bibinfo {year} {2016})}\BibitemShut
  {NoStop}%
\bibitem [{\citenamefont {Van~Acoleyen}\ \emph {et~al.}(2016)\citenamefont
  {Van~Acoleyen}, \citenamefont {Bultinck}, \citenamefont {Haegeman},
  \citenamefont {Marien}, \citenamefont {Scholz},\ and\ \citenamefont
  {Verstraete}}]{VanAcoleyen2016}%
  \BibitemOpen
  \bibfield  {author} {\bibinfo {author} {\bibfnamefont {K.}~\bibnamefont
  {Van~Acoleyen}}, \bibinfo {author} {\bibfnamefont {N.}~\bibnamefont
  {Bultinck}}, \bibinfo {author} {\bibfnamefont {J.}~\bibnamefont {Haegeman}},
  \bibinfo {author} {\bibfnamefont {M.}~\bibnamefont {Marien}}, \bibinfo
  {author} {\bibfnamefont {V.~B.}\ \bibnamefont {Scholz}}, \ and\ \bibinfo
  {author} {\bibfnamefont {F.}~\bibnamefont {Verstraete}},\ }\href {\doibase
  10.1103/PhysRevLett.117.131602} {\bibfield  {journal} {\bibinfo  {journal}
  {Phys. Rev. Lett.}\ }\textbf {\bibinfo {volume} {117}},\ \bibinfo {pages}
  {131602} (\bibinfo {year} {2016})}\BibitemShut {NoStop}%
\bibitem [{\citenamefont {Aoki}\ \emph {et~al.}(2017)\citenamefont {Aoki},
  \citenamefont {Iizuka}, \citenamefont {Tamaoka},\ and\ \citenamefont
  {Yokoya}}]{Aoki2017}%
  \BibitemOpen
  \bibfield  {author} {\bibinfo {author} {\bibfnamefont {S.}~\bibnamefont
  {Aoki}}, \bibinfo {author} {\bibfnamefont {N.}~\bibnamefont {Iizuka}},
  \bibinfo {author} {\bibfnamefont {K.}~\bibnamefont {Tamaoka}}, \ and\
  \bibinfo {author} {\bibfnamefont {T.}~\bibnamefont {Yokoya}},\ }\href
  {\doibase 10.1103/PhysRevD.96.045020} {\bibfield  {journal} {\bibinfo
  {journal} {Phys. Rev. D}\ }\textbf {\bibinfo {volume} {96}},\ \bibinfo
  {pages} {045020} (\bibinfo {year} {2017})}\BibitemShut {NoStop}%
\bibitem [{\citenamefont {Kogut}\ and\ \citenamefont
  {Susskind}(1975)}]{Kogut1975}%
  \BibitemOpen
  \bibfield  {author} {\bibinfo {author} {\bibfnamefont {J.}~\bibnamefont
  {Kogut}}\ and\ \bibinfo {author} {\bibfnamefont {L.}~\bibnamefont
  {Susskind}},\ }\href {\doibase 10.1103/PhysRevD.11.395} {\bibfield  {journal}
  {\bibinfo  {journal} {Phys. Rev. D}\ }\textbf {\bibinfo {volume} {11}},\
  \bibinfo {pages} {395} (\bibinfo {year} {1975})}\BibitemShut {NoStop}%
\bibitem [{\citenamefont {Hamer}(1977)}]{Hamer1977}%
  \BibitemOpen
  \bibfield  {author} {\bibinfo {author} {\bibfnamefont {C.}~\bibnamefont
  {Hamer}},\ }\href {\doibase 10.1016/0550-3213(77)90334-0} {\bibfield
  {journal} {\bibinfo  {journal} {Nucl. Phys. B}\ }\textbf {\bibinfo {volume}
  {121}},\ \bibinfo {pages} {159 } (\bibinfo {year} {1977})}\BibitemShut
  {NoStop}%
\bibitem [{Note1()}]{Note1}%
  \BibitemOpen
  \bibinfo {note} {Notice that the values of $j$ represent the total angular
  momentum corresponding to the quantum rigid rotor on that link and, thus, are
  positive.}\BibitemShut {Stop}%
\bibitem [{\citenamefont {Nielsen}\ and\ \citenamefont
  {Chuang}(2004)}]{Nielsen2004}%
  \BibitemOpen
  \bibfield  {author} {\bibinfo {author} {\bibfnamefont {M.~A.}\ \bibnamefont
  {Nielsen}}\ and\ \bibinfo {author} {\bibfnamefont {I.~L.}\ \bibnamefont
  {Chuang}},\ }\href@noop {} {\emph {\bibinfo {title} {{Q}uantum {C}omputation
  and {Q}uantum {I}nformation ({C}ambridge {S}eries on {I}nformation and the
  {N}atural {S}ciences)}}}\ (\bibinfo  {publisher} {Cambridge University
  Press},\ \bibinfo {year} {2004})\BibitemShut {NoStop}%
\bibitem [{\citenamefont {Perez-Garcia}\ \emph {et~al.}(2007)\citenamefont
  {Perez-Garcia}, \citenamefont {Verstraete}, \citenamefont {Wolf},\ and\
  \citenamefont {Cirac}}]{Perez-Garcia2007}%
  \BibitemOpen
  \bibfield  {author} {\bibinfo {author} {\bibfnamefont {D.}~\bibnamefont
  {Perez-Garcia}}, \bibinfo {author} {\bibfnamefont {F.}~\bibnamefont
  {Verstraete}}, \bibinfo {author} {\bibfnamefont {M.~M.}\ \bibnamefont
  {Wolf}}, \ and\ \bibinfo {author} {\bibfnamefont {J.~I.}\ \bibnamefont
  {Cirac}},\ }\href {http://www.rintonpress.com/xqic7/qic-7-56/401-430.pdf}
  {\bibfield  {journal} {\bibinfo  {journal} {Quantum Inf. Comput.}\ }\textbf
  {\bibinfo {volume} {7}},\ \bibinfo {pages} {401} (\bibinfo {year}
  {2007})}\BibitemShut {NoStop}%
\bibitem [{\citenamefont {Verstraete}\ \emph {et~al.}(2008)\citenamefont
  {Verstraete}, \citenamefont {Murg},\ and\ \citenamefont
  {Cirac}}]{Verstraete2008}%
  \BibitemOpen
  \bibfield  {author} {\bibinfo {author} {\bibfnamefont {F.}~\bibnamefont
  {Verstraete}}, \bibinfo {author} {\bibfnamefont {V.}~\bibnamefont {Murg}}, \
  and\ \bibinfo {author} {\bibfnamefont {J.}~\bibnamefont {Cirac}},\ }\href
  {\doibase 10.1080/14789940801912366} {\bibfield  {journal} {\bibinfo
  {journal} {Adv. Phys.}\ }\textbf {\bibinfo {volume} {57}},\ \bibinfo {pages}
  {143} (\bibinfo {year} {2008})}\BibitemShut {NoStop}%
\bibitem [{\citenamefont {Schollw\"{o}ck}(2011)}]{Schollwoeck2011}%
  \BibitemOpen
  \bibfield  {author} {\bibinfo {author} {\bibfnamefont {U.}~\bibnamefont
  {Schollw\"{o}ck}},\ }\href {\doibase 10.1016/j.aop.2010.09.012} {\bibfield
  {journal} {\bibinfo  {journal} {Ann. Phys.}\ }\textbf {\bibinfo {volume}
  {326}},\ \bibinfo {pages} {96 } (\bibinfo {year} {2011})},\ \bibinfo {note}
  {january 2011 Special Issue}\BibitemShut {NoStop}%
\bibitem [{\citenamefont {Or{\'u}s}(2014)}]{Orus2014a}%
  \BibitemOpen
  \bibfield  {author} {\bibinfo {author} {\bibfnamefont {R.}~\bibnamefont
  {Or{\'u}s}},\ }\href {\doibase 10.1016/j.aop.2014.06.013} {\bibfield
  {journal} {\bibinfo  {journal} {Ann. Phys.}\ }\textbf {\bibinfo {volume}
  {349}},\ \bibinfo {pages} {117 } (\bibinfo {year} {2014})}\BibitemShut
  {NoStop}%
\bibitem [{\citenamefont {Verstraete}\ \emph {et~al.}(2004)\citenamefont
  {Verstraete}, \citenamefont {Porras},\ and\ \citenamefont
  {Cirac}}]{Verstraete2004}%
  \BibitemOpen
  \bibfield  {author} {\bibinfo {author} {\bibfnamefont {F.}~\bibnamefont
  {Verstraete}}, \bibinfo {author} {\bibfnamefont {D.}~\bibnamefont {Porras}},
  \ and\ \bibinfo {author} {\bibfnamefont {J.~I.}\ \bibnamefont {Cirac}},\
  }\href {\doibase 10.1103/PhysRevLett.93.227205} {\bibfield  {journal}
  {\bibinfo  {journal} {Phys. Rev. Lett.}\ }\textbf {\bibinfo {volume} {93}},\
  \bibinfo {pages} {227205} (\bibinfo {year} {2004})}\BibitemShut {NoStop}%
\bibitem [{\citenamefont {Stathopoulos}\ and\ \citenamefont
  {McCombs}(2010)}]{Stathopoulos2010}%
  \BibitemOpen
  \bibfield  {author} {\bibinfo {author} {\bibfnamefont {A.}~\bibnamefont
  {Stathopoulos}}\ and\ \bibinfo {author} {\bibfnamefont {J.~R.}\ \bibnamefont
  {McCombs}},\ }\href {\doibase 10.1145/1731022.1731031} {\bibfield  {journal}
  {\bibinfo  {journal} {ACM Trans. Math. Softw.}\ }\textbf {\bibinfo {volume}
  {37}},\ \bibinfo {pages} {21:1} (\bibinfo {year} {2010})}\BibitemShut
  {NoStop}%
\bibitem [{\citenamefont {Wu}\ \emph {et~al.}(2016)\citenamefont {Wu},
  \citenamefont {Romero},\ and\ \citenamefont {Stathopoulos}}]{Wu2016}%
  \BibitemOpen
  \bibfield  {author} {\bibinfo {author} {\bibfnamefont {L.}~\bibnamefont
  {Wu}}, \bibinfo {author} {\bibfnamefont {E.}~\bibnamefont {Romero}}, \ and\
  \bibinfo {author} {\bibfnamefont {A.}~\bibnamefont {Stathopoulos}},\ }\href
  {https://arxiv.org/abs/1607.01404} {\bibfield  {journal} {\bibinfo  {journal}
  {arXiv:1607.01404}\ } (\bibinfo {year} {2016})}\BibitemShut {NoStop}%
\bibitem [{\citenamefont {McCulloch}(2007)}]{McCulloch2007}%
  \BibitemOpen
  \bibfield  {author} {\bibinfo {author} {\bibfnamefont {I.~P.}\ \bibnamefont
  {McCulloch}},\ }\href {\doibase 10.1088/1742-5468/2007/10/P10014} {\bibfield
  {journal} {\bibinfo  {journal} {J. Stat. Mech.}\ }\textbf {\bibinfo {volume}
  {2007}},\ \bibinfo {pages} {P10014} (\bibinfo {year} {2007})}\BibitemShut
  {NoStop}%
\bibitem [{\citenamefont {Steinhardt}(1980)}]{Steinhardt1980}%
  \BibitemOpen
  \bibfield  {author} {\bibinfo {author} {\bibfnamefont {P.~J.}\ \bibnamefont
  {Steinhardt}},\ }\href {\doibase 10.1016/0550-3213(80)90065-6} {\bibfield
  {journal} {\bibinfo  {journal} {Nucl. Phys. B}\ }\textbf {\bibinfo {volume}
  {176}},\ \bibinfo {pages} {100 } (\bibinfo {year} {1980})}\BibitemShut
  {NoStop}%
\bibitem [{\citenamefont {Calabrese}\ and\ \citenamefont
  {Cardy}(2004)}]{Calabrese2004}%
  \BibitemOpen
  \bibfield  {author} {\bibinfo {author} {\bibfnamefont {P.}~\bibnamefont
  {Calabrese}}\ and\ \bibinfo {author} {\bibfnamefont {J.}~\bibnamefont
  {Cardy}},\ }\href {http://stacks.iop.org/1742-5468/2004/i=06/a=P06002}
  {\bibfield  {journal} {\bibinfo  {journal} {J. Stat. Mech.}\ }\textbf
  {\bibinfo {volume} {2004}},\ \bibinfo {pages} {P06002} (\bibinfo {year}
  {2004})}\BibitemShut {NoStop}%
\bibitem [{\citenamefont {Rothe}(2006)}]{Rothe2006}%
  \BibitemOpen
  \bibfield  {author} {\bibinfo {author} {\bibfnamefont {H.~J.}\ \bibnamefont
  {Rothe}},\ }\href@noop {} {\emph {\bibinfo {title} {{L}attice gauge theories:
  an introduction}}}\ (\bibinfo  {publisher} {World Scientific Lecture Notes in
  Physics},\ \bibinfo {year} {2006})\BibitemShut {NoStop}%
\bibitem [{\citenamefont {Laflorencie}\ \emph {et~al.}(2006)\citenamefont
  {Laflorencie}, \citenamefont {S\o{}rensen}, \citenamefont {Chang},\ and\
  \citenamefont {Affleck}}]{Laflorencie2006}%
  \BibitemOpen
  \bibfield  {author} {\bibinfo {author} {\bibfnamefont {N.}~\bibnamefont
  {Laflorencie}}, \bibinfo {author} {\bibfnamefont {E.~S.}\ \bibnamefont
  {S\o{}rensen}}, \bibinfo {author} {\bibfnamefont {M.-S.}\ \bibnamefont
  {Chang}}, \ and\ \bibinfo {author} {\bibfnamefont {I.}~\bibnamefont
  {Affleck}},\ }\href {\doibase 10.1103/PhysRevLett.96.100603} {\bibfield
  {journal} {\bibinfo  {journal} {Phys. Rev. Lett.}\ }\textbf {\bibinfo
  {volume} {96}},\ \bibinfo {pages} {100603} (\bibinfo {year}
  {2006})}\BibitemShut {NoStop}%
\bibitem [{\citenamefont {Calabrese}\ \emph {et~al.}(2010)\citenamefont
  {Calabrese}, \citenamefont {Campostrini}, \citenamefont {Essler},\ and\
  \citenamefont {Nienhuis}}]{Calabrese2010}%
  \BibitemOpen
  \bibfield  {author} {\bibinfo {author} {\bibfnamefont {P.}~\bibnamefont
  {Calabrese}}, \bibinfo {author} {\bibfnamefont {M.}~\bibnamefont
  {Campostrini}}, \bibinfo {author} {\bibfnamefont {F.}~\bibnamefont {Essler}},
  \ and\ \bibinfo {author} {\bibfnamefont {B.}~\bibnamefont {Nienhuis}},\
  }\href {\doibase 10.1103/PhysRevLett.104.095701} {\bibfield  {journal}
  {\bibinfo  {journal} {Phys. Rev. Lett.}\ }\textbf {\bibinfo {volume} {104}},\
  \bibinfo {pages} {095701} (\bibinfo {year} {2010})}\BibitemShut {NoStop}%
\bibitem [{\citenamefont {Buyens}\ \emph {et~al.}(2017)\citenamefont {Buyens},
  \citenamefont {Montangero}, \citenamefont {Haegeman}, \citenamefont
  {Verstraete},\ and\ \citenamefont {Van~Acoleyen}}]{Buyens2017}%
  \BibitemOpen
  \bibfield  {author} {\bibinfo {author} {\bibfnamefont {B.}~\bibnamefont
  {Buyens}}, \bibinfo {author} {\bibfnamefont {S.}~\bibnamefont {Montangero}},
  \bibinfo {author} {\bibfnamefont {J.}~\bibnamefont {Haegeman}}, \bibinfo
  {author} {\bibfnamefont {F.}~\bibnamefont {Verstraete}}, \ and\ \bibinfo
  {author} {\bibfnamefont {K.}~\bibnamefont {Van~Acoleyen}},\ }\href {\doibase
  10.1103/PhysRevD.95.094509} {\bibfield  {journal} {\bibinfo  {journal} {Phys.
  Rev. D}\ }\textbf {\bibinfo {volume} {95}},\ \bibinfo {pages} {094509}
  (\bibinfo {year} {2017})}\BibitemShut {NoStop}%
\bibitem [{\citenamefont {Dalmonte}\ and\ \citenamefont
  {Montangero}(2016)}]{Dalmonte2016}%
  \BibitemOpen
  \bibfield  {author} {\bibinfo {author} {\bibfnamefont {M.}~\bibnamefont
  {Dalmonte}}\ and\ \bibinfo {author} {\bibfnamefont {S.}~\bibnamefont
  {Montangero}},\ }\href {\doibase 10.1080/00107514.2016.1151199} {\bibfield
  {journal} {\bibinfo  {journal} {Contemp. Phys.}\ }\textbf {\bibinfo {volume}
  {57}},\ \bibinfo {pages} {388} (\bibinfo {year} {2016})}\BibitemShut
  {NoStop}%
\bibitem [{\citenamefont {Ligterink}\ \emph {et~al.}(2000)\citenamefont
  {Ligterink}, \citenamefont {Walet},\ and\ \citenamefont
  {Bishop}}]{Ligterink2000}%
  \BibitemOpen
  \bibfield  {author} {\bibinfo {author} {\bibfnamefont {N.}~\bibnamefont
  {Ligterink}}, \bibinfo {author} {\bibfnamefont {N.}~\bibnamefont {Walet}}, \
  and\ \bibinfo {author} {\bibfnamefont {R.}~\bibnamefont {Bishop}},\ }\href
  {\doibase 10.1006/aphy.2000.6070} {\bibfield  {journal} {\bibinfo  {journal}
  {Ann. Phys.}\ }\textbf {\bibinfo {volume} {284}},\ \bibinfo {pages} {215 }
  (\bibinfo {year} {2000})}\BibitemShut {NoStop}%
\bibitem [{\citenamefont {Deutsch}(1999)}]{Deutsch1999}%
  \BibitemOpen
  \bibfield  {author} {\bibinfo {author} {\bibfnamefont {E.}~\bibnamefont
  {Deutsch}},\ }\href {\doibase 10.1016/S0012-365X(98)00371-9} {\bibfield
  {journal} {\bibinfo  {journal} {Discrete Mathematics}\ }\textbf {\bibinfo
  {volume} {204}},\ \bibinfo {pages} {167 } (\bibinfo {year}
  {1999})}\BibitemShut {NoStop}%
\end{thebibliography}%
\end{document}